\documentclass[11pt]{article}
\pdfoutput=1
\usepackage{jheppub}
\usepackage{amsmath,amssymb,mathabx,bbm}
\usepackage{enumerate}
\usepackage[table]{xcolor}
\usepackage{graphics,placeins}
\usepackage{tikz-feynman}
\usepackage{subcaption}
\usepackage{multirow}
\usepackage{mathtools}

\newcommand{\G}{  \bar{G} }

\newcommand{\con}{  \bar{c}_{\phi} }

\newcommand{\comment}[1]{}
\newcommand{\LambdaP}{\tilde{\Lambda}}

\newcommand{\ccell}{\cellcolor{blue!8}}

\begin{document}

\title{Scalar Fields Near Compact Objects: \\ Resummation versus UV Completion}

\author{Anne-Christine Davis}
\author{and Scott Melville}
\affiliation{DAMTP, Center for Mathematical Sciences, University of Cambridge, CB3 0WA, UK}

\emailAdd{a.c.davis@damtp.cam.ac.uk}
\emailAdd{scott.melville@damtp.cam.ac.uk}

\abstract{
Low-energy effective field theories containing a light scalar field are used extensively in cosmology, but often there is a tension between embedding such theories in a healthy UV completion and achieving a phenomenologically viable screening mechanism in the IR. \\
Here, we identify the range of interaction couplings which allow for a smooth resummation of classical non-linearities (necessary for kinetic/Vainshtein-type screening), and compare this with the range allowed by unitarity, causality and locality in the underlying UV theory.
The latter region is identified using positivity bounds on the $2\to2$ scattering amplitude, and in particular by considering scattering about a non-trivial background for the scalar we are able to place constraints on interactions at all orders in the field (beyond quartic order). 
We identify two classes of theories can both exhibit screening and satisfy existing positivity bounds, namely scalar-tensor theories of $P(X)$ or quartic Horndeski type in which the leading interaction contains an odd power of $X$. 
Finally, for the quartic DBI Galileon (equivalent to a disformally coupled scalar in the Einstein frame),
the analogous resummation can be performed near two-body systems and imposing positivity constraints introduces a non-perturbative ambiguity in the screened scalar profile. 
These results will guide future searches for UV complete models which exhibit screening of fifth forces in the IR. 
}


\maketitle

\section{Introduction}
\label{sec:intro}

Light scalar fields are an essential building block of theoretical cosmology.
Since General Relativity (GR) is only an effective description of gravity at low energies (much below the Planck scale), and suffers from the well-known cosmological constant problem \cite{Weinberg:1988cp} when accounting for the observed late-time acceleration \cite{Riess:1998cb, Perlmutter:1998np}, it cannot be the fundamental description of our Universe.
This has led to much interest in theories which go beyond the tensor polarisations of GR by including additional light scalar degrees of freedom \cite{Capozziello:2007ec, Capozziello:2011et, Clifton:2011jh, Joyce:2014kja, Bull:2015stt, Koyama:2015vza}. Such scalar-tensor theories are versatile enough to construct a diverse range of models for the dark sector (for instance dark energy \cite{Chiba:1999ka, ArmendarizPicon:2000dh, ArmendarizPicon:2000ah, Boisseau:2000pr, Copeland:2006wr, Bamba:2012cp} and dark matter \cite{Sin:1992bg, Hu:2000ke, Burgess:2000yq, Bekenstein:2004ne}), and form the basis of model-independent explorations of modified gravity effects in linear cosmology \cite{Gubitosi:2012hu,Bloomfield:2012ff,Gleyzes:2014rba,Bellini:2014fua}.

~
 
However, any coupling between an additional light scalar field and matter generically introduces a long-range fifth force which is not observed in solar system tests of gravity \cite{Will:2001mx, Bertotti:2003rm, Williams:2004qba}. 
The resolution often proposed is to exploit a screening mechanism which suppresses the scalar field on small scales \cite{Vainshtein:1972sx, Damour:1994zq, Khoury:2003aq, Khoury:2003rn, Brax:2004qh, Nicolis:2008in, Babichev:2009ee, Khoury:2010xi}, evading solar system tests but allowing large effects on cosmological scales (see \cite{Burrage:2016bwy,Burrage:2017qrf, Sakstein:2018fwz,Baker:2019gxo} for modern reviews).  
Central to the majority of these screening mechanisms is a non-linear self-interaction, which dominates the scalar's equation of motion when sufficiently close to any source. 
In the language of classical perturbation theory, in which one expands order by order about the linearised solution, this corresponds to resumming an infinite number of tree-level Feynman diagrams. 
This resummation is not always possible: whether or not the linearised solution at large distances can smoothly interpolate to the screened non-linear solution at small distances (i.e. whether or not the perturbative series can be analytically continued beyond its radius of convergence) depends on the sign of the self-interaction coupling. For instance, an interaction $\mathcal{L} \supset + \lambda_2 (\partial \phi)^{4}$ in the Lagrangian can only provide screening if $\lambda_2 < 0$ \cite{Dvali:2012zc, Brax:2014gra}. 

~

As a quantum theory, the non-renormalisable scalar self-interactions which lead to screening also lead to a violation of perturbative unitarity at high energies where the theory becomes strongly coupled, i.e. where loop corrections become as large as the tree-level diagrams. 
From the viewpoint of a low-energy effective field theory (EFT), 
the interactions arise as a result of removing (integrating out) heavy UV physics above some cutoff, and  in order to probe scales near or above the cutoff one must reintroduce that UV physics.
But just as it is not always possible to resum tree-level diagrams and find a smooth field configuration that interpolates between large and small distances, it is also not always possible to resum all quantum corrections into a consistent UV completion. 
What could seem a perfectly consistent (low-energy) EFT may not have any healthy (high-energy) UV completion. To ensure that the underlying UV theory respects fundamental properties---such as unitarity, causality and locality---the low-energy EFT must satisfy ``positivity bounds'', which place constraints on the signs of the interaction couplings (see \cite{Adams:2006sv,Jenkins:2006ia,  Adams:2008hp, Nicolis:2009qm, Bellazzini:2014waa, Bellazzini:2015cra, Baumann:2015nta,  Cheung:2016yqr,Bonifacio:2016wcb,deRham:2017avq,deRham:2017zjm,deRham:2018qqo,Bellazzini:2016xrt,deRham:2017imi,Bellazzini:2017fep,deRham:2017xox, Melville:2019tdc, Alberte:2019xfh, Alberte:2019zhd,Bellazzini:2020cot,Tolley:2020gtv,Caron-Huot:2020cmc, Arkani-Hamed:2020blm, Herrero-Valea:2020wxz, Alberte:2020jsk, Alberte:2020bdz, Wang:2020xlt} for many recent advances in these EFT bounds and their consequences for dark energy and modified gravity). For instance, an interaction $\mathcal{L} \supset + \lambda_2 (\partial \phi )^4$ is only compatible with a standard  UV completion if $\lambda_2 > 0$ \cite{Adams:2006sv}. 

~

In this work, we bring these two strands together and compare the signs of EFT coefficients which allow for classical resummation/screening, and the signs which could be compatible with a UV completion of the quantum theory. 
Clearly, for the simple self-interaction $\lambda_2 ( \partial \phi )^4$, positivity bounds from the UV ($\lambda_2 > 0$) are orthogonal to the requirement for screening in the IR theory ($\lambda_2 < 0$) \cite{Dvali:2012zc}, but is this always the case?  
This question is crucial if we are to understand the possible extensions beyond GR, since it sheds light on the coupling constants of the dark sector both theoretically (i.e. that they are consistent with UV unitarity, causality and locality) and phenomenologically (i.e. that they allow for screened fifth forces in the solar system). 
Since the scale at which the theory becomes strongly coupled and the scale at which classical non-linearities dominate the equation of motion are generically separated \cite{deRham:2014wfa}, there are three distinct regimes: at large distances the scalar field is described by its linearised classical equation of motion (and can have significant effects on cosmology), at small distances the scalar is described by its non-linear classical equation of motion (and can be efficiently screened by the non-linearities), and at very small distances the EFT breaks down and we must use the full UV-complete theory.  
Our goal is to identify the region of parameter space in which it is possible to smoothly connect these three regimes. 

~

Throughout we will be working within a particular class of scalar-tensor EFTs known as Horndeski \cite{Horndeski:1974wa,Deffayet:2009wt} (or Weakly Broken Galileon \cite{Pirtskhalava:2015nla}) theories. Specifically, we focus on, 
\begin{align}\label{quarticA}
S = \int d^4x \sqrt{-g} \Big\{   G_{4}  M_P^2 R   +  M_P \Lambda^3 \left( G_{2}   -  2 G_{4,X}  \frac{ \phi^\mu_\mu \phi^\nu_\nu - \phi_\mu^\nu \phi^\mu_\nu }{\Lambda^6}  \right)    + \mathcal{L}_{\rm matter} [ \psi , C(\phi) g_{\mu\nu} ] \Big\}\,,
\end{align}
where $G_2$ and $G_4$ are dimensionless functions of the ratio $X \equiv \phi_\mu \phi^\mu/(M_P\Lambda^3)$ with $\phi_\mu \equiv \nabla_\mu \phi$ and $\phi_{\mu\nu} \equiv \nabla_{\mu} \nabla_\nu \phi$. 
$\mathcal{L}_{\rm matter} [ \psi , C(\phi) g_{\mu\nu} ]$ indicates the Lagrangian for matter fields $\psi$ and we allow for a conformal coupling to the scalar field $\phi$ in this frame. 
$M_P$ and $\Lambda$ are constant scales which characterise the EFT\footnote{
While the multi-messenger event GW170817 \cite{TheLIGOScientific:2017qsa,Monitor:2017mdv,GBM:2017lvd} constrains the speed of gravitational waves at LIGO frequencies ($\sim 10^2$Hz) to be that of light within one part in $10^{15}$, which would place exceptionally tight constraints on $G_4$  \cite{Lombriser:2015sxa,Lombriser:2016yzn,Creminelli:2017sry, Sakstein:2017xjx, Ezquiaga:2017ekz, Baker:2017hug, Akrami:2018yjz, Heisenberg:2017qka, BeltranJimenez:2018ymu}, typically for dark energy the scale $\Lambda$ at which the EFT breaks down is close to this LIGO scale \cite{deRham:2018red}, and so here we consider \eqref{quarticA} as an acceptable low-energy EFT for the sub-LIGO scales relevant for dark energy and the late-time acceleration of the Universe.
}, and the power counting in \eqref{quarticA} is protected by Galileon invariance---although this symmetry is broken by gravitational corrections, since these are suppressed by factors of $M_P$ the hierarchy $ \Lambda \ll M_P$ is radiatively stable \cite{Luty:2003vm,Nicolis:2004qq,deRham:2010eu,Burrage:2010cu}.

~

Scalar-tensor theories of the form \eqref{quarticA} are particularly well-studied. For instance, the $G_2$ function alone captures $P(X)$ ($K$-essence) theories \cite{ArmendarizPicon:1999rj, Garriga:1999vw}, and when the non-linearities in $\partial \phi$ become large these theories exhibit the $K$-mouflage screening mechanism \cite{Babichev:2009ee}. The $G_4$ Horndeski function captures the (quartic) covariant Galileon \cite{Nicolis:2008in, Deffayet:2009wt}, and when the non-linearity in $\partial \partial \phi$ becomes large this interaction can lead to Vainshtein screening \cite{Vainshtein:1972sx, Koyama:2013paa}. Positivity bounds have also recently been applied to \eqref{quarticA} using scattering around a flat Minkowski background to place constraints on $G_{2}$ and $G_{4}$  \cite{Melville:2019wyy, deRham:2021fpu} (see also \cite{Kennedy:2020ehn}). This makes \eqref{quarticA} a natural arena within which to investigate the interplay between screening in the IR and consistency (unitarity, causality, locality) in the UV.

~

To date, the vast majority of positivity-type arguments which connect the UV and the IR have relied on properties of the $2$- and $4$-particle scattering amplitudes, and therefore are limited to the lowest derivatives $G_{2,XX}$, $G_{4,X}$ and $G_{4,XX}$ only.
Exploring the higher-order $X$ dependence of these functions is particularly important in the context of screening, where these higher-point interactions can have a significant effect.  
In this work, we pioneer a new way of constraining higher-point interactions: using the 4-particle amplitude  \emph{around a non-trivial background}. 
For instance, expanding $\phi \, = \, \alpha t + \varphi$, the resulting $\varphi \varphi \to \varphi \varphi$ amplitude can be used to constrain higher-point interactions, thanks to the positivity bounds recently developed in \cite{Grall:2021xxm} for such (boost-breaking) backgrounds. 
By exploiting these bounds, we are able to place bounds on \emph{all} higher order derivatives of $G_2$ and $G_4$---these are summarised in Table~\ref{table:summary}.  
This approach is complementary to the recent $n$-point positivity bounds of \cite{Chandrasekaran:2018qmx}\footnote{
see also \cite{Logunov:1977xb, Elvang:2012st} and more recently \cite{Herrero-Valea:2021dry}.
}, which have also been used to constrain single $X^N$ interactions in a $P(X)$ theory. 

\bgroup
\setlength{\tabcolsep}{15pt}
\begin{table}
\small\centering
\begin{tabular}{ | r | ccccc |}
\hline
$P(X) \sim$ 
&  
& $\lambda_2 X^2$ 
& $\lambda_3 X^3$ 
& ... 
& $\lambda_n X^n$ 
\\
\hline
Screening
&
& $\lambda_2 < 0$ 
& $\lambda_3 < 0$ 
& ... 
& $\lambda_n < 0$ 
\\
UV Completion
&
& $\lambda_2 > 0$ 
& $\lambda_3 < 0$ 
&... 
& $ (-1)^n \lambda_n > 0$ 
\\
Subluminal SWs
&
& $\lambda_2 > 0$ 
& $\lambda_3 < 0$ 
&...
& $ (-1)^n \lambda_n > 0$ 
\\
\hline  \hline
$G_4 (X) \sim$ 
& $- \beta_1 X$   
& $-\beta_2 X^2$ 
& $-\beta_3 X^3$ 
& ... 
& $-\beta_n X^n$ 
\\
\hline 
Screening
&\ccell $\beta_1 \gtrless 0^* $
& $\beta_2 < 0$
& $\beta_3 < 0$
& ... 
& $\beta_n < 0$
\\
UV Completion
& \ccell $\beta_1 > 0$
& $\beta_2 > 0$
& $\beta_3 < 0$
& ...
&  $(-1)^n \beta_n > 0$
\\
Subluminal GWs
& \ccell $\beta_1 < 0$ 
& $\beta_2 > 0$ 
& $\beta_3 < 0$ 
& ... 
& $ (-1)^n \beta_n > 0$ 
\\
\hline
\end{tabular}
\caption{
Signs required for screening, standard UV completion and subluminal scalar/gravitational waves (SWs/GWs) in $P(X)$ and quartic Horndeski theories when a single interaction dominates. 
The highlighted $\beta_1 X$ interaction does not conform to the $(-1)^n \beta_n$ pattern of the $n > 1$ positivity bounds, and only provides screening around two or more compact objects (this screened profile is smooth for either sign but is only unique if $\beta_1 < 0$). 
}
\label{table:summary}
\end{table}

\subsubsection*{Summary of Results} 
\begin{itemize}

\item[(i)] For a simple $P(X)$ theory in which one particular $X^N$ dominates, i.e. \eqref{quarticA} with 
 $G_2 (X) = -\tfrac{1}{2} X + \lambda_N X^N$ (and $G_4 = 0$)\footnote{
or a constant $G_4 (X)$ and working to leading order in the decoupling limit $M_P\to\infty$ with $\Lambda$ fixed. 
}, we show that,

\begin{itemize}

\item classical perturbation theory near a compact object can only be resummed into a smooth field profile when $\lambda_N < 0$ (this is consistent with similar observations made in \cite{Brax:2014gra} from a different perspective).

\item positivity bounds from $2\to2$ scattering on the background $X=-\alpha^2$  require that $(-1)^N \lambda_N > 0$ when $\alpha$ is small, which coincides with the bound from $N \to N$ scattering on the $X=0$ background \cite{Chandrasekaran:2018qmx}, and for a general $P(X)$ theory at finite $\alpha$ this bound becomes  \eqref{eqn:pos_PX}.

\item it is therefore impossible to UV complete a $K$-mouflage screening mechanism which relies on a large $X^{2n}$ interaction (without sacrificing one of the basic positivity assumptions), however there is no such obstruction for a large $X^{2n+1}$ interaction.

\end{itemize}

\item[(ii)] For the scalar-tensor theory \eqref{quarticA} with the following $G_4$ function,
\begin{align}
 G_4 (X) = \sqrt{ 1 - \beta_1 X  - \beta_{N} X^{N}  } + \mathcal{O} \left( X^{N+1} \right)  \; .
 \label{eqn:G4sq_intro}
\end{align}
where the higher order $X^{N+1}$ terms are suppressed by the weakly broken Galileon power counting, we show that, 

\begin{itemize} 

\item the square root structure ensures a precise cancellation between the scalar self-interactions and scalar-tensor interactions leading to a strong coupling scale $\Lambda_{\rm sc}^{4N-2} \sim M_P^{N-2} \Lambda^{3N}$ set by an effective interaction $\beta_{N} X^{N-1} \left( \phi^{\mu}_{\mu} \phi^\nu_\nu - \phi_\mu^\nu \phi_\nu^\mu  \right)$ for any value of $\beta_1$ (this can be understood as a disformal field redefinition of the $\beta_1 = 0$ theory, $G_4 = 1 - \tfrac{1}{2} \beta_N X^N + \mathcal{O} \left( X^{N+1} \right)$).

\item this $\beta_N$ interaction can be resummed near a compact object of mass $m$ only if $\beta_{N} < 0$, which leads to Vainshtein screening inside a radius $ \Lambda_{\rm sc}^{-1} \left( m/M_P \right)^{N/(4N-2)} $.

\item positivity bounds from $2\to2$ scattering on the background $X=-\alpha^2$  require that $(-1)^N \beta_N > 0$ when $\alpha$ is small, which coincides with the bound from $2 \to 2$ scattering on the $X=0$ background when $N=2$ \cite{Melville:2019wyy}, and for a general $G_4 (X)$ theory at finite $\alpha$ this bound becomes \eqref{eqn:pos_G4XX}.

\item it is therefore impossible to UV complete a Vainshtein screening mechanism which relies on a large $X^{2n}$ interaction in $G_4$ (without sacrificing one of the basic positivity assumptions), however there is no such obstruction for a large $X^{2n+1}$ interaction.

\end{itemize}

\item[(iii)] Finally, we consider \eqref{eqn:G4sq_intro} when $\beta_N = 0$ (which has the highest strong coupling scale, $M_P \Lambda^3$). This $G_4(X) = \sqrt{1 - \beta_1 X}$ theory is also known as the (quartic) DBI Galileon, and can be recast in the Einstein frame as a disformal coupling to matter. 
It was shown recently in \cite{Davis:2019ltc} that resummation can take place in two-body systems and lead to ``ladder screening''. Here we show that,

\begin{itemize}

\item this resummation can only be unique if $\beta_1 < 0$, otherwise there is a non-perturbative correction whose value is not fixed by the boundary condition at infinity,

\item positivity bounds from scalar-matter scattering require that $\beta_1 > 0$ \cite{deRham:2021fpu}, and for a general $G_4 (X)$ theory at finite $\alpha$ this bound becomes \eqref{eqn:pos_G4X}.

\item therefore a disformally coupled scalar EFT must contain a non-perturbative ambiguity near binary systems to be compatible with unitarity, causality and locality in the UV. 

\end{itemize} 

\end{itemize}

We will begin in section~\ref{sec:PX} by analysing a $P(X)$ theory (neglecting any coupling to gravity), reviewing how the resummation of a perturbative series solution about a single compact object leads to the $K$-mouflage screening mechanism for particular signs of the EFT couplings, which can be in conflict with the positivity bounds required for a standard (Lorentz invariant, unitary, causal, local) UV completion.
Then in section~\ref{sec:horndeski} we turn to the quartic Horndeski theory~\eqref{quarticA}, identifying the strong coupling scale (taking account of scalar-tensor mixing) in section~\ref{sec:cutoff}, resumming the classical perturbative series near a one- (/two-)body system to produce Vainshtein (/ladder) screening in section~\ref{sec:screening}, and finally derive new positivity bounds required of $G_4$ for a standard UV completion in section~\ref{sec:positivity}. We conclude in section~\ref{sec:disc} and collect algebraic details of the scattering amplitudes, positivity bounds and disformal field redefinitions in the Appendices.
Throughout we will be considering a flat Minkowski spacetime background and will neglect any backreaction from the scalar field on this geometry (this amounts to keeping the scalar background $X = -\alpha^2$ sufficiently small)---the effects of a cosmological background metric will be discussed elsewhere.

\section{$P(X)$ Theories}
\label{sec:PX}

%
%

We begin by considering simple effective field theories for a single scalar field with derivative self-interactions which have at most one derivative per field, namely Lagrangians of the form $\mathcal{L} = P(X)$, where in this section $X = \eta^{\mu\nu} \phi_\mu \phi_\nu / \LambdaP^4$ is the canonical kinetic term on a fixed Minkowski background and $\LambdaP^4$ represents the EFT cutoff ($=M_P \Lambda^3$ in the context of \eqref{quarticA}). 
Such EFTs have been used extensively in theoretical cosmology, for instance $K$-inflation models of the early Universe \cite{ArmendarizPicon:1999rj, Garriga:1999vw}, $K$-essence models of the late Universe \cite{Chiba:1999ka, ArmendarizPicon:2000dh, ArmendarizPicon:2000ah}, as well as the ghost condensate \cite{ArkaniHamed:2003uy}.
Since a general $P(X)$ theory can be viewed as the leading terms in a derivative expansion of any scalar field theory with a shift symmetry ($\phi \to \phi + c$) they naturally arise in a variety of other contexts as well: for instance as the EFT of a Nambu–Goldstone mode or as the effective action of a superfluid \cite{Greiter:1989qb, Son:2002zn}. 
 
In this section, we revisit the kinetic screening (or ``$K$-mouflage'') mechanism that occurs in $P(X)$ theories from the perspective of resumming a perturbative series expansion, 
and compare this with the constraints placed on the EFT couplings by the existence of a unitarity, causal and local UV completion. 
While many of the intermediate results have appeared elsewhere, the overall conclusion that $K$-mouflage screening can only be UV completed for odd powers of $X$ is novel and has important implications for future model-building.

\subsection{Strong Coupling and Classical Non-linearity}
\label{sec:PX_cutoff}

We must first distinguish carefully between two scales: the scale at which the theory becomes strongly coupled (dominated by quantum effects), and the scale at which the theory becomes non-linear (dominated by classical non-linearities). 
To illustrate the key ideas as simply as possible, we will focus on the effect of a single $X^{N+1}$ term in the Lagrangian, i.e. 
\begin{align}
P(X) = -\tfrac{1}{2} X + \lambda_{N+1} X^{N+1} \; , 
\label{eqn:PN}
\end{align}
with $N \geq 1$ (a general $P(X)$ theory is discussed in Appendix~\ref{app:pos_PX}).

\paragraph{Power Counting.}
While one could simply assume that the corrections from any other term in $P(X)$ are small, a systematic way to quantify this is to adopt a power counting in which higher-order interactions are suppressed by a small parameter $\epsilon$. For instance, 
\begin{align}
\mathcal{L}   =  - \frac{1}{2} X + \frac{ \LambdaP^4}{\epsilon^2} K \left(  \epsilon X  \right)   \equiv -\frac{1}{2} X +  \frac{ \LambdaP^4}{\epsilon^2}  \sum_{n=2}^\infty  \lambda_n \left( \epsilon X \right)^n   \; ,
\label{eqn:Kexp}
\end{align}
where each coupling constant $\lambda_n$ is order unity or smaller.
In contrast to a basic dimensional power counting (i.e. $P(X) = - \tfrac{1}{2} X + \sum_n \lambda_n X^n$ with order unity $\lambda_n$) this more general power counting also captures UV completions in which there is some hierarchy that leads to the $X^n$ interactions appearing at different scales. 
This mimics the power counting of the $G_4$ terms in the action \eqref{quarticA}, where the weakly broken Galileon symmetry leads to a hierarchy between the different interactions\footnote{
As discussed in \cite{Goon:2016ihr}, a power counting scheme of the  form \eqref{eqn:Kexp} is radiatively stable against quantum corrections since any loop must introduce at least four additional derivatives, so only $\partial^4 X^n$ and higher-derivative terms are renormalised in this EFT.  
}. 
The simple $P(X) = - \tfrac{1}{2} X + \lambda_{N+1} X^{N+1}$ theory that we consider below can be viewed as an EFT of the form \eqref{eqn:Kexp} subject to a finite number of tunings (i.e. $\lambda_n =0$ for $2 \leq n \leq N$ to remove the lower-order terms and $\epsilon \ll 1$ to suppress the higher-order terms)---this language is useful because it closely parallels the $G_4$ interactions that we consider in section~\ref{sec:horndeski}.

\paragraph{Strong Coupling Scale.}
Beyond the scale $\LambdaP$, the size of the quantum corrections to $\mathcal{L}$ become comparable to $\mathcal{L}$ itself---the theory becomes strongly coupled and requires UV completion (or an infinite resummation of loops) to be predictive, see e.g. \cite{deRham:2014wfa}.
With the power counting \eqref{eqn:Kexp}, when $\epsilon \ll 1$ it is the loops of the $\lambda_2 X^2$ interaction which lead to strong coupling at $\LambdaP$, but note that if this interaction were removed (by setting $\lambda_2 = 0$) then the next-to-leading $\lambda_3 X^3$ interaction would lead to strong coupling at a parametrically higher scale $\LambdaP^4 / \sqrt{\epsilon}$ ($\gg \LambdaP^4$ when $\epsilon \ll 1$). 
Explicitly, each $(\partial \phi)^{2n}$ non-linearity is suppressed by the scale $\epsilon^{- \frac{n-2}{n-1}} \LambdaP^4$, so the strong coupling scale can be systematically raised by tuning to zero successive couplings,
\begin{align}
 \lambda_2 = \lambda_3 = ... = \lambda_{N} = 0 \;\;\;\; \Rightarrow \;\;\;\; \text{Strong coupling at } \Lambda_{\rm sc}^4 \sim  \LambdaP^4 \epsilon^{- 1 + 1/N } \; ,   
 \label{eqn:KRaise}
\end{align} 
where the $\sim$ indicates that we have neglected order one combinatoric factors and couplings. 
This is a somewhat trivial observation from the point of view of \eqref{eqn:PN}, which becomes strongly coupled at $\Lambda_{\rm sc}^{4N-4} \sim \LambdaP^{4N-4} / \lambda_n$ (so making $\lambda_n$ smaller by factors of $\epsilon$ will seem to ``raise'' the strong coupling scale), but we wish to highlight that in the context of a power counting like \eqref{eqn:Kexp} the strong coupling scale can be raised from $\LambdaP^4$ as high as $\LambdaP^4 / \epsilon$ by turning off the lowest lying interactions, since this is the analogue of raising the strong coupling scale from $\Lambda^4$ to $M_P \Lambda^3$ for the weakly broken Galileon \eqref{quarticA} that we will see in section~\ref{sec:horndeski} (where $\Lambda/M_P \ll 1$ plays the role of $\epsilon$).

\paragraph{Coupling to Matter.}
Now consider adding to \eqref{eqn:Kexp} a conformal coupling to matter, $\frac{ \phi }{ M } \eta^{\mu\nu} T_{\mu\nu}$. In particular, for a static, spherically symmetric compact object we can model the stress-energy as that of a point particle, with trace $\eta^{\mu\nu} T_{\mu\nu} = \frac{m}{4 \pi} \delta^3 \left( \mathbf{r} \right) $. 
In terms of the radial coordinate $r = |\mathbf{r}|$ (the spatial distance from the object in its rest frame), the field sourced by this stress-energy can be written as,
\begin{align}
 \phi' (r) = - \frac{m}{4 \pi M r^2} G_\phi (r)
 \label{eqn:Gphi_def}
\end{align}
representing the usual Newtonian force law modulated by an effective coupling $G_\phi (r)$, which is determined by the equation of motion for $\phi$,
\begin{align}
 G_\phi \left[ 1  -  \frac{2}{\epsilon} K'  \left(   \tfrac{ \epsilon R_K^4 }{ r^4 } G_\phi^2  \right) \right]   =   1 
\label{eqn:KGphi}
\end{align} 
where we have introduced the length scale $R_K$, defined by,
\begin{align}
 R_K^2 \LambdaP^2 = \frac{m}{4 \pi M} \; .
 \label{eqn:Rstar}
\end{align}
At distances  $r$ much smaller than $R_K$, the non-linear terms in $K$ can become large and dominate the classical equation of motion. 
Note that since typically $m / M \gg 1$ (e.g. for $M$ the Planck mass and $m$ the mass of an astrophysical body), $R_K^2 \gg 1/\LambdaP^2 $ and there is a regime in which the theory is dominated by these non-linearities and yet remains weakly coupled from the point of view of the quantum theory \cite{deRham:2014wfa}.

\paragraph{Classical Perturbation Theory.}
For the simple case of $P(X) = -\tfrac{1}{2} X + \lambda_{N+1} X^{N+1}$, the equation of motion \eqref{eqn:KGphi} becomes,
\begin{align}
 G_\phi - z \, G_{\phi}^{2N+1}   =   1 
\label{eqn:KGphiEom}
\end{align} 
where we have introduced the dimensionless parameter,
\begin{align}
z =  \frac{ \lambda_{N+1} }{\epsilon} \left( \frac{\epsilon R_K^4}{r^4} \right)^N \; .
\label{eqn:z_PX}
\end{align}
At sufficiently large distances from the compact object, $r \gg R_K$, then this parameter $z$ is small and can be used to organise a perturbative expansion,  $G_\phi = \sum_{n=0}^\infty G_\phi^{(n)}$, around the linearised solution $G_\phi^{(0)} = 1$ (which corresponds to the usual Newtonian profile for $\phi (r)$). This series solution can be depicted as summing over tree-level (one-point) Feynman diagrams, as shown in Figure~\ref{fig:Vainshtein}(a) for the $N=1$ case, for which the first few coefficients are given by,
\begin{align}
G_\phi^{(1)} &= z  G_\phi^{(0)} G_\phi^{(0)} G_\phi^{(0)} 
,   &\Rightarrow \;\; G_\phi^{(1)} &= z   \nonumber \\
 G_\phi^{(2)} &= z\,  3 G_\phi^{(1)} G_\phi^{(0)}G_\phi^{(0)}  
 ,   &\Rightarrow \;\; G_\phi^{(2)} &= 3 z^2   \nonumber \\ 
 G_\phi^{(3)} &=z \left( 3 G_\phi^{(1)} G_\phi^{(1)}G_\phi^{(0)} + 3 G_\phi^{(2)} G_\phi^{(1)}G_\phi^{(0)} \right) 
 ,   &\Rightarrow \;\; G_\phi^{(3)} &=  12  z^3  \; .  \label{eqn:Gseries_X2} 
\end{align}
Of course, in this simple theory \eqref{eqn:KGphiEom} can be solved algebraically for $G_\phi (r)$ at any $r$, but we focus on the series solutions for three reasons: (i) while \eqref{eqn:KGphiEom} is algebraic, the equations of motion we will encounter for two-body systems in section~\ref{sec:horndeski} are not, and this series expansion approach allows us to treat these different cases in a uniform way, (ii) in order to integrate $G_\phi$ for the scalar field $\phi (r)$, it is more convenient to integrate the series solution term by term rather than attempt to integrate the exact algebraic solution to \eqref{eqn:KGphiEom}, (iii) phenomenologically, \eqref{eqn:Gseries_X2} is a simple description of how the scalar behaves far from sources which in many cases is more convenient than a lengthy algebraic solution.

\paragraph{Breakdown of Perturbative Series.}
The series solution $\sum_{n} G_\phi^{(n)}$ relied on the parameter $z$ being much less than one, and clearly breaks down at small $r$ when the higher $G_\phi^{(n)}$ corrections become comparable to the linearised solution, $G_\phi^{(0)}$. When that happens, one can no longer truncate the series at any finite order and must include \emph{all} series coefficients, which in the example of \eqref{eqn:KGphiEom} are given by,
\begin{align}
 G_\phi^{(n)} =  \frac{ (n + 2 n N)! }{n! ( 1 + 2 n N )}  \;  z^n
 \label{eqn:Gseries_XN}
\end{align}
More precisely, the perturbative series \eqref{eqn:Gseries_X2} no longer converges when the ratio of successive terms $| G^{(n+1)}_\phi / G^{(n)}_\phi |$ exceeds 1, which first happens for the large $n$ terms when,
\begin{align}
|z| > \frac{ (2N)^{2N} }{ (2N+1)^{2N+1} }  \;\; \Rightarrow \;\; \text{Perturbative series breaks down}  .
\label{eqn:znl}
\end{align}
Physically, this defines a critical radius $R_{\rm nl}$ at which the non-linearities dominate the classical equation of motion and a perturbative series solution is no longer valid, e.g. \eqref{eqn:z_PX} implies,
\begin{align}
R_{\rm nl}^{4} =  \epsilon R_K^{4}  \left( \tfrac{|\lambda_{N+1} |}{ \epsilon} \tfrac{(2N+1)^{2N+1}}{(2N)^{2N}}  \right)^{\frac{1}{N}} .
\label{eqn:Rnl}
\end{align}
For instance, a $\lambda_2 X^2$ self-interaction can be described with the perturbative series \eqref{eqn:Gseries_X2} of tree-level Feynman diagrams providing $r^4 > \tfrac{27}{4} |\lambda_2| R_K^4 $, but in contrast a very high order interaction $X^N$ can be described perturbatively for $r^4 \gtrsim \epsilon R_K^4$. 
This is in line with the strong coupling scale being raised from $\Lambda^4$ to $\Lambda^4/\epsilon$ as in \eqref{eqn:KRaise}, and so $R_{\rm nl}^2 \Lambda_{\rm sc}^2 \sim  R_K^2 \Lambda^2 = m/(4\pi M)$ for any $N$ or $\epsilon$, as one might expect. This guarantees that, providing $m/M \gg 1$, there is always a range of $r$ over which the classical non-linearities dominate and yet quantum corrections can be neglected.

\subsection{Resummation and $K$-mouflage Screening}
\label{sec:PX_screening}

In the region $r \gg R_K$ ($\gg 1/\LambdaP$), the perturbative series \eqref{eqn:Gseries_X2} is a good approximation to the scalar field profile $\phi (r)$ around a compact object. 
However, in the regime $1/ \LambdaP^4 \ll r^4 \ll R_K^4$, the theory is weakly coupled and yet classical non-linearities in the equation of motion are large, invalidating a perturbative series solution. 
In order to describe $\phi (r)$ at these scales, one must ``resum'' the entire infinite series $\sum_{n=0}^\infty G_\phi^{(n)}$. This amounts to finding an analytic continuation of the series beyond its radius of convergence. 

\paragraph{Resummation of $X^2$.}
In order to go beyond $r = R_{\rm nl}$, one needs to find an analytic continuation of the perturbative series, namely a smooth function $G_{\phi} (z)$ whose Taylor series about $z=0$ reproduces $\sum_{n} G_{\phi}^{(n)}$. 
For instance, for $P(X) = -\frac{1}{2} X + \lambda_2 X^2$, the series coefficients \eqref{eqn:Gseries_X2} can be resummed into elementary functions, 
\begin{align}
 G_\phi &= 1 +    \frac{ \lambda_2 R_K^4 }{r^4}  +   3 \left( \frac{ \lambda_2  R_K^4 }{r^4 } \right)^2  + 12 \left(\frac{ \lambda_2  R_K^4 }{r^4 } \right)^3 +  ...  
 \nonumber \\
 &= \begin{cases}
  \frac{ 3 r^2 }{ R_{\rm nl}^2 } \sin \left( \frac{1}{3}   \text{arcsin} \left(  \frac{R_{\rm nl}^2 }{r^2}  \right) \right) &\text{when } \lambda_2 > 0 \\[10pt]
  \frac{ 3 r^2 }{ R_{\rm nl}^2 } \sinh \left( \frac{1}{3}  \text{arcsinh} \left(  \frac{R_{\rm nl}^2 }{r^2}   \right)  \right) &\text{when } \lambda_2 < 0  \; .   
  \end{cases}
 \label{eqn:Gresum_X2} 
\end{align}
The dependence on $R_{\rm nl}^2 \sim \sqrt{|\lambda_2|}$ indicates that this is a non-perturbative expression (does not correspond to any Feynman diagram with an integer number of vertices), and the sign of the EFT coupling determines the branch.  
Crucially, while the perturbative series is well-defined for either sign of $\lambda_2$, the resummation beyond $r =R_{\rm nl}$ is only possible for $\lambda_2 < 0$, since for positive values of $\lambda_2$ the resummed $G_\phi(r)$ becomes complex when $R_{\rm nl}/r > 1$.

\paragraph{Resummation of $X^{N+1}$.}
That the resummation is only possible for certain signs of the EFT couplings turns out to be very general. When the interaction $\lambda_{N+1} X^{N+1}$ dominates, the perturbative series coefficients \eqref{eqn:Gseries_XN} fall off like $G^{(n)}_\phi \sim n^{-3/2} ( \pm R_{\rm nl}^{4}/r^{4} )^{nN}$ at large $n$, where the sign corresponds to the sign of $\lambda_{N+1}$. 
When $\lambda_{N+1} > 0$, at $r = R_{\rm nl}$ this series develops a branch cut singularity at which the second derivative $\phi''( R_{\rm nl} )$ diverges (since its series solution $\sim \sum_n n^{-1/2}$ at large $n$). 
On the other hand, when $\lambda_{N+1} < 0$, at $r = R_{\rm nl}$ there is no singularity since the alternating sign in $\sum_n (-1)^n n^{-3/2}$ improves convergence, and so $G_{\phi} (z)$ can be smoothly continued\footnote{
For concreteness, we note that the explicit resummation of \eqref{eqn:Gseries_XN} can be written in terms of the hypergeometric function ${}_{2N} F_{2N+1}$,
\begin{align}
\sum_{n=0}^{\infty} G_\phi^{(n)} =  {}_{2N} F_{2N-1} \left( \tfrac{1}{2N+1} , \tfrac{2}{2N+1} , ... , \tfrac{2N}{2N+1} ; \tfrac{2}{2N} , \tfrac{3}{2N} , ... , \tfrac{2N-1}{2N} , \tfrac{2N+1}{2N}   ;  \hat{z} \right) \; , 
\label{eqn:Gresum_XN}
\end{align}
where $\hat{z} = \pm R_{\rm nl}/r$. This indeed becomes complex for any $z > +1$ but is smooth for all negative $z < 0$. 
} to any value of $r$. 
A smooth resummation of the series of tree-level diagrams can only  take place when the coupling has the right sign, namely $\lambda_{N+1} < 0$.

\paragraph{$K$-mouflage Screening.}
Phenomenologically, the resummed solutions can display qualitatively different behaviour to the perturbative solution---most notably, when $r \ll R_{\rm nl}$, the scalar field profile is greatly suppressed, $G_{\phi} (r) \ll 1$.  
Putting aside numerical factors, 
\begin{align}
 G_\phi (r \ll R_{\rm nl} ) \sim \left( \frac{r^4}{R_{\rm nl}^4} \right)^{\frac{N}{1+2N} } \text{ if } \lambda_{N+1} < 0 \; .
 \label{eqn:Gphi_screened}
\end{align}
This is the $K$-mouflage (or ``kinetic'') screening mechanism \cite{Babichev:2009ee}. 
Note that as $N$ increases, the screening becomes more efficient, tending to $G_\phi \sim r^2/ R_{\rm nl}^2$ at large $N$.
This screening mechanism is radiatively stable against quantum corrections from light degrees of freedom \cite{deRham:2014wfa, Brax:2016jjt}, allows for novel cosmologies \cite{Brax:2014wla,Brax:2014yla} and
has also been observed in the strong gravity regime in numerical simulations \cite{terHaar:2020xxb, Bezares:2021yek}. 

Equation~\eqref{eqn:Gphi_screened} is the small $r$ behaviour of the resummed solution (i.e. \eqref{eqn:Gresum_X2} for $N=1$ and \eqref{eqn:Gresum_XN} for general $N$) but can also be understood as a separate series expansion of $G_{\phi} (z)$ about $z=\infty$,
\begin{align}
 G_\phi =   \left(   - z  \right)^{\frac{-1}{1+2N}} - \tfrac{1}{1+2N} \left( - z  \right)^{\frac{-2}{1+2N}} +  \mathcal{O} \left( \left(  - z  \right)^{\frac{-3}{1+2N}} \right) \; .
 \label{eqn:Gn_XN_near}
\end{align}
Note that this series has a complementary radius of convergence $r/R_{\rm nl} < 1$, and when resummed coincides with the profile found by resumming the large $r$ series \eqref{eqn:Gseries_XN}.  
It is clear from \eqref{eqn:Gn_XN_near} that this screened profile only exists if $\lambda_{N+1} < 0$ (namely $z<0$), since otherwise $G_\phi$ is complex due to the fractional powers $n/(1+2N)$.
Had one started from the original equation of motion \eqref{eqn:Gseries_XN}, it may have appeared that $G_\phi = - ( + z )^{-1/(1+2N)}+...$ is an acceptable real solution when $\lambda_{N+1} > 0$, but this solution is not smoothly connected to the boundary condition at infinity. 
This can also be seen by inspecting the discriminant of the equation of motion polynomial \eqref{eqn:Gseries_XN}, which changes sign at $r= R_{\rm nl}$ when $\lambda_{N+1}>0$, signalling multiple branches of solution (the singularity of $\phi''(r)$ at $r = R_{\rm nl}$ is the bifurcation of two such branches).  
Screening on small scales can only take place near compact objects (given a Newtonian $\phi(r) \sim 1/r$ boundary condition at large $r$) if $\lambda_{N+1} < 0$.

\paragraph{Resummation and Matching.}
Finally, note that while we have focussed on $G_\phi (r)$, this is straightforward to integrate for the scalar $\phi (r)$. 
We close this section by remarking that, had we simply solved the equation of motion in two separate limits $r \gg R_{\rm nl}$ and $r \ll R_{\rm nl}$ (without performing any resummation), then this integration would have introduced \emph{two} undetermined constants of integration, one in each expansion region,  
\begin{align}
\phi = \begin{cases}
\frac{m}{4 \pi M r} \left[  \frac{r}{R_{\rm nl}} C_{\rm far} + 1 + \mathcal{O} \left( \frac{R_{\rm nl}}{r}  \right) \right] &\text{ when } r \gg R_{\rm nl}  \; , \\[10pt]
\frac{m}{4 \pi M r} \left[
\frac{r}{ R_{\rm nl} }  C_{\rm near}   + \mathcal{O} \left(  \left( \frac{r}{R_{\rm nl}} \right)^{\frac{4N}{1+2N} }   \right)  \right]   &\text{ when } r \ll R_{\rm nl} \; . 
\end{cases}
\end{align}
Imposing the desired boundary condition at infinity, namely $\phi \to 0$ as $r \to \infty$, fixes $C_{\rm far} = 0$, but in order to fix $C_{\rm near}$ one must match these two expansions at $r=R_{\rm nl}$. 
This is straightforward to do from the viewpoint of the resummation described above, since resummed solutions like \eqref{eqn:Gresum_X2} smoothly propagate boundary conditions from infinity to small $r$. For instance, for $P(X) = -\tfrac{1}{2} X + \lambda_{N+1} X^{N+1}$, the resummation gives\footnote{
For concreteness, the resummed field profile around a point-like mass in the presence of a $\lambda_{N+1} X^{N+1}$ self-interaction can be written in terms of the hypergeometric function ${}_{2N+1} F_{2N}$, 
\begin{align}
\sum_{n=0}^\infty \phi^{(n)} = \tfrac{m}{4 \pi M r}\; \times \;  {}_{2N+1} F_{2N} \left( \tfrac{1}{4N} , \tfrac{1}{2N+1} , \tfrac{2}{2N+1} , ... , \tfrac{2N}{2N+1}    ; \tfrac{2}{2N} , \tfrac{3}{2N}, ... , \tfrac{2N-1}{2N} , \tfrac{4N+1}{4N}  , \tfrac{2N+1}{2N}    ; \hat{z} \right)
\label{eqn:phiresum}
\end{align}
where $\hat{z}= \pm R_{\rm nl}/r$, with sign determined by the sign of $\lambda_{N+1}$. This gives a general expression for the near-zone integration constant \eqref{eqn:Cnear},
\begin{align}
c_{N+1} = \frac{ 1 }{ \text{sinc} \left( \pi /4N \right) }   \frac{  \prod_{k=2}^{2N+1} \Gamma \left( \tfrac{k}{2N} \right)  \prod_{j=1}^{2N} \Gamma \left( \tfrac{j}{1+2N} - \tfrac{1}{4N} \right)  }{ 
\prod_{k=2}^{2N+1} \Gamma \left( \tfrac{k}{2N} - \tfrac{1}{4N} \right)  \prod_{j=1}^{2N} \Gamma \left( \tfrac{j}{1+2N}  \right) }  \; .
\label{eqn:CnearExact}
\end{align}
},
\begin{align}
C_{\rm far} = 0 \;\;\;\; \Rightarrow \;\;\;\; 
C_{\rm near} = c_{N+1} \left( - \frac{\lambda_{N+1} }{|\lambda_{N+1}|} \right)^{\frac{1}{4N}}  \; .
\label{eqn:Cnear}
\end{align}
where $c_{N+1}$ is a numerical coefficient that begins at $c_2  \approx 6$ and decreases monotonically to $c_{N+1} \approx 2$ at large $N$. 
Again we see that it is not possible to find a real solution for $\phi(r)$ at small $r$ satisfying the perturbative boundary conditions if $\lambda_{N+1} > 0$ has the wrong sign. 
Being able to straightforwardly fix this integration constant in the screened region is yet another reason why viewing the small-scale behaviour of $\phi (r)$ as due to a resummation of its large-scale perturbative series can be more useful than viewing it as a separate expansion of the non-linear equation of motion.

\subsection{Positivity and UV Completion}
\label{sec:PX_positivity}

Resumming the perturbative series solution produces a field profile $\phi (r)$ which is a good description of our system on all scales $r \gg 1/\Lambda_{\rm sc}$. But beyond $1/\Lambda_{\rm sc}$, the field theory becomes strong coupled: there is no longer any hierarchy between tree- and loop-level diagrams,  and the interactions can no longer be treated classically. 
In the absence of a fully non-perturbative computation to all loop orders, the only way to probe smaller radii is to UV complete the theory by introducing new fields. 
However, this is not always possible. There are some low-energy EFTs which admit no UV completion: while seemingly consistent as a purely low-energy theory, they do not have any physical small-scale description. 
We will now use positivity arguments to assess whether such a $P(X)$ theory (i.e. \eqref{eqn:PN} with $\lambda_{N+1} < 0$ to  allow for screening) could ever be embedded into a consistent UV complete theory.

\paragraph{Positivity of $X^2$.}
The bridge that we will use to connect the low-energy EFT to properties of its underlying  UV completion is the elastic 2-particle scattering amplitude, $\mathcal{A} (s,t)$, a complex function of the centre-of-mass energy $s$ and momentum transfer $t$.
In \cite{Adams:2006sv}\footnote{see also \cite{Pham:1985cr,Ananthanarayan:1994hf,Pennington:1994kc,Comellas:1995hq} for earlier discussion of this constraint in chiral perturbation theory, and also \cite{Jenkins:2006ia,Dvali:2012zc}.}, 
it was shown that the basic properties of unitarity, causality and locality in the UV require any Lorentz-invariant EFT to obey the positivity bound,
\begin{align}
 \partial_s^2 \mathcal{A}_{\rm EFT} (s, t) |_{t=0}  > 0 \; .
 \label{eqn:pos_ddA}
\end{align}
If the EFT scattering amplitude violates \eqref{eqn:pos_ddA}, then it signals that this effective theory can never arise from a UV completion with these standard properties, which are described in more detail in Appendix~\ref{app:positivity}. 

The leading $\lambda_2 X^2$ interaction was considered in \cite{Adams:2006sv, Dvali:2012zc}, where the forward limit amplitude $\mathcal{A}_{\rm EFT} (s,0) \sim \lambda_2 s^2$ led to the conclusion,
\begin{align}
 \text{Positivity requires } \lambda_2 > 0 \; .
\end{align}
This was a powerful result: in particular, since no standard UV completion can produce a $P(X)$ theory with $\lambda_2 <0$, there is no way to UV complete an EFT which exhibits $K$-mouflage screening due to a large $ \lambda_2 X^2$ interaction.

\paragraph{Positivity of $X^{N+1}$.}
However, since the higher-point interactions $\lambda_N X^N$ (with $N > 2$) give no tree-level contribution to the $2\to 2$ amplitude $\mathcal{A} (s,t)$, their coefficients are not so readily bounded by the traditional arguments.
More recently, \cite{Chandrasekaran:2018qmx} was able to apply similar positivity arguments to $N \to N$ scattering in $P(X)$ theories in which a single $\lambda_N X^N$ interaction dominates, and used a variety of arguments to conclude that the analogous bound should be $(-1)^N \lambda_N > 0$. 
Here, we confirm this result from a complementary direction, using the observation that once $\phi$ is expanded around a non-trivial background, such as $\phi = \alpha t + \varphi$, then the $2 \to 2$ amplitude for $\varphi$ fluctuations receives contributions from any $\lambda_N X^N$ interaction (which $\sim \lambda_N (-\alpha^2)^{N-2} (\partial \varphi)^4 $, schematically). 
Applying the recent positivity bounds of \cite{Grall:2021xxm}, which allow for such non-trivial backgrounds, we find that when $\alpha$ is small, 
\begin{align}
  \text{Positivity requires } (-1)^{N+1} \lambda_{N+1} > 0 \; ,
\label{eqn:lambda_pos}
\end{align}
and indeed the coefficients are required to have an alternating sign. 
In effect, we are using positivity arguments to probe when the low-energy EFT for fluctuations about a vacuum solution which is arbitrarily close to the trivial solution ($\phi = \alpha t$ at arbitrarily small $\alpha$) can be UV completed. 
We carefully list the UV assumptions which underpin this bound in Appendix~\ref{app:pos_derivation} (the analogue of unitarity, causality and locality for boost-breaking amplitudes), and give the full $\varphi \varphi \to \varphi \varphi$ amplitude and the corresponding positivity bound for a general $P(X)$ in Appendix~\ref{app:pos_PX}.

\paragraph{Consequences for $K$-Mouflage.}
The positivity bound \eqref{eqn:lambda_pos} shows that $K$-mouflage screening from a large $X^{N}$ interaction can only be embedded in a standard UV completion if $N$ is odd. 
Happily, this seems to point in right direction for the existence of a well-defined Cauchy problem, see e.g. \cite{Akhoury:2011hr, Leonard:2011ce, Bernard:2019fjb, Figueras:2020dzx, Bezares:2020wkn}. 
In light of these bounds (and the further bounds on a general $P(X)$ theory given in Appendix~\ref{app:positivity}) and their relation to classical resummation, it will be interesting to renew the search for potential UV completions which can exhibit $K$-mouflage screening in the IR. 

\paragraph{Subluminality of Scalar Waves.}
Finally, note that the sound speed of these scalar perturbations around a time-like background ($X < 0$) is given by,
\begin{align}
c_s^2  = \frac{ - 2 P_{,X} }{ -2 P_{,X} - 4 X P_{,XX}  } = 1 + 4 N (N+1) \lambda_{N+1} X^N + ... \; , 
\end{align}
and we see that the positivity bound $ (-1)^N \lambda_N > 0$ (causality in the UV) is precisely the condition for $c_s^2$ to be subluminal (below 1) in the IR. The effect of integrating out unitarity, causal, local physics is to push these scalar waves inside the light-cone (at least for weak backgrounds, $|X| \ll 1$)\footnote{
See also the discussion in \cite{Adams:2006sv} and more recently in \cite{Chandrasekaran:2018qmx}, where the bound $(-1)^N \lambda_N > 0$ is related directly to causality via the null dominant energy condition. 
} .
We emphasis this here because in the Horndeski theory that we consider next it will no longer be the case that positivity and subluminality always coincide (due to the gravitational degrees of freedom).

~\\
To sum up, in the simple theory $P(X)  = - \frac{1}{2} X +  \lambda_N X^N$ (which can be viewed as a general expansion \eqref{eqn:Kexp} in which a small parameter $\epsilon$ introduces a separation of scales such that $\lambda_N X^N$ is the dominant interaction), the scale at which the theory becomes strongly coupled \eqref{eqn:KRaise} and the radius at which classical perturbation theory near a compact object \eqref{eqn:Rnl} are related by $\Lambda_{\rm sc}^2 R_{\rm nl}^2 \sim  m/M$ ($\gg 1$, typically). In the regime $1/\Lambda_{\rm sc} \ll r \ll R_{\rm nl}$, classical non-linearities can be resummed providing $\lambda_N < 0$, and this leads to $K$-mouflage screening. Positivity bounds require that $(-1)^N \lambda_N > 0$, and so screening is only compatible with UV completion for such theories if the power of $X^N$ is odd. 
%

\section{Horndeski Theories}
\label{sec:horndeski}

Now we turn to the scalar-tensor theory \eqref{quarticA}, and similarly ask for what values of the EFT couplings is there an obstruction to resummation in the classical theory or to UV completion in the quantum theory?  
Scalar-tensor theories in this Horndeski class (and its generalisations) form the basis of recent model-independent parameterised approaches that systematically explore modified gravity effects in linear cosmology  \cite{Gubitosi:2012hu,Bloomfield:2012ff,Gleyzes:2014rba,Bellini:2014fua,Gleyzes:2013ooa,Kase:2014cwa,DeFelice:2015isa,Langlois:2017mxy,Frusciante:2019xia,Renevey:2020tvr,Lagos:2016wyv,Lagos:2017hdr}, resulting in various cosmological constraints on deviations from GR \cite{Noller:2018wyv,Bellini:2014fua,Hu:2013twa,Raveri:2014cka,Gleyzes:2015rua,Kreisch:2017uet,Zumalacarregui:2016pph,Alonso:2016suf,Arai:2017hxj,Frusciante:2018jzw,Reischke:2018ooh,Mancini:2018qtb,Brando:2019xbv,Arjona:2019rfn,Raveri:2019mxg,Perenon:2019dpc,SpurioMancini:2019rxy,Baker:2020apq,Joudaki:2020shz,Noller:2020lav,Noller:2020afd}.
\eqref{quarticA} is also the theory previously studied in \cite{Melville:2019wyy, deRham:2021fpu} and has the convenient feature that positivity bounds can be mapped directly onto constraints on the effective parameters which control linearised cosmological perturbations \cite{Bellini:2014fua}. 

The structure of this section will parallel that of the simpler $P(X)$ theory above: we will begin by identifying the scale at which the theory becomes strongly coupled ($\Lambda_{\rm sc}$) and the scale at which classical non-linearities dominate ($R_{\rm nl}$), and then move on to discuss resummation to go beyond $R_{\rm nl}$ in~\ref{sec:screening} and finally use positivity bounds to assess whether one could ever go beyond $1/\Lambda_{\rm sc}$ via standard UV completion in~\ref{sec:positivity}.

\subsection{Strong Coupling and Classical Non-linearity}
\label{sec:cutoff}

The important qualitative distinction with a simple $P(X)$ theory is that \eqref{quarticA} contains both scalar and metric degrees of freedom and a non-trivial $G_4 (X)$ function mixes these fluctuations. 
In particular, the analogue of $P(X) = - \frac{1}{2} X + \lambda_{N+1} X^{N+1}$ that we will consider is\footnote{
Note that with our normalisation for $G_4(0)$ the Einstein-Hilbert term is $\sqrt{-g} M_P^2 R$, which differs by a factor of 2 from some other conventions (which simply amounts to a rescaling of $M_P$). 
}, \\[-15pt]
\begin{align}
 G_4 (X) = \sqrt{ 1 - \beta_1 X  - \beta_{N+1} X^{N+1} } \; .
 \label{eqn:G4sq}
\end{align}
for two constant couplings $\beta_1$ and $\beta_{N+1}$. 
Since the linear $\beta_1 X$ term can be removed by a field redefinition (which unmixes the scalar and tensor fluctuations), it does not affect the strong coupling scale of the theory (as we show below), at least neglecting any matter fields. We will reintroduce the matter sector at the end of this subsection, and show that $\beta_1$ determines the effective (disformal) coupling between $\phi$ and matter.

\paragraph{Power Counting.}
With the weakly broken Galileon power counting, \eqref{quarticA} contains scalar self-interactions $X^N (\partial \partial \phi)^2$ which are suppressed by the scales $M_P^{N-1} \Lambda^{3N+3}$. 
Since $M_P$ can be much larger than $\Lambda$, the hierarchy $\delta = \Lambda/M_P \ll 1$ separates these interaction scales\footnote{
For scalar-tensor dark energy, typically $\Lambda^3$ is chosen close to $M_P H_0^2$, where $H_0$ is the  Hubble rate today---in terms of the notation $\Lambda_k = \left( M_P H_0^{k-1} \right)^{1/k}$ often used in this context, these scales correspond to $\Lambda_{2 + 1/(2N-1)}$, beginning at $\Lambda_3^3 = M_P H_0^2$ and increasing to $\Lambda_2^4 = M_P^2 H_0^2$. 
},
\begin{align}
\Lambda^4  \, \gg \,  \Lambda^4 \, \delta^{-2/5} \, \gg \,   ... \,  \gg   \Lambda^4 \, \delta^{- \frac{4N-4}{4N+2}}   \, \gg \,   ... \,  \gg \,   \Lambda^4 \, \delta^{-1}  \; . 
\end{align}
We can therefore view \eqref{eqn:G4sq} as the general theory,
\begin{align}
 G_4^2 (X) = 1 - \sum_{n=1} \beta_n X^n 
 \label{eqn:G4squared}
\end{align}
subject to a finite number of tunings ($\beta_n = 0$ for all $2 \leq n \leq N$) and at leading order in $\delta$ (i.e. in $1/M_P$). 
In the absence of any tuning, it is the lowest of these scales ($\Lambda$) that sets the strong coupling scale of the theory, and the dominant interaction ($\beta_2 X^2$) corresponds to the quartic Galileon $X ( \phi^\mu_\mu \phi^\mu_\mu - \phi^\mu_\nu \phi^\nu_\mu )$ in \eqref{quarticA}. 
In the $P(X)$ example of section~\ref{sec:PX_cutoff}, the only way to remove the lowest-lying scalar self-interactions was to tune the $\lambda_n$ coefficients to zero. 
However, these scalar self-interactions are not the only interactions in \eqref{quarticA}, there are also interactions that mix scalar and tensor fluctuations. This opens up a new possibility: raising the strong coupling scale by arranging a cancellation between the scalar self-interactions and scalar-tensor mixing. We are now going to show that \eqref{eqn:G4sq}, thanks to its square root structure, achieves such a cancellation and thus has a parametrically raised strong coupling scale.
Put another way, tuning each $\beta_{n \geq 2}$ to zero in \eqref{eqn:G4squared} parametrically raises the strong coupling scale, 
\begin{align}
 \beta_2 = \beta_3 = ... = \beta_{N} = 0 \;\;\;\; \Rightarrow \;\;\;\; \text{Strong coupling at } \Lambda_{\rm sc}^4 \sim  \Lambda^4 \, \delta^{- \frac{4N-4}{4N+2}}   \; ,
 \label{eqn:G4Raise}
\end{align}
for any value of $\beta_1$, despite $G_{4,X} (X)$ containing apparently lower-order terms.

\paragraph{Leading Interactions.}
Expanding $g_{\mu\nu} = \eta_{\mu\nu} + h_{\mu\nu}/M_P$, the leading interactions at $\Lambda$ are,
\begin{equation}
\mathcal{L}  \supset \, \delta^{\mu \alpha \rho}_{\nu \beta \sigma} \left[  \frac{ \G_{4,X}}{\Lambda^3}  \phi  \phi_\mu^\nu  \partial_\alpha \partial^\beta h_{\rho}^{\sigma}  + \frac{ \G_{4,XX}}{\Lambda^6} \phi  \phi_\mu^\nu \phi_\alpha^\beta  \phi_\rho^\sigma   \right]  
\label{eqn:Horndeski_vertex_LO}
\end{equation}
where the overbar indicates that the function has been evaluated at $X=0$. 
We immediately see that tuning both $\bar{G}_{4,X} = 0$ and $\bar{G}_{4,XX} = 0$ would remove these interactions and lift the strong  coupling scale above $\Lambda$.  
However, comparing the scalar and metric equations of motion, \\[-12pt]
\begin{align}
\frac{\delta S}{\delta \phi} &\supset
 2 \delta^{\mu \alpha \rho}_{\nu \beta \sigma} \left[  \frac{ \bar{G}_{4,X} }{ \Lambda^3 } \partial_{\alpha} \partial^{\sigma} h_{\rho}^{\;\; \beta} -  \frac{2 \bar{G}_{4,XX} }{\Lambda^6}   \phi_\alpha^\beta \phi_\rho^\sigma   \right] \phi_\mu^\nu  
 \label{eqn:phi_eom_L} \\ 
\frac{\delta S}{\delta h^{\;\;\nu}_{\mu}} &\supset
\frac{1}{2} \delta_{\nu \beta \sigma}^{\mu \alpha \rho} \left[  \bar{G}_4  \partial_\alpha \partial^\sigma h_{\rho }^{\;\; \beta} + \frac{2 \bar{G}_{4,X}}{ \Lambda^3} \phi_\alpha^\beta \phi_\rho^\sigma  \right]     \, ,
\label{eqn:metric_eom_L}
\end{align}
we see that \eqref{eqn:metric_eom_L} can be used to remove $h_{\mu\nu}$ from \eqref{eqn:phi_eom_L}, leaving an effective scalar self-interaction,
\begin{align}
\frac{\delta S}{\delta \phi}  \supset  - 4 \left(  \frac{ \bar{G}_{4,X}^2 + \bar{G}_4 \bar{G}_{4,XX} }{ \bar{G}_4} \right) \, \frac{ \delta^{\mu \alpha \rho}_{\nu \beta \sigma} \phi_\mu^\nu \phi_\alpha^\beta  \phi_\rho^\sigma }{  \Lambda^6 }   \; . 
\label{eqn:phi_eom_L2}
\end{align}
The single tuning $\beta_2 = - \tfrac{1}{2} \left(  \bar{G}_{4,X}^2 + \bar{G}_4 \bar{G}_{4,XX} \right) = 0$ is therefore enough to eliminate this quartic self-interaction at $\Lambda$, for any value of $\bar{G}_{4,X}$. 
While the metric equation \eqref{eqn:metric_eom_L} appears to contain a further interaction  
at $\Lambda$ (which could lead to strong coupling in the tensor sector), this is harmless since it can be removed completely by redefining the metric fluctuation,
\begin{equation}
\tilde{h}_{\rho}^{\;\; \beta} = h_{\rho}^{\;\; \beta} + \frac{2 \bar{G}_{4,X} }{ \bar{G}_4 \Lambda^3} \phi_\rho \phi^\beta \, ,
 \label{eqn:metric_redef_L}
\end{equation}
which leads to free propagation of $\tilde{h}_\rho^{\;\; \beta}$ at this order. 
Since physical observables are insensitive to such field redefinitions (we will comment on the effect that \eqref{eqn:metric_redef_L} has on the coupling to matter below), the strong coupling scale is set by the scalar interaction \eqref{eqn:phi_eom_L2} and can be raised above $\Lambda$ by setting $\beta_2 = 0$ only (since this leads to interactions \eqref{eqn:Horndeski_vertex_LO} which can be completely removed by \eqref{eqn:metric_redef_L}).

\paragraph{Higher-Order Interactions.}
Since $h_{\mu\nu}/M_P  \sim X / (M_P \Lambda^3)$, if we are to capture all interactions at scales up to (and including) $\Lambda^4 \delta^{-1} = M_P \Lambda^3$ we may no longer expand in these metric fluctuations (since truncating this expansion at any finite order means throwing away interactions at $M_P \Lambda^3$). 
Rather, we must use the non-linear equations of motion for $g_{\mu\nu}$,
\begin{align}
 \frac{\delta S}{\delta \phi}
 &= \left[  - 2 G_{2,X} g^{\mu\nu} - 4 \frac{ G_{2,XX} }{ M_P \Lambda^3 } \phi^\mu \phi^\nu  + 4 G_{4,X} \frac{M_P G^{\mu\nu}}{\Lambda^3}    -  4 \frac{ G_{4,XX}  }{\Lambda^6} I^{\mu\nu} -  \frac{8 G_{4,XXX}}{3 M_P \Lambda^{9} }  J^{\mu\nu}   \right] \phi_{\mu\nu}   \nonumber \\
 \frac{\delta S}{\delta g_{\mu \nu} } &= M_P \Lambda^3 \left[  \frac{G_2}{2} g^{\mu\nu} - \frac{ G_{2,X} }{ M_P \Lambda^3}  \phi^\mu \phi^\nu  - G_4 \frac{ M_P  G^{\mu\nu} }{\Lambda^3}  - \frac{ G_{4,X} }{\Lambda^6} I^{\mu\nu}  - \frac{ 2 G_{4,XX} }{  M_P \Lambda^{9}}  J^{\mu\nu} \right] 
\label{eqn:metric_eom_NL1}
\end{align}
where we have introduced two independent tensor structures,
\begin{align}
 I^{\mu\nu} &=    \delta^{\alpha \rho (\mu}_{ \beta \sigma \nu'}  g^{\nu) \nu'}   \left( \nabla_\alpha \left( \phi^\beta \phi_\rho^\sigma  \right) -   \frac{1}{2} \phi_\alpha^\beta \phi_\rho^\sigma   \right)    - G^{\alpha (\mu} \phi^{\nu)} \phi_\alpha   \; ,  \nonumber \\
 J^{\mu\nu} &=   \delta^{\alpha \rho (\mu}_{ \beta \sigma \nu'} \left(   
 g^{\nu) \nu'}  \phi^\beta \phi_\gamma \phi_\alpha^\gamma
+ \frac{1}{2} \phi^{\nu)} \phi^{\nu'}    \phi_\alpha^\beta  
  \right)  \phi_\rho^\sigma    \; ,
\end{align}
which are symmetrised using $T^{(\mu\nu)} = T^{\mu\nu} + T^{\nu\mu}$. 
Using the metric equation of motion \eqref{eqn:metric_eom_NL1}, we can replace the $M_P G^{\mu\nu} / \Lambda^3$ in the scalar equation of motion, leaving,
\begin{align}
G_4 \frac{\delta S}{\delta \phi}
 &= \left[ - 2 G_4^2 \partial_X \left( \frac{G_2}{G_{4}}  \right) g^{\mu\nu} - 4  \partial_X \left( G_{2,X} G_{4} \right)  \frac{\phi^\mu \phi^\nu}{ M_P \Lambda^3} 
  -   2 \partial_X^2 G_{4}^2  \frac{  I^{\mu\nu} }{\Lambda^6} - \frac{4}{3}  \partial_X^3 G_{4}^2    \frac{  J^{\mu\nu}}{ M_P \Lambda^{9} }   \right] \phi_{\mu\nu}  \; .
  \label{eqn:phi_eom_NL}
\end{align}
\eqref{eqn:phi_eom_NL} makes it clear that it is the function $G_4^2$ which controls the $\phi$ self-interactions that lie below $M_P \Lambda^3$ (the $I^{\mu\nu} \phi_{\mu\nu}$ and $J^{\mu\nu} \phi_{\mu\nu}$ terms).
Once the tuning \eqref{eqn:G4Raise} is performed, $G_4^2 \supset -\beta_{N+1} X^{N+1}$ is the dominant interaction and leads to strong coupling at $\Lambda^4 \delta^{- \frac{4N-4}{4N+2} }$ due  to the $(\partial \phi)^{2N-2} I^{\mu\nu} \phi_{\mu\nu} / M_P^{N-1} \Lambda^{3N+3}$ and $(\partial \phi)^{2N-4} J^{\mu\nu} \phi_{\mu\nu} / M_P^{N-1} \Lambda^{3N+3}$ interactions in \eqref{eqn:phi_eom_NL}.

The metric equation of motion \eqref{eqn:metric_eom_NL1} apparently contains interactions at the same order as \eqref{eqn:phi_eom_NL}. 
However, as before these can be removed with a disformal field redefinition\footnote{
Note that a more general, $X$-dependent, disformal field redefinition would introduce three further tensor structures in the metric equation of motion, which correspond to the beyond Horndeski interactions \cite{Gleyzes:2014dya, Gleyzes:2014qga, Crisostomi:2016tcp}. Including quartic beyond Horndeski terms in the original Lagrangian, one finds that there are analogous tunings which can be applied to raise the cutoff, and the unique choice which raises the cutoff all the way to $M_P \Lambda^3$ corresponds to the disformal field redefinition of the Einstein-Hilbert term. 
},
\begin{align}
 g_{\mu\nu} = \tilde{g}_{\mu\nu} + \beta_1 \;  \frac{ \phi_\mu \phi_\nu}{ M_P \Lambda^3} 
\label{eqn:dis_redef_NL}
\end{align}
which we describe in more detail in Appendix~\ref{app:disformal}. We find that, supposing $G_4^2$ has been tuned to remove the scalar interactions up to $\Lambda^4 \, \delta^{\frac{4N-4}{4N+2}}$, 
then the $\tilde{I}^{\mu\nu}$ and $\tilde{J}^{\mu\nu}$ interactions in the metric equation of motion both begin at $(\partial \phi)^{2N} ( \partial^2 \phi )^2 / M_P^{N} \Lambda^{3N+ 6}$ and
are one power of $M_P$ suppressed compared with the interactions in the scalar equation of motion \eqref{eqn:phi_eom_NL}, and so it is the scalar self-interactions that sets the strong coupling scale in this frame.

\paragraph{Raising the Strong Coupling Scale.}
Altogether, we  conclude that the tuning \eqref{eqn:G4Raise} leads to a parametrically raised strong coupling scale, which is set by the effective scalar self-interaction,
\begin{align}
\mathcal{L}_{N+1}  =  +  \partial_X \left(  \beta_{N+1} X^{N+1} \right) \frac{ \phi^\mu_\mu \phi^\nu_\nu - \phi_\mu^\nu \phi^\mu_\nu }{\Lambda^6}   \; .
\label{eqn:SRaise}
\end{align}
In hindsight, this result is not surprising: we can think of this as starting with the simple theory $G_4 (X) = 1 - \tfrac{1}{2} \beta_{N+1} X^{N+1} + ...$ (whose scalar interactions in \eqref{quarticA} clearly start at the scale $M_P^{N-1} \Lambda^{3N+3}$) and then performing the disformal field redefinition \eqref{eqn:dis_redef_NL}, which maps this $G_4$ to \eqref{eqn:G4sq} (up to subleading corrections in $X$) without affecting the strong coupling scale.
But in the language of perturbative scattering amplitudes, the tuning \eqref{eqn:G4Raise} corresponds to a non-trivial cancellation of different Feynman diagrams in the original Horndeski frame,  \\[15pt]
\begin{tabular}{r c l}
\multirow{2}{*}{ $\displaystyle \frac{\beta_2}{\bar{G}_4} = \bar{G}_{4,XX} +  \frac{ \bar{G}_{4,X}^2}{ \bar{G}_4}  = 0 $}  & \multirow{2}{*}{$\Rightarrow$} &
\multirow{2}{*}{$0=$}     \multirow{2}{*}{ \includegraphics[width=0.25\textwidth]{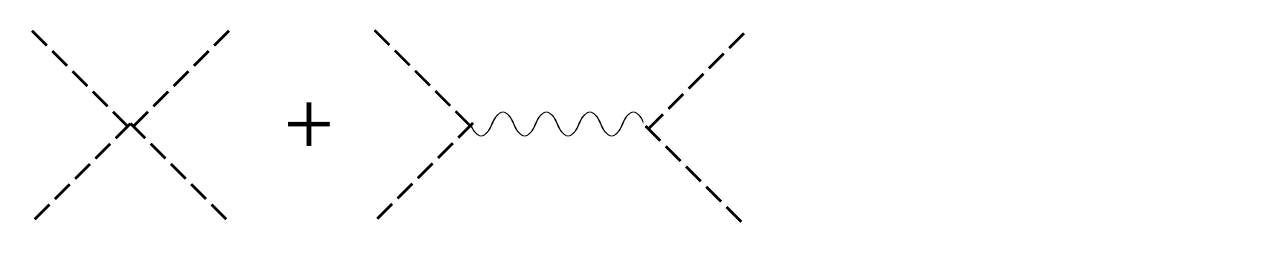}  }
 \\[10pt]
&   \\[10pt]
\multirow{2}{*}{ $\displaystyle \frac{3 \beta_3}{ \bar{G}_4} =  \bar{G}_{4,XXX}  + \frac{3 \bar{G}_{4,X} \bar{G}_{4,XX} }{ \bar{G}_4}  =0 $}   & \multirow{2}{*}{$\Rightarrow$} &
\multirow{2}{*}{$0=$}     \multirow{2}{*}{ \includegraphics[width=0.25\textwidth]{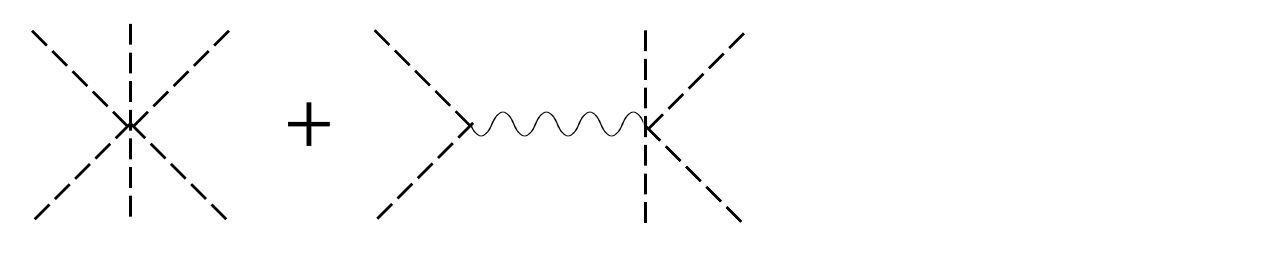}  }
 \\[10pt]
&   \\[10pt]
\multirow{2}{*}{ $\displaystyle \tfrac{12 \beta_4}{ \bar{G}_4} = \bar{G}_{4,XXXX}  +  \tfrac{4 \bar{G}_{4,X} \bar{G}_{4,XXX} }{ \bar{G}_4} + \tfrac{3 \bar{G}_{4,XX}^2 }{ \bar{G}_4} = 0 $}  & \multirow{2}{*}{$\Rightarrow$} &
\multirow{2}{*}{$0=$}     \multirow{2}{*}{ \includegraphics[width=0.25\textwidth]{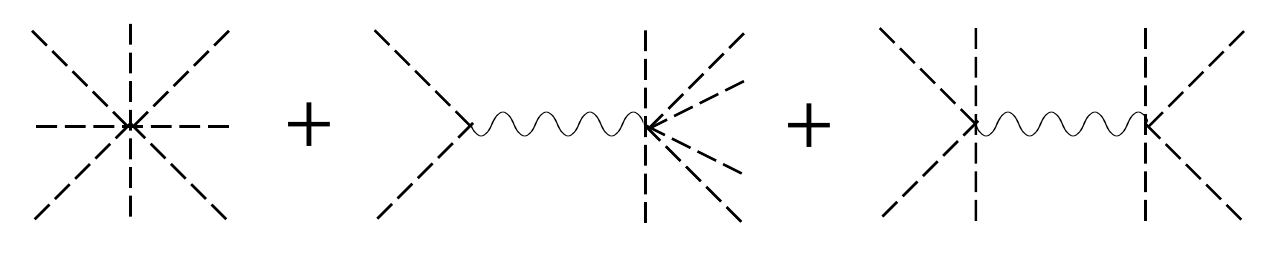} }
 \\[10pt]
&  
\end{tabular}
and so on. This is analogous to the cancellation which leads to improved soft behaviour in theories with exceptional/non-linearly realised symmetries \cite{Cheung:2016drk, Padilla:2016mno, Guerrieri:2017ujb}.

\paragraph{DBI Galileon.}
The highest strong coupling scale, $M_P \Lambda^3$, can be achieved by tuning all $\beta_{N+1} =0$ for $N \geq 1$, leaving simply $G_{4}^2 = 1 -  \beta_1 X$. 
This theory is known as the quartic DBI Galileon \cite{deRham:2010eu}. 
Performing the field redefinition \eqref{eqn:dis_redef_NL} brings this theory to an Einstein frame, 
\begin{align}
S = \int d^4 x \, \sqrt{-\tilde{g}} \left( M_P^2 R + M_P \Lambda^3 \, P(\tilde{X}) \right) \;\; \text{with} \;\;  P(\tilde{X} ) =  \sqrt{ 1 +  \beta_1 \tilde{X} } \; \; G_2 \left( \frac{ \tilde{X} }{ 1 + \beta_1 \tilde{X}}   \right) \; , 
\label{eqn:DBIGal}
\end{align}
and now all scalar interactions manifestly take place at the scale $M_P \Lambda$. 
The choice,
\begin{align}
 G_2 (X) = \frac{ - \tfrac{1}{2} X}{ \sqrt{ 1  -  \beta_1  X } } \; , 
 \label{eqn:G2_DBI}
\end{align}
in \eqref{quarticA} corresponds to a free scalar field $P(\tilde{X} )= -\tfrac{1}{2} \tilde{X}$ in the Einstein frame \eqref{eqn:DBIGal}.
It would be interesting to re-interpret the cancellation that occurs between the different Feynman diagrams shown above in terms of an approximate (weakly broken) DBI symmetry, which could naturally explain the higher strong coupling scale and offer some insight into whether the tuning $\beta_2 = \beta_3 = ... = \beta_N = 0$ is protected against quantum corrections. We leave these directions for the future, and move on to discuss the coupling to matter. 

\paragraph{Coupling to Matter.}
In \eqref{quarticA}, we have included the possibility that matter couples to an effective metric $C(\phi) g_{\mu\nu}$. 
When including matter fields, the field redefinition \eqref{eqn:dis_redef_NL} that was required to remove the leading $\Lambda$ interactions in the metric equation of motion introduces a disformal coupling between $\phi$ and matter,
\begin{align}
 S \supset \int d^4 x \sqrt{-\tilde{g}}  \;  \frac{C (\phi)}{2} \left( \tilde{g}_{\mu\nu} + \beta_1  \frac{\phi^\mu \phi^\nu}{ M_P \Lambda^3}  \right) T^{\mu\nu}
 \label{eqn:Smat}
\end{align}
where $\sqrt{-\tilde{g}} \, T^{\mu\nu} = 2 \delta S / \delta g_{\mu\nu}$ is the stress-energy tensor with respect to the Horndeski frame metric $g_{\mu\nu}$. 
Once the tuning \eqref{eqn:G4Raise} has been performed to raise the strong coupling scale, the scalar profile near a matter distribution $T^{\mu\nu}$ is given by the variation of \eqref{eqn:SRaise} and \eqref{eqn:Smat},
\begin{align}
  \phi^\mu_\mu  +   \frac{ 2 N (N+1) \beta_{N+1} }{   \Lambda^{6} } \delta^{\mu \alpha \rho}_{\nu \beta \sigma}  \partial_\mu \left(  \phi^\mu \phi_\alpha^\beta \phi_\rho^\sigma  X^{N-1}   \right) = \frac{1}{M_P} T^{\mu\nu} \left( -  \frac{\con}{2} \tilde{g}_{\mu\nu}  +   \beta_1 \frac{ \phi_{\mu\nu} }{\Lambda^3}   \right)   \; , 
 \label{eqn:phi_eom_N}
\end{align}
where we have set $\bar{G}_{2,X} = -1/2$ for a canonically normalised field and kept only the leading terms, for instance expanding $C (\phi) = 1 + \con \; \phi/M_P + \mathcal{O} \left( 1/ M_P^2 \right)$\footnote{
We have also dropped the tildes over all $\phi_{\mu\nu}$ since the difference $\phi_{\mu\nu} - \tilde{\phi}_{\mu\nu}$ is suppressed by $\delta$, and similarly neglected a term in $\tilde{\nabla}_{\mu} T^{\mu\nu}$ which to leading order in $\delta$ is $\nabla_{\mu} T^{\mu\nu} = 0$. 
}.
The characteristic scale $\mathcal{M}^4 = M_P \Lambda^3/ |\beta_1|$ of the disformal coupling between matter and a light scalar field has been constrained via a number of astrophysical and terrestrial experiments \cite{Koivisto:2008ak, Zumalacarregui:2010wj, Koivisto:2012za, vandeBruck:2013yxa, Neveu:2014vua, Sakstein:2014isa, Sakstein:2014aca, Ip:2015qsa, Sakstein:2015jca, vandeBruck:2015ida, vandeBruck:2016cnh, Kaloper:2003yf, Brax:2014vva, Brax:2015hma, Brax:2012ie, vandeBruck:2012vq, Brax:2013nsa}.
For typical dark energy values, $(\Lambda^3 \sim M_P H_0^2)$, the coupling is much too large and such a theory would be ruled out by these observations unless some screening mechanism takes place.

\subsection{Resummation and Screening}
\label{sec:screening}

We will now show that the scalar equation of motion can be solved around a compact object, and that classical non-linearities can be resummed \emph{for a particular sign of $\beta_{N+1}$}, leading to screening of the scalar field profile. 
We first consider a single compact object, and show that the successive tunings \eqref{eqn:G4Raise} to raise the strong coupling scale do not affect the efficiency of Vainshtein screening (but do decrease the $R_{\rm nl}$ at which it becomes effective). 
Then we consider the DBI Galileon limit, where all interactions below $M_P \Lambda^3$ are turned off, and show that a resummation of ladder diagrams can lead to screening near two-body systems, and in particular that the sign of $\beta_1$ plays an important role  in the uniqueness of this resummation.

\subsubsection{One-Body System (Vainshtein)}
\label{sec:Vainshtein}

\begin{figure}
 \centering
\qquad \qquad
     \begin{subfigure}[b]{0.3\textwidth}
         \centering
\includegraphics[width=1.0\textwidth]{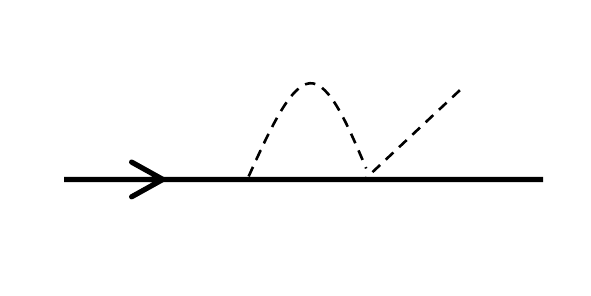}
         \caption{Galileon frame.}
     \end{subfigure}
     \hfill
     \begin{subfigure}[b]{0.3\textwidth}
         \centering
\includegraphics[width=1.0\textwidth]{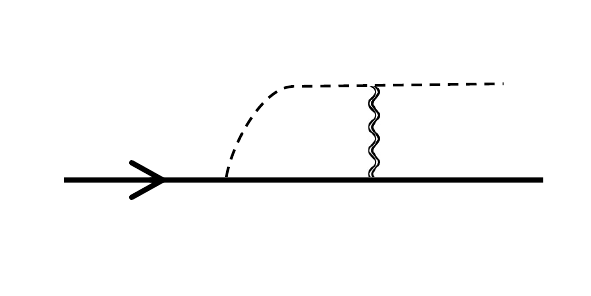}
         \caption{Horndeski frame.}
     \end{subfigure}
     \qquad\qquad
\caption{
Near a single compact object, the $\beta_1 X$ interaction contributes only to self-energy divergences at leading order, shown diagrammatically in both the original Horndeski frame (b) and the Galileon frame (a) following the field redefinition \eqref{eqn:dis_redef_NL}, which trades scalar-tensor mixing for a disformal coupling between $\phi$ and matter. 
}
\label{fig:selfenergy}
\end{figure}

For a static, spherically symmetric point-like source, $T^{\mu\nu} = \frac{m}{4 \pi} u^{\mu} u^{\nu} \delta^{(3)} (\mathbf{r})$ (normalised so that $\tilde{g}_{\mu\nu}  u^{\mu} u^{\nu} = -1$), the disformal interaction contributes only a self-energy divergence $\phi_{\mu\nu} u^\mu u^\nu \, |_{r=0}$ to the scalar equation of motion, depicted in Figure~\ref{fig:selfenergy}. This term will simply renormalise the point-particle EFT (see e.g. \cite{Kuntz:2019zef}) and so we neglect it at this order. 
If we again express $\phi$ as a Newtonian potential modulated by an effective coupling $G_\phi$ \eqref{eqn:Gphi_def}, then the equation of motion \eqref{eqn:phi_eom_N} for $G_\phi$ near a single compact object has precisely the same form as in the $P(X)$ theory \eqref{eqn:KGphiEom}, only now the parameter $z$ is given by,
\begin{align}
 z =  4 N (N+1) \beta_{N+1} \frac{R_V^6}{r^{6}} \left(  \frac{R_K^4}{r^4}   \right)^{N-1} \; , 
 \label{eqn:z_G4}
\end{align}
where we have introduced the scales,
\begin{align}
 R_V^3 \Lambda^3 = \frac{m}{4 \pi M_P} = R_K^2 \sqrt{ M_P \Lambda^3 } \; . 
\end{align}
At large $r$, this $z$ acts as a small expansion parameter and the corresponding series solution $G_\phi = \sum_{n} G_\phi^{(n)}$ is given in \eqref{eqn:Gseries_XN}.
Classical perturbation theory breaks down when $|z|$ exceeds \eqref{eqn:znl}, just like in the $P(X)$ example, where now using \eqref{eqn:z_G4} implies a non-linear scale, 
\begin{align}
 R_{\rm nl}^{2+4N} =   R_V^6 R_K^{4N-4} \left(  \tfrac{  (2N+1)^{2N+1} }{ (2N)^{2N} }  4 N (N+1) | \beta_{N+1} |  \right) \; .
\end{align}
Since only $z < 0$ allows for a smooth continuation of the boundary condition to small $r$, we see that resummation for the Horndeski theory \eqref{eqn:G4sq} requires $\beta_{N+1} < 0$.

\paragraph{Vainshtein Screening.}
The resummation of the full $G_\phi^{(n)}$ series leads to a screening of $G_\phi$, 
\begin{align}
G_{\phi} (r \ll R_{\rm nl} ) \sim  \frac{r^2}{R_{\rm nl}^2 }  \; \text{if } \beta_{N+1} < 0 \; . 
\end{align}
This is the Vainshtein screening mechanism \cite{Vainshtein:1972sx, Koyama:2013paa}. 
Note that as $N$ is increased  (raising the strong coupling scale \eqref{eqn:G4Raise}), the scale $R_{\rm nl}$ decreases from $R_V$ towards $R_K$, but the functional form of $G_{\phi} (r)$ remains unchanged---the Vainshtein mechanism is equally efficient for any $X^N$ non-linearity in $G_4$. 
In contrast, for simple $P(X)$ theories when $X^2$ is the dominant interaction $K$-mouflage screening results in $G_\phi \sim (r/R_{\rm nl})^{4/3}$, and only for $X^N$ for very large $N$ do we approach  a screening as efficient as $(r/R_{\rm nl})^2$.  

The full resummed expression for $G_\phi$ is given in \eqref{eqn:Gresum_XN}, and the corresponding $\phi (r)$ profile given in \eqref{eqn:phiresum}, with the understanding that now $z$ is replaced by \eqref{eqn:z_G4}. 
Again we point out that these fully resummed expressions are necessary if one is to determine the constant part of $\phi$ in the screened regime (by matching onto the boundary condition at large $r$), and the $C_{\rm near}$ needed here coincides with what we have determined for general $N$ in \eqref{eqn:CnearExact} for $P(X)$.

\paragraph{Resummation in Horndeski Frame.}
The resummation which leads to Vainshtein screening can be depicted diagrammatically, as shown in Figure~\ref{fig:Vainshtein}. Ultimately, following the field redefinition of section~\ref{sec:cutoff}, the problem of determining $\phi$ near a compact object in this scalar-tensor theory \eqref{quarticA} has reduced to solving the simple algebraic equation~\eqref{eqn:KGphiEom}.
It is perhaps worth emphasising that the problem would have seemed far more involved had we remained in the original Horndeski frame, in which there is mixing between the scalar and metric fluctuations.  
Take for instance the case where the leading interaction is from the $\beta_2/\bar{G}_4 = \bar{G}_{4,XX} + \bar{G}_{4,X}^2/\bar{G}_4$ interaction. 
When solving for the $G_{\phi}^{(n)}$ above, we are computing the diagrams shown in \ref{fig:Vainshtein}(a), where the worldline of the compact object undergoes $2n+1$ conformal emissions of the scalar field, which then combine via $n$ quartic non-linearities to source $\phi^{(n)}$. 
Note that had we worked directly in the original Horndeski frame, in which there is a cubic interaction $\bar{G}_{4,X} \phi^2 h$ between scalar and metric fluctuations, then the analogous computation would have involved the diagrams shown in Figure~\ref{fig:Vainshtein}(b), in which $G_\phi^{(n)}$ also receives contribution from graviton exchange. These diagrams are individually more challenging computationally, but always organise into factors of $\beta_2$, since they must reproduce the same result as the Galileon frame $\tilde{g}_{\mu\nu}$ calculation.

\begin{figure}[htbp!]
     \begin{subfigure}[b]{\textwidth}
\includegraphics[width=0.9\textwidth]{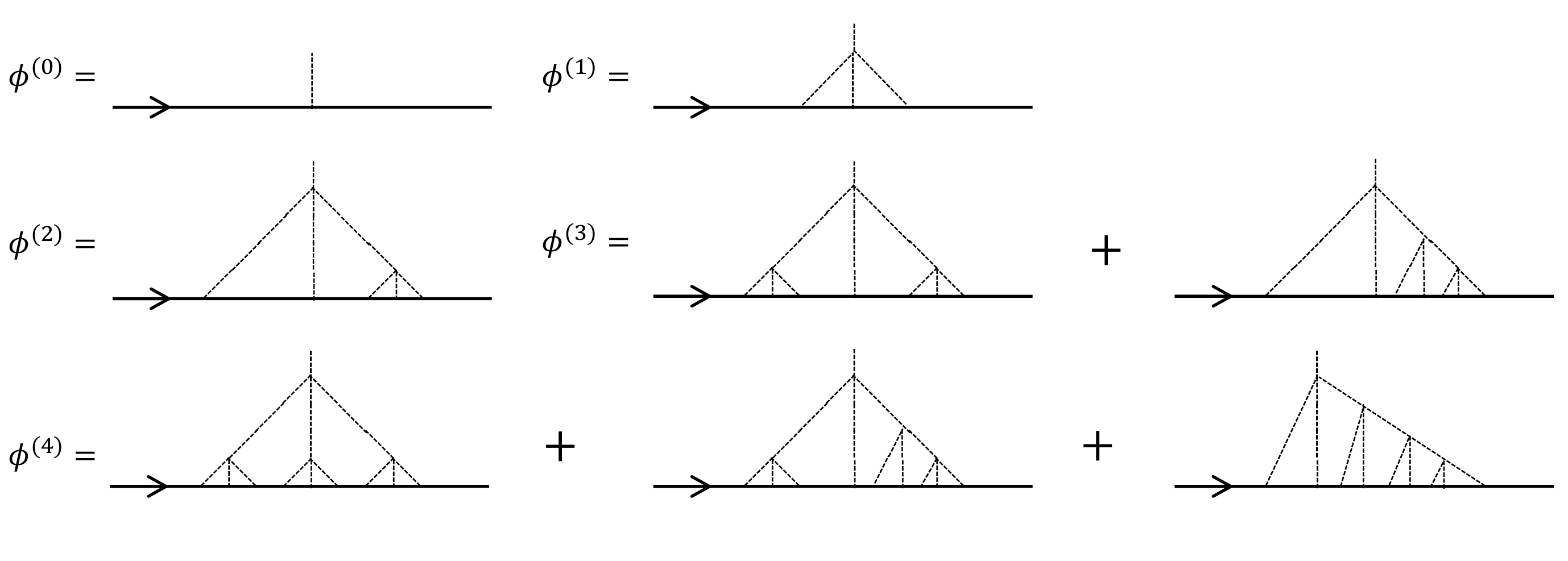}
         \caption{Galileon frame.}
     \end{subfigure}
     \hfill
     \begin{subfigure}[b]{\textwidth}
         \centering
\includegraphics[width=1.0\textwidth]{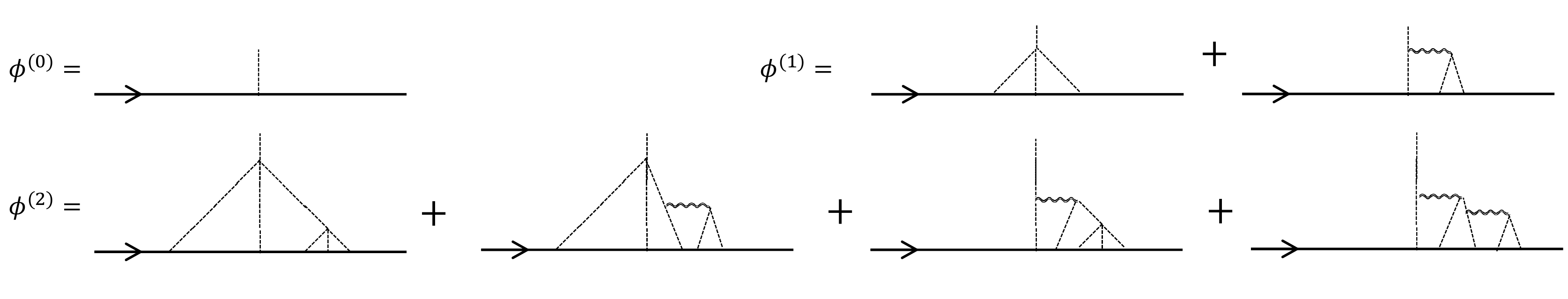}
         \caption{Horndeski frame.}
     \end{subfigure}
\caption{
The first few terms in perturbation theory for the scalar field profile near a compact object's worldline. In the Horndeski frame, there are diagrams involving graviton exchange, but these can all be removed by a field redefinition to the Galileon frame. 
}
\label{fig:Vainshtein}
\end{figure}

\paragraph{Vainshtein in Motion.}
Next we are going to consider a two-body system, in which a pair of compact objects move with a non-relativistic relative velocity. Since the full scalar-tensor theory \eqref{quarticA} is Lorentz-invariant, it is only the relative velocity between the bodies that can have any physical effect (and not the ``absolute'' speed of either object).  
As a segue to this two-body case, let us end our discussion of Vainshtein screening around single objects by showing explicitly that the screening is unaffected by the motion of the object. 

In the rest frame of the source, deep inside the Vainshtein radius at $r \ll R_{\rm nl}$ we have,
\begin{equation}
 \phi ( r) \sim  \frac{r}{ R_{\rm nl}^2 }  \; , 
\end{equation}
up to the constant of integration $C_{\rm near}$. The field is efficiently screened compared with the Newtonian $\phi^{(0)} (r) \sim 1/r$. 
Now suppose we boost to a Lorentz frame in which the particle is in motion, with instantaneous 4-velocity $u^\mu$ and 4-position $x_p^\mu$. Using the tensor $P_{\mu \nu}  = g_{\mu\nu} - u_\mu u_\nu / u^2$, which projects onto space-like components in the instantaneous rest frame of the particle, the scalar profile in this frame is simply,
\begin{equation}
 \phi ( x ) \sim  \frac{ \sqrt{ r_\mu P^{\mu\nu} r_\nu }  }{R_{\rm nl}^2} 
 \label{eqn:Vainshtein}
\end{equation}
where $r^\mu = x^\mu - x_p^\mu$ and $\sqrt{ r_\mu P^{\mu\nu} r_\nu }$ represents the retarded spatial distance from the particle. \eqref{eqn:Vainshtein} is valid for any velocity,
but for a point mass at $\mathbf{x}_{p} (t)$ moving with non-relativistic velocity $\mathbf{v}$ it becomes,
\begin{equation}
 \phi ( \mathbf{x}, t ) \sim \frac{ | \mathbf{x} - \mathbf{x}_p (t) | }{ R_{\rm nl}^2 }  \, \left[ 1 + \frac{ ( \mathbf{v} \cdot ( \mathbf{x} - \mathbf{x}_p (t) ) )^2  }{2 | \mathbf{x} - \mathbf{x}_p (t) |^2 } + ...    \right] 
 \label{eqn:VainshteinPN}
\end{equation}
subject to corrections suppressed by $\mathcal{O} (v^2)$ 
or by $\mathcal{O} ( r/ R_{\rm nl}  )$.
This is to emphasise that providing the separation from the body ($\sqrt{ r_\mu P^{\mu\nu} r_\nu }$ in general) is much smaller than $R_{\rm nl}$, then $\phi$ is screened by the Vainshtein mechanism and the nature of this screening is unaffected by any absolute motion of the body.

\subsubsection{Two-Body System (Ladder)}
\label{sec:ladder}


We will now discuss the DBI Galiileon tuning $G_4 = \sqrt{1- \beta_1 X}$, in which the cutoff \eqref{eqn:G4Raise} is raised to its maximum value of $M_P \Lambda^3$. 
As remarked in section~\ref{sec:cutoff}, this theory corresponds to a disformally coupled scalar in the Einstein frame. This is the system studied in \cite{Davis:2019ltc}, where it was shown that a certain class of Feynman diagrams in two-body systems can be resummed, leading to a ``ladder screening'' suppression of the scalar field. 
In this subsection, we briefly review this ladder resummation, 
and by considering when the perturbative solution can be smoothly continued beyond its radius of convergence we are led to conclude that: 
\begin{quotation}
\centering
Ladder resummation can only be unique when the disformal coupling $\beta_1 < 0$, 
\end{quotation}
which we demonstrate explicitly in the simple example of two masses colliding head-on. 
For simplicity we will focus on the particular $G_2 (X)$ given in \eqref{eqn:G2_DBI}, which corresponds to a canonical kinetic term in the Einstein frame, but as we will show below for hard scattering processes any scalar self-interactions at the energy scale $M_P \Lambda^3$ (the length scale $R_K$) are suppressed relative to the $\beta_1$ disformal coupling to the binary.

\paragraph{Field Sourced by Binary System.}
Consider a binary system composed of two compact objects at positions $\mathbf{x}_1 (t)$ and $\mathbf{x}_2 (t)$, moving with a non-relativistic relative speed. The scalar field profile in this system is sourced by,  
\begin{align}
 T^{\mu\nu} =  u_1^\mu u_1^\nu  \delta \left( \mathbf{x} - \mathbf{x}_1 (t) \right) + u_2^\mu u_2^\nu  \delta \left( \mathbf{x} - \mathbf{x}_2 (t) \right)
\end{align}
where $u_1$ and $u_2$ are the corresponding 4-velocities of the objects (normalised so that $u_A^2 = -1$). 
When considering multiple point-like sources, it is very difficult to determine the non-linear scalar field profile (though see \cite{Dar:2018dra, Kuntz:2019plo, Brax:2020ujo, Renevey:2021tcz} for recent progress). 
Although the leading term in a perturbative series solution of \eqref{eqn:phi_eom_N} is simply the sum of two Newtonian potentials,
\begin{align}
\phi^{(0)} = \frac{\con}{4 \pi M_P} \left[  \frac{ m_1}{  | \mathbf{x} - \mathbf{x}_1 (t) |} + \frac{ m_2}{ | \mathbf{x} - \mathbf{x}_2 (t) |} \right]  \; ,
\label{eqn:phi0}
\end{align}
the corrections at $\mathcal{O} ( \beta_{N+1} )$ quickly become complicated, and no exact analytic solution is known. 
However, as pointed out in \cite{Davis:2019ltc}, the $\beta_1$ disformal coupling is special because it appears \emph{linearly} in the scalar field \eqref{eqn:phi_eom_N}, which allows this particular interaction to be solved exactly.
In particular, the first correction to \eqref{eqn:phi0} in the series solution is given by \cite{Brax:2018bow, Brax:2019tcy},
\begin{align}
\Box \phi^{(1)} = \frac{\beta_1}{M_P \Lambda^3} \phi^{(0)}_{\mu\nu} T^{\mu\nu} \;\; \Rightarrow \;\; \phi^{(1)} =  \frac{ \con \beta_1 }{ 32 \pi^2 M_P^2 \Lambda^3}   \partial_t^2 \left( \frac{ m_1 m_2 }{| \mathbf{x}_1 (t) - \mathbf{x}_2 (t) |}  \right) \left[
\frac{1}{| \mathbf{x} - \mathbf{x}_1 (t) |} + \frac{1}{ | \mathbf{x} - \mathbf{x}_2 (t) |}
 \right]   
 \label{eqn:phi1}
\end{align}
where note that again we have neglected divergent self-energy corrections (diagrams of the form shown in Figure~\ref{fig:selfenergy}).  
Comparing \eqref{eqn:phi0} and \eqref{eqn:phi1}, we see that the scale controlling the size of the $\beta_1$ correction to each object's Newtonian potential is its ``ladder radius'', $R_{L_A}$, 
\begin{align}
 R_{L_A}^3  \, \Lambda^3  = \frac{ v^2 m_A}{4 \pi M_P} = v^2 R_{V_A}^3  \Lambda^3 \; .
\end{align}
Since we are working with non-relativistic velocities, an object's ladder radius $R_L$ is always smaller than its characteristic Vainshtein radius $R_V$. 
In fact, as commented in \cite{Kuntz:2019zef}, for virialised systems $v^2 \sim G_N m/r$, which suggests that\footnote{
This is often an over-estimate due to projection effects---these factors of $v$ arise from $\partial_t | \mathbf{x}_{1} - \mathbf{x}_2 |$ and thus correspond to the relative \emph{radial} velocity, which for bounds orbit is smaller than $G_N m /r$ by a factor of the orbital eccentricity \cite{Davis:2019ltc}.  
} $R_L^4 \sim \delta R_V^4= R_K^4$. In that case, the $\beta_1$ correction \eqref{eqn:phi1} is never more relevant than the $\beta_{N+1}$ non-linearities (which become important at distances $\delta^{\frac{4N-4}{4N+2}} R_V^3 > \delta R_V^4$). 
However, for hard scattering processes $v^2$ can be much larger $G_N m/r$ (and yet remain non-relativistic), and in that case $R_L$ can be larger than $R_K^4$. Furthermore, for the DBI Galileon tuning, in which all other $\beta_{N+1}$ are parametrically small, it is always these $\beta_1$ interactions that dominate.

\paragraph{Ladder Resummation.}
In \cite{Davis:2019ltc}, the entire perturbative series $\phi^{(n)}$ was computed for $G_4 = \sqrt{1-\beta_1 X}$. 
It is convenient to parametrise the field in terms of its free Newtonian potential \eqref{eqn:phi0} modulated by effective time-dependent couplings $G_1$ and $G_2$, 
\begin{align}
\phi = \frac{ \con}{4\pi M_P} \left[  \frac{m_1 G_1 (t)}{ | \mathbf{x} - \mathbf{x}_1 (t) }  +  \frac{m_2 G_2 (t)}{ | \mathbf{x} - \mathbf{x}_1 (t) }      \right] 
\end{align}
and write separate series solutions for $G_1$ and $G_2$. 
The series coefficients are given by, 
\begin{align}
 G_1^{(n)} = \begin{cases}
  \left( \hat{z}_{2} \hat{z}_{1} \right)^{n/2} &\text{when $n$ even,} \\
   \left( \hat{z}_{2} \hat{z}_{1} \right)^{ (n-1)/2} \hat{z}_{2} \;\; &\text{when $n$ odd} \; .   
   \end{cases}
 \label{eqn:Gseries}
\end{align}
and similarly for $G_2$, where we have introduced the dimensionless differential operators,
\begin{align}
\hat{z}_{A} \left[ f (t) \right] =   \beta_1 R_{V_A}^3  \frac{ \partial^2 }{  \partial t^2} \left(  \frac{f(t)}{ | \mathbf{x}_1 (t) - \mathbf{x}_2 (t) | }   \right)   \; .
\label{eqn:zhat}
\end{align}
Note that in \eqref{eqn:Gseries} terms like $(\hat{z}_2 \hat{z}_1 )^2$ should be understood as $\hat{z}_2 [ \hat{z}_1 [ \hat{z_2} [ \hat{z}_1 [ 1 ] ] ] ]$.
The first few $G_\phi^{(n)}$ are shown diagrammatically in Figure~\ref{fig:Ladder_a}---each $\hat{z}_A$ represents the disformal vertex factor of body $A$, and the factor $(\hat{z}_2 \hat{z}_1 )^{n/2}$ represents $n$ ``bounces'' of an intermediate scalar between bodies 1 and 2. 
\eqref{eqn:Gseries} can be resummed into a coupled system of second-order equations, 
\begin{align}
G_1 -  \hat{z}_{2} G_2 &= 1 \;\; \nonumber  \\
G_2 - \hat{z}_{1} G_1 &= 1   \;\; 
\label{eqn:ladder_eqn}
\end{align}
which are the analogue of \eqref{eqn:KGphiEom} for binary systems---in particular, note that $\hat{z}_A \sim  \beta_1 R_{L_A}^3 / r^3$ are the ``small parameters'' that controls the series expansion at large $r$.

\paragraph{Ladder Screening.}
The equations \eqref{eqn:ladder_eqn} (as well as their relativistic counterparts) were studied in detail in \cite{Davis:2019ltc}. 
While it is difficult to find exact solutions for arbitrary particle trajectories (arbitrary $\mathbf{x}_1(t)$ and $\mathbf{x}_2 (t)$ in \eqref{eqn:zhat}), it was found in various simple examples that \eqref{eqn:ladder_eqn} leads to a suppression of the scalar profile at small $r = | \mathbf{x}_1 - \mathbf{x}_2 |$,
\begin{align}
 G_A ( t )  \sim \frac{  r^3 }{ R_{L_A}^3 } \;\; \text{when} \;\; r \ll R_{L_1} \; \text{and} \;  R_{L_2}  \; .
\label{eqn:G_ladder}
\end{align} 
which can be viewed as a complementary series expansion of \eqref{eqn:ladder_eqn} in powers of $1/\hat{z}_A$. 
We refer to this suppression in binary systems due to the disformal interaction as ``ladder screening''. Note that since the perturbative series \eqref{eqn:Gseries} is controlled by the product $\hat{z}_2 \hat{z}_1$, the characteristic distance\footnote{
Note that strictly speaking this is an \emph{asymptotic series}: at any finite $\hat{z}_2 \hat{z}_1$ there is an optimal (finite) number of terms to include and the series as a whole only formally converges when $\hat{z}_2 \hat{z}_1$ is zero. 
} at which the perturbative series gives way to resummation is,
\begin{align} 
 R_{\rm nl}^3 = | \beta_1 | \sqrt{ R_{L_1}^3 R_{L_2}^3 } \; . 
\end{align}
A sufficient condition for ladder screening is therefore that $r \ll R_{\rm nl}$, and in fact schematically \eqref{eqn:ladder_eqn} implies that the suppression can be even greater than \eqref{eqn:G_ladder} for intermediate values of $r$ when there is a hierarchy $m_1 \gg m_2$ (i.e. $R_{L_2} \ll R_{\rm nl}$)\footnote{
Although note that when $m_1 \gg m_2$ there is also a region $R_{\rm nl} \ll r \ll R_{L_1}$ in which perturbation theory is valid and $G_2 \sim R_{L_1}^3 / r^3$ can be \emph{enhanced} by the disformal coupling, but not by more than a factor of $\sqrt{m_1/m_2}$.  
},
 \begin{align}
 G_1 ( t )  \sim \frac{  r^6 }{ R_{\rm nl}^6 } \;\; \text{when} \;\; R_{L_2} \ll r \ll R_{\rm nl} \; ,
\end{align} 
which follows from treating $\hat{z}_2 \hat{z}_1 [ G_1 ] \gg G_1$ but $\hat{z}_2 [1] \ll 1$. 

The new observation that we make here is the importance of the \emph{sign} of the disformal coupling $\beta_1$. In particular, while the resummation leads to a smooth scalar profile for either sign of $\beta_1$, it is only unique if $\beta_1 < 0$.
To exemplify the general idea, we will show explicitly what happens in the simple example of a head-on collision between the two compact objects.

\begin{figure}[htbp!]
 \centering
     \begin{subfigure}[b]{\textwidth}
         \centering
\includegraphics[width=1.0\textwidth]{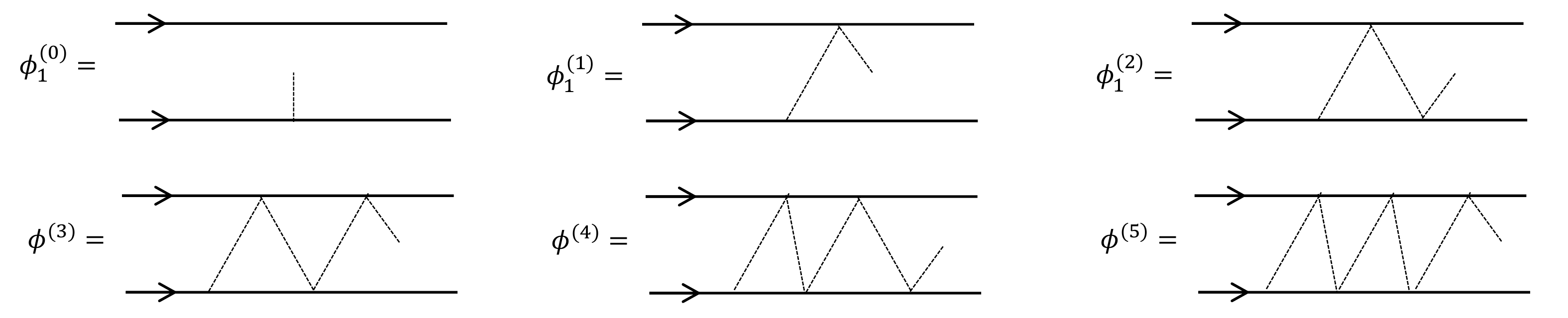}
         \caption{Einstein frame.}
         \label{fig:Ladder_a}
     \end{subfigure}
     \hfill ~\\[20pt]
     \begin{subfigure}[b]{\textwidth}
         \centering
\includegraphics[width=1.0\textwidth]{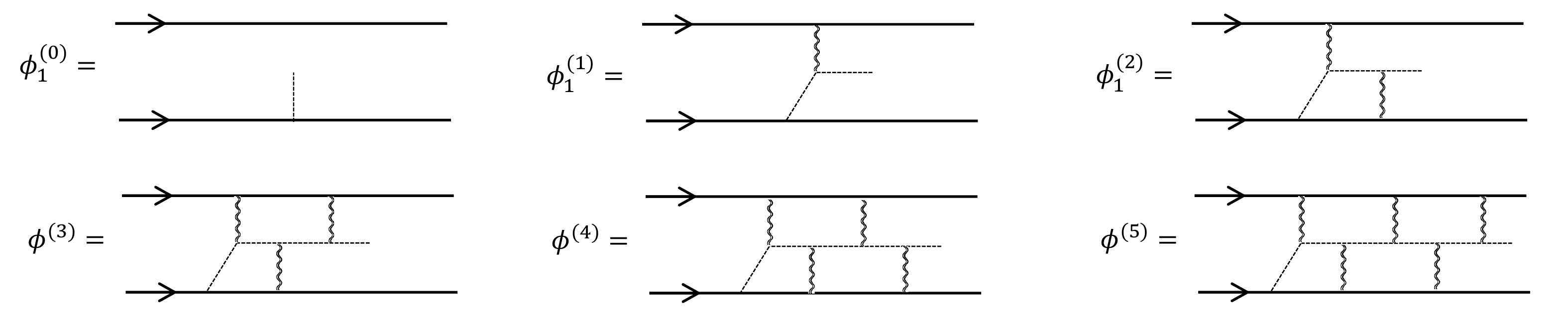}
         \caption{Horndeski frame.}
     \end{subfigure}
\caption{
The first few terms in perturbation theory for the scalar field profile near a two-body system.
In the Horndeski frame, there are diagrams involving graviton exchange, but these can all be removed by a field redefinition to the Einstein frame, where they become disformal contact interactions.  
} 
\label{fig:Ladder}
\end{figure}

\paragraph{Equal Masses Colliding.}
Consider two identical particles with mass $m$ colliding head-on with a relative velocity $v$ (which is non-relativistic, $v^2 \ll 1$, and yet sufficiently large that the backreaction from the field on the particle motion can be ignored). In this case, $G_1 = G_2$ (since bodies identical) and can be expressed in terms of the relative separation $r (t) = | \mathbf{x}_1 (t) - \mathbf{x}_2 (t) |$ (since $\partial_t r (t) = v$ is approximately constant). The resummed scalar profile is then given by solving \eqref{eqn:ladder_eqn}, which is now simply,
\begin{align}
 G - \beta_1  R_L^3 \;   \frac{\partial^2}{ \partial r^2} \left( \frac{G}{r} \right)  =  1 \; .
 \label{eqn:Collision_Eom}
\end{align}
It is tempting to view this differential equation for $G$ in the same light as the algebraic equations \eqref{eqn:KGphiEom} encountered in $K$-mouflage or Vainshtein resummation---in particular, one might imagine that a smooth resummation of the series \eqref{eqn:Gseries} (i.e. a smooth solution of \eqref{eqn:Collision_Eom} with boundary condition $G = 1 + ...$ at large $r$) can only be found if the differential operator $\hat{z} = \beta_1 R_L^3 \partial_r^2 r^{-1}$ is ``negative'' (otherwise there is a ``singularity'' of the form $G \sim 1/(1- \hat{z})$). However, this is not quite the case. The equation \eqref{eqn:Collision_Eom} can actually be solved for either sign of $\beta_1$, and screening takes place in either case \cite{Davis:2019ltc}. 
Rather, the sign of $\beta_1$ controls whether the \emph{homogeneous} equation $(1 - \hat{z}) G = 0$ has real solutions. When $\beta_1 < 0$, there are no real solutions which obey the asymptotic boundary condition, and so the resummed scalar field profile which solves \eqref{eqn:Collision_Eom} is unique. But when $\beta_1 > 0$, 
there are real solutions to the homogeneous equation, and the resummation becomes ambiguous: the boundary condition at large $r$ is not enough to fully determine $\phi (r)$, and there is a non-perturbative correction which appears at small $r$ with an undetermined constant of integration. 

We will now show this explicitly by solving \eqref{eqn:Collision_Eom}, since exact solutions to this equation are known (namely the Scorer functions, or inhomogeneous Airy functions, $\text{Gi} (z)$ and $\text{Hi} (z)$). Given the boundary condition that $G = 1 + ...$ at large $r$ (so that the Newtonian potential is recovered), one particular solution to \eqref{eqn:Collision_Eom} is, 
\begin{align}
 G (r) =  \frac{\pi r}{R_{\rm nl} } \times  \begin{cases}
    \text{Gi} \left(  \frac{r}{R_{\rm nl} }   \right)  \quad &\text{when } \beta_1 > 0 \; ,  \\
   \text{Hi} \left( - \frac{r}{R_{\rm nl} }   \right)  \quad &\text{when } \beta_1 < 0 \; . 
   \end{cases}
   \label{eqn:Scorer}
\end{align}
In contrast the Vainshtein and $P(X)$ examples like \eqref{eqn:Gresum_X2}, these entire functions smoothly extrapolate between the ladder expansion when $r \gg R_L$ and the small distance expansion at $r \ll R_L$ for either sign of $\beta_1$. 
However, \eqref{eqn:Scorer} is not the most general solution to \eqref{eqn:Collision_Eom}. We can also add to $G$ any additional $\delta G$ which obeys the homogeneous equation, which in this case is the Airy equation, $f''(x)=x f(x) $. \eqref{eqn:Scorer} can therefore be shifted by,
\begin{align}
 \delta G  (r) = \frac{r}{R_{\rm nl} } \times \begin{cases}
 c_1 \text{Ai} \left( \frac{r}{R_{\rm nl} } \right) + c_2 \text{Bi} \left(  \frac{r}{R_{\rm nl}}  \right)  \quad &\text{when } \beta_1 > 0 \; , \\ 
 c_1 \text{Ai} \left( - \frac{r}{R_{\rm nl} } \right) + c_2 \text{Bi} \left( - \frac{r}{R_{\rm nl} }  \right)  \quad &\text{when } \beta_1 < 0 \;  ,   
 \end{cases}  
\end{align}
where $c_1$ and $c_2$ are constants of integration. 
Now comes the importance of $\text{sign} (\beta_1)$. When $\beta_1 < 0$, these Airy functions are not consistent with the asymptotic boundary condition (since both $\text{Ai} (-r)$ and $\text{Bi} (-r)$ have an oscillatory fall-off like $ \sim r^{-1/4}$ at large $r$) and so we must set $c_1 = c_2 = 0$. 
However, when $\beta_1 > 0$, it is only $\text{Bi} (r)$ ($\sim e^{2 r^{3/2}/3}/r^{1/4}$ at large $r$) which is inconsistent with the boundary condition---the asymptotic expansion of $\text{Ai} (r) \sim e^{-2r^{3/2}/3}/r^{1/4}$ is \emph{invisible} in perturbation theory, and so \emph{any} choice of $c_1$ is a good solution to \eqref{eqn:Scorer} and coincides with the perturbative series at large $r$. 
Only when $\beta_1 < 0$ do we have a unique resummation of the perturbative series\footnote{
The freedom to add this new function with undetermined coefficient $c_1$ to the field profile stems from the ambiguity in ``going around'' the pole, $G_\phi \sim 1/(1-\hat{z})$.
This can be made explicit using the Borel resummation, $G_\phi (r) = \int_0^\infty dw \, \mathcal{I} (w, r)$ with $\mathcal{I} (w, r) = \exp \left( - \frac{R_{\rm nl}}{ r } w \pm w^3/3 \right)$ \cite{Davis:2019ltc}, which has a singularity at $w = \infty$ when $\beta_1 > 0$ (for which $\mathcal{I} \sim e^{+w^3/3}$) and therefore the resummation is ambiguous.  
}. 

Note that, in spite of this non-perturbative ambiguity, the ladder screening mechanism takes place regardless of the sign of $\beta_1$: since $\text{Gi} (r)$, $\text{Hi} (r)$ and $\text{Ai} (r)$ all $\sim r^0$ at small $r$, we have that $\phi (r) \sim r^0/ R_{\rm nl}$ when $r \ll R_{\rm nl}$, which is suppressed relative to the Newtonian potential $\phi \sim 1/r$ of the free theory.

\paragraph{Unequal Masses Colliding.}
Before moving on from this simple example, let us relax one of the assumptions and consider the head-on collision of two \emph{distinguishable} particles. In that case, $G_1$ and $G_2$ no longer coincide, and so \eqref{eqn:ladder_eqn} now implies a fourth-order equation for $G_1$, 
\begin{align}
G_1 - \hat{z}_2 \hat{z}_1 G_1 = 1 + \hat{z}_2 \; .
\label{eqn:Collision_Eom2}
\end{align}
Again, it is tempting draw parallels with the algebraic equations \eqref{eqn:KGphiEom} encountered in $K$-mouflage or Vainshtein resummation, and conclude that resummation requires the sign of $\hat{z}_2 \hat{z}_1$ to be ``negative'' (otherwise there is a singularity $G_1 \sim 1/(1-\hat{z}_2 \hat{z}_1)$). However, na\"{i}vely $\hat{z}_2 \hat{z}_1 \sim \beta_1^2 R_{L_1}^3 R_{L_2}^3$ is always ``positive''! 
This would suggest that the ladder resummation is never unique, since the homogeneous equation $(1- \hat{z}_2 \hat{z}_1 ) G_1 = 0$ always has real (non-perturbative) solutions that can be added to the field profile.
While this is true, one thing this schematic argument misses is the role played by the equivalence principle. We will now show that, when $\beta_1 < 0$, the equivalence principle removes any non-perturbative correction and guarantees a unique resummation of ladder diagarms.   

Exact solutions to \eqref{eqn:Collision_Eom2} are again Scorer functions and Airy functions. One particular solution of \eqref{eqn:Collision_Eom2} which coincides with perturbative series at large $r$ is,  
\begin{align}
 G_1 (r)  =  \frac{\pi r}{2 R_{\rm nl}} \begin{cases}
  \left( 1 + \sqrt{\frac{m_2}{m_1}} \right) \text{Gi} \left(  \frac{r}{R_{\rm nl} }   \right) + \left( 1 - \sqrt{\frac{m_2}{m_1}} \right) \text{Hi} \left( - \frac{r}{R_{\rm nl} }   \right)  \quad &\text{when } \beta_1 > 0 \; ,  \\[10pt]
   \left( 1 + \sqrt{\frac{m_2}{m_1}} \right)   \text{Hi} \left( - \frac{r}{R_{\rm nl}}   \right) +  \left( 1 - \sqrt{\frac{m_2}{m_1}} \right)  \text{Gi} \left(  \frac{r}{R_{\rm nl}}   \right) \quad &\text{when } \beta_1 < 0 \; ,
   \end{cases}
\end{align}
and similarly for $G_2(r)$ (with $m_2 \leftrightarrow m_1$). 
These entire functions again extrapolate smoothly between the ladder expansion at $r \gg R_L$ and the screened regime $r \ll R_L$ for either sign of $\beta_1$. 
However, we can also add any additional $\delta G_1 (r)$ which obeys the homogeneous equation (which now has four solutions, $\{  \text{Ai} ( \pm r) , \text{Bi} (\pm r) \}$ for either sign of $\beta_1$), and as before the only addition which is consistent with the boundary condition at large $r$ is the Airy function $\text{Ai} (r)$,
\begin{align}
\delta G_1 (r) = C m_2  \; \tfrac{ r }{ R_{\rm nl} } \,  \text{Ai} \left( \tfrac{r}{R_{\rm nl}}  \right) \; .
\end{align}
While this is a valid non-perturbative solution for either sign of $\beta_1$, when we use \eqref{eqn:ladder_eqn} to infer $\delta G_2$ we find that,
\begin{align}
 \delta G_2 (r)  =  \frac{r}{R_{\rm nl}} \times  \begin{cases}
  + C m_1  \;  \,  \text{Ai} \left( \frac{r}{R_{\rm nl}}  \right)   \quad &\text{when } \beta_1 > 0 \; ,  \\[10pt]
  - C m_1 \; \text{Ai} \left( \frac{r}{R_{\rm nl}}  \right)  \quad &\text{when } \beta_1 < 0 \; . 
   \end{cases}
\end{align}
When $\beta_1 > 0$, the additions $\delta G_1$ and $\delta G_2$ must have opposite signs. This means that the scalar profile sourced by particle 1 of mass $m_1$ ($\phi_1 (m_1)$) does not match the profile sourced by particle 2 of mass $m_2$ ($\phi_2 (m_2)$) if the masses were to be exchanged (i.e. $\phi_1 (m_2)  \neq \phi_2 (m_1) $). There must therefore be some kind of additional ``charge'', beyond the mass of the particle, which determines whether $\delta G_A > 0$ or $<0$, and this violates the equivalence principle unless $C=0$. 
On the other hand, when $\beta_ 1 >0$ then $\delta G_A > 0$ for all particles (or $<0$ for all particles), and is consistent with the equivalence principle for any value of $C$.

\paragraph{Resummation in Jordan Frame.}
Finally, we close this discussion of the two-body system with a comment about the importance of choosing the right frame for these calculations. As in the Vainshtein example above, performing a metric field redefinition to remove the mixing between $\phi$ and metric fluctuations has led to a simpler equation \eqref{eqn:ladder_eqn} in terms of $\phi$ only, albeit with a disformal coupling to matter. 
Had we instead worked in the original Horndeski frame \eqref{quarticA}, in which there is no disformal coupling, we would have found that these ladder diagrams are replaced by the ones shown in Figure~\ref{fig:Ladder}(b), in which graviton emissions from the  compact objects mix with a conformally emitted scalar fluctuation.  
One can verify by explicit (and laborious!) computation that these diagrams match the simpler Einstein frame diagrams in Figure~\ref{fig:Ladder}(a). 
In particular, note that the cubic $h^2 \phi$ vertices are proportional to the metric equation of motion, and these vertex factors effectively cancel the graviton propagators---there are no graviton poles in these diagrams, as expected from the fact that they can be removed via a field redefinition.  

\noindent Altogether, we have now shown that $\beta_{N+1} < 0$ is required in order for a smooth resummation of classical non-linearities and corresponding screening mechanism around a single compact object, and that $\beta_1 < 0$ is required for a unique resummation in two-body systems.   
Note that in the case of a single compact object, comparing $R_{\rm nl}$ and $\Lambda_{\rm sc}$ we find that $R_{\rm nl}^3 \Lambda_{\rm sc}^3 \sim m / M_P$ for any $N$ and $\delta$, while for a virialised two-body system we find $R_{\rm nl}^2 \Lambda_{\rm sc}^2 \sim m /M_P$ (analogous to the $P(X)$ scales of section~\ref{sec:PX}). Interestingly, the non-linear radius for a general two-body system obeys $R_{\rm nl}^3 \Lambda_{\rm sc}^3 \sim v^{2} \delta^{-3/4} \, m/M_P$ and seems as though it can be significantly larger or smaller than $m/M_P$ depending on the sizes of $v^2$ and $\delta$. It would be interesting to revisit this point in future, replacing our inference of $\Lambda_{\rm sc}$ in a vacuum with a more careful consideration of the strong coupling scale near a disformally coupled binary system.

\subsection{Positivity and UV Completion}
\label{sec:positivity}

It is an open question whether an EFT that exhibits Vainshtein screening in the IR can ever be UV completed \cite{Kaloper:2014vqa, Keltner:2015xda, Padilla:2017wth, deRham:2017imi, deRham:2017xox, Burrage:2020bxp}, particularly since the massless Galileon interactions violate the positivity bounds required for a standard Wilsonian UV completion \cite{Adams:2006sv}.
As in section~\ref{sec:PX_positivity}, we will now apply positivity bounds to the higher-order interactions and show that there can be no standard UV completion of screening from an even $X^{2n}$ interaction in $G_4 (X)$. 
Intriguingly, we find that theories which admit screening due to a large $X^{2n+1}$ interaction seem to satisfy the positivity constraints which rule out their even counterparts, suggesting these odd theories are more amenable to UV  completion.

\paragraph{Positivity of $X^2$.}
Let us begin with the leading self-interaction $-\beta_2 X^2$ in $G_4^2$, which contributes at tree level to the $\phi \phi \to \phi \phi$ scattering amplitude. This amplitude (about the background $\phi = 0$) 
was computed in \cite{Melville:2019wyy} and is reproduced here in Appendix~\ref{app:pos_WBG}.
Since this amplitude vanishes in the forward limit, $\partial_s^2 \mathcal{A} |_{t=0} = 0$, the simplest positivity bound \eqref{eqn:pos_ddA} simply places a constraint on $G_2 (X)$---to bound $G_4$ one must go beyond forward limit scattering. As described in \cite{deRham:2017avq} (see also \cite{Vecchi:2007na, Nicolis:2009qm}),
the same basic UV properties of unitarity, causality and locality require that a Lorentz-invariant EFT obeys, 
\begin{align}
 \partial_t \partial_s^2 \mathcal{A}_{\rm EFT} (s,t) |_{t=0}  \geq  \frac{3}{2 s_b} \partial_s^2 \mathcal{A}_{\rm EFT} (s,t) |_{t=0} \; , 
 \label{eqn:pos_dddA}
\end{align}
where $s_b$ is the scale up to which the EFT can be used to reliably compute the amplitude in the complex $s$ plane\footnote{
Strictly speaking the $\mathcal{A}_{\rm EFT}$ appearing in \eqref{eqn:pos_dddA} is the amplitude with all branch cuts subtracted up to $s_b$, but this distinction is unimportant at the (tree-level)  order at which we are working.
} (see Appendix~\ref{app:pos_derivation}). 
This leads to the bound \cite{Melville:2019wyy},
\begin{equation}
 \partial_t \partial_s^2 \mathcal{A}_{\rm EFT} (s,t) |_{t=0}  \; \propto  \;  - 2 \left(  \G_{4,X}^2 +  \G_4 \G_{4,XX} \right)  =   \beta_2 > 0 \, .
 \label{eqn:pos_b2}
\end{equation}
Notice that the scattering amplitude depends only on the coefficient $\beta_2$ of $G_4^2$, and is insensitive to $\beta_1$. 
If we set $\beta_1 =  0$ to remove any scalar-metric mixing (i.e. $\bar{G}_{4,X} = 0$), then this bound becomes simply $\bar{G}_{4,XX} < 0$ and coincides with the bound on the quartic Galileon \cite{deRham:2017imi}.   
In fact, there are even stronger positivity constraints on $\beta_2$ which can be derived by using additional information from the 1-loop EFT amplitude \cite{deRham:2017imi, Bellazzini:2017fep, Bellazzini:2019xts}} or crossing symmetry \cite{Tolley:2020gtv}, but we shall postpone those to the end of this section, and for the moment turn our attention to finding the analogue of \eqref{eqn:pos_b2} for the higher-point $X^{N+1}$ interactions.

\paragraph{Positivity of $X^{N+1}$.}
Since the higher-order interactions $-\beta_{N} X^{N}$ in $G_4^2$ do not contribute to the $2 \to 2$ scattering amplitude, they cannot be constrained directly using traditional positivity arguments. 
Furthermore, although they do contribute to the higher-point $N \to N$ scattering amplitude, these contributions \emph{always vanish} for the special forward-limit kinematics used in \cite{Chandrasekaran:2018qmx} to derive  $(-1)^N \lambda_N > 0$ for the $P(X)$ theory, so to date there have been no constraints placed on $\beta_N$ for $N>2$.

To constrain these higher-order interactions, we consider scattering fluctuations about a non-trivial background, $\phi = \alpha  t + \varphi$, and employ the positivity bounds of \cite{Grall:2021xxm} on the resulting $\varphi \varphi \to \varphi \varphi$ amplitude, which is given explicitly in Appendix~\ref{app:pos_WBG}. This amplitude also vanishes in the forward limit, but the analogue of \eqref{eqn:pos_dddA} provides a constraint on $G_4 ( - \alpha^2)$.
We find that this strategy for constraining higher-point interactions gives the analogous bound to the $P(X)$ theory, namely when $\alpha$ is taken to be very small,
\begin{align}
 \text{Positivity requires } (-1)^{N+1} \beta_{N+1} > 0 \; ,
\label{eqn:beta_pos}
\end{align}
for the $G_4(X)$ in \eqref{eqn:G4sq} and the sign required by UV positivity alternates. 
Again, notice that while no constraint is placed on the disformal coupling $\beta_1$ at this order, if we simply  set  $\beta_1 = 0$ to remove any scalar-tensor mixing then \eqref{eqn:beta_pos} becomes $(-1)^N \partial_X^N G_{4} |_{X=0} < 0$ for $N>1$. 

If the positivity bound \eqref{eqn:beta_pos} is violated, then the scalar-tensor EFT~\eqref{quarticA} can have no UV completion with the basic properties listed in Appendix~\ref{app:pos_derivation}, which are the direct analogues of unitarity, causality and locality for boost-breaking amplitudes. 
When this same alternating pattern was found in $P(X)$ theories, \cite{Chandrasekaran:2018qmx} were able to show explicitly that a fairly general class of tree-level UV completions could never generate a $\lambda_n$ with the ``wrong'' sign---it would be interesting in future to similarly study how \eqref{eqn:beta_pos} comes about for particular simple classes of UV completion.

\paragraph{Positivity of Disformal Coupling.}
Finally, we turn our attention to disformal coupling, $\beta_1$. Notice that this particular interaction, the linear term in $G_4^2$, does not affect any scalar scattering amplitude or positivity bound---this is because, as explained in section~\ref{sec:cutoff}, it represents a scalar-tensor mixing that can be removed via a field redefinition. However, this field redefinition changes the coupling to matter fields.
We can therefore place a positivity constraint on $\beta_1$ by considering a scattering process $\phi \psi \to \phi \psi$ between $\phi$ and any matter field $\psi$ appearing in \eqref{quarticA}, as proposed in \cite{deRham:2021fpu}. The explicit amplitude is given in Appendix~\ref{app:pos_WBG}, and applying the positivity bound~\eqref{eqn:pos_ddA} gives,
\begin{align}
\text{Positivity requires } \beta_1 > 0 \; ,
\label{eqn:pos_b1}
\end{align}
and the disformal coupling to matter must be \emph{positive} to be compatible with unitarity, causality and locality in the UV \cite{deRham:2021fpu}. 
Since $\beta_1 = - 2 \bar{G}_{4,X} \bar{G}_4$, this can also be written as simply $\bar{G}_{4,X} < 0$.
Note that since the $\beta_{N+1} X^{N+1}$ non-linearities do not contribute to the $\phi \psi \to \phi \psi$ amplitude at this order, the bound \eqref{eqn:pos_b1} applies to \emph{any} theory of the form \eqref{quarticA}, even one without any weakly broken Galileon tuning. 
It is particularly interesting that the bound on $\beta_1$ is something of an outlier---it does not conform to the pattern $(-1)^N \beta_N > 0$ of the higher ($N>1$) coefficients. 
One possible explanation of this is that only $\beta_1$ directly affects the causal structure (effective metric) that matter fields ``see''---in particular, it changes the sound speed of matter relative to the metric.

\paragraph{Subluminality of GWs.}
As pointed out in \cite{deRham:2021fpu}, this peculiar bound $\beta_1 > 0$ can imply that gravitational waves travel faster than any matter field, including light, on a background that spontaneously breaks Lorentz invariance. In particular, the speed of gravitational waves (relative to matter\footnote{
In the original Horndeski frame, neglecting the conformal coupling we have $c_{\rm mat} = 1$ and it is $c_{\rm GW}$ which is modified by the scalar-tensor mixing. But in the Einstein frame $c_{\rm GW} = 1$ and it is the speed of matter which is modified by the disformal coupling to $\phi$. Only the ratio $c_{\rm GW}/c_{\rm mat}$ is a frame-independent observable. 
}) in this quartic Horndeski theory \eqref{quarticA} with \eqref{eqn:G4sq} is,
\begin{align}
 c_{\rm GW}^2 = \frac{G_4^2}{G_4^2 -  X \partial_X G_{4}^2 }  =  1 - \beta_1 X  - (N+1) \beta_{N+1} X^{N+1}  + ... 
 \label{eqn:cGW}
\end{align}
on a time-like background ($X < 0$) when $|X| \ll 1$. 
In non-gravitational theories, positivity bounds can often coincide with the requirement that EFT does not admit superluminal propagation around simple backgrounds \cite{Adams:2006sv}, and indeed for the scalar self-interactions $\beta_{N+1} X^{N+1}$ with $N>1$ indeed that is what we find---the condition $(-1)^N \beta_N > 0$ is pushing $c_{\rm GW}^2$ below the matter speed in \eqref{eqn:cGW}. 
However, in gravitational theories the connection with subluminality is more subtle \cite{deRham:2019ctd, deRham:2021fpu}, and in particular we see that the bound $\beta_1 > 0$ on the disformal coupling actually pushes $c_{\rm GW}^2$ towards superluminal values (see also \cite{Shore:2007um, Babichev:2007dw})). 
It would be interesting to further explore the connection between positivity (i.e. causality in the UV) and the relative sound speeds in the low-energy EFT, particularly in theories in with a non-trivial scalar tensor mixing and in which Einstein and Jordan frame metrics do not coincide.

\paragraph{Consequences for Vainshtein Screening.}
In theories dominated by a $\beta_{N+1} X^{N+1}$ scalar interaction in $G_4^2(X)$, we observed in section~\ref{sec:screening} that Vainshtein resummation and screening can only take place around compact objects when $\beta_{N+1} < 0$. 
Comparing that with the positivity requirement $(-1)^{N+1} \beta_{N+1} > 0$, we find that theories with an even power of $X$ (including the quartic covariant Galileon) could never exhibit screening and be compatible with standard UV completion. On the other hand, theories with an \emph{odd} power of $X$ can simultaneously support screened scalar profiles in the IR and satisfy this positivity requirement for unitarity, causality and locality in the UV.

Of course, constructing an explicit UV theory which produces screening in the IR remains a difficult problem. In particular, in order to trust the classical resummation deep within the Vainshtein radius one is assuming that corrections from higher-derivative interactions are suitably suppressed (see for instance the discussion in \cite{deRham:2017xox}) and this is often not the case in explicit UV models (see for instance \cite{Burrage:2020bxp} for an example of UV completion which does not preserve the screening). 
While these positivity constraints by no means guarantee that any particular EFT can be UV completed in a way consistent with resummation and screening in the IR, they are a powerful way of removing large classes of models from consideration---there is no longer any need to search for UV completions which produce screening due to an $G_4^2 \sim \beta_N X^N$ interaction with $N$ even, since we have shown that none can exist.    

One important caveat is that we have focussed on scalar-tensor theories of the form~\eqref{quarticA}, with quartic Horndeski interactions only. 
As shown in \cite{deRham:2017imi}, the inclusion of other interactions (such as a cubic Galileon term) can open up a small region of parameter space in which positivity bounds are satisfied and Vainshtein screened solutions exist. 
Here, our goal was to demonstrate that there are several simple theories in which classical resummation and UV completion (at least at the level of existing positivity bounds) can co-exist peacefully. 
In future, it would be interesting to repeat the analysis performed here for more general scalar-tensor theories to search for further candidate theories which may admit standard UV completions and be phenomenologically viable---we will return to this point in section~\ref{sec:disc} below.

\paragraph{Consequences for Ladder Screening.}
For the DBI Galileon $G_4 = \sqrt{1 - \beta_1 X}$ (or equivalently, a disformal coupling to matter in Einstein frame), no Vainshtein screening takes place near single compact objects since the equation of motion \eqref{eqn:phi_eom_N} is linear in $\phi$. 
However, near multiple sources, the disformal coupling can provide a non-linear effect and an analogous resummation can lead  to ``ladder screening'' \cite{Davis:2019ltc} (see section~\ref{sec:ladder}). 
Comparing with the positivity bound $\beta_1  > 0$, we see that theories in which the ladder screened profile is unique have no standard UV completion. 
Instead, positivity requires that the resummation contains a non-peturbative ambiguity.
This clearly shows that these non-perturbative corrections are not some theoretical curiosity living in an unphysical region of parameter space, but rather are necessary consequences of unitarity, causality and locality in the UV. 
For disformally coupled scalar fields to exhibit a phenomenologically viable screening mechanism, it is essential that this feature of their resummation is better understood.

\paragraph{Stronger Positivity Bounds.}
Finally, let us remark that we have focussed on the simplest positivity bounds: \eqref{eqn:pos_ddA} and \eqref{eqn:pos_dddA} for the $X =0$ amplitudes (\eqref{eqn:posLO2} and \eqref{eqn:posNLO2} for the $X = -\alpha^2$ amplitudes). 
These fix the overall sign of the $\beta_N$ coefficients and already this is enough to rule out many low-energy EFTs that exhibit Vainshtein screening from ever having a standard UV completion. 
But there has been much recent progress developing even stronger positivity bounds which could also be applied to these theories.
For instance, how weakly coupled the UV completion must be can be quantified by subtracting the EFT loops from the bounds, as suggested in \cite{Bellazzini:2016xrt, Nicolis:2009qm} (see also \cite{deRham:2017avq, deRham:2017imi}).
This was carried out in \cite{Bellazzini:2017fep, Bellazzini:2019xts} for the cubic Galileon with Galileon symmetry weakly broken by an $X^2$ correction and indeed a very weak coupling, or low cutoff, was required (this is also the conclusion for massive gravity \cite{Cheung:2016yqr, Bellazzini:2017fep, deRham:2017xox, deRham:2018qqo}).
More recently, two-sided positivity bounds from full crossing symmetry have been derived  \cite{Tolley:2020gtv,Caron-Huot:2020cmc, Sinha:2020win} and these forbid a weak breaking of Galileon symmetry in the $\phi \phi \to \phi \phi$ amplitude \cite{Tolley:2020gtv}. 
For the simple $G_4 (X)$ we have considered here \eqref{eqn:G4sq}, this amounts to requiring that $\beta_2 \sim \delta^{3/2}$ when $N=1$, so that the $G_4$ and $G_2$ interactions both enter at $M_P \Lambda^3$.
At present there is no analogue of these fully crossing bounds for boost-breaking backgrounds, and so there is no known obstruction to having a higher-point $\beta_{N} X^N$ interaction ($N>2$) at a much lower scale than $M_P \Lambda^3$.
It would be interesting to explore in future whether these higher $n$-point interactions can also be constrained using full crossing symmetry for scattering amplitudes on a non-trivial background.

\section{Discussion}
\label{sec:disc}

In summary, we have investigated the constraints placed on scalar-tensor theories by demanding a viable screening mechanism (namely a smooth resummation of classical non-linearities) and also by requiring a standard UV completion (namely unitarity, causality and locality at high energies). 
In the context of $P(X)$ or quartic Horndeski theories in which a single interaction dominates, we have shown that a theory which exhibits screening can only be UV completed if the interaction is odd in $X$. 
We have also shown that, once metric and scalar fluctuations have been unmixed, it is the behaviour of $G_4^2 (X)$ which determines the strong coupling scale of the theory, and in particular $G_4^2 (X) = 1 - \beta_1 X$ corresponds to the highest possible cutoff (of $M_P \Lambda^3$, given the weakly broken Galileon power counting). 
This theory is equivalent to a disformally coupled scalar in the Einstein frame, and by reconsidering the ladder resummation near binary systems recently proposed in \cite{Davis:2019ltc}\footnote{
It would be interesting to compare this with the coupling between $\phi$ and matter which is induced when using the Galileon duality \cite{deRham:2013hsa, DeRham:2014wnv} to map a particular quintic Galileon to a free kinetic term, since this is potentially another example of a theory in which screening arises due to derivative couplings to matter rather than scalar self-interactions.
} we have shown that the screened profile in this theory is only unique for a particular sign of the disformal coupling.  
These results open up a number of interesting directions which can now be pursued. 

\paragraph{Speed of Gravitational Waves.}
Since the EFT \eqref{quarticA} breaks down at the scale $\Lambda$, which for typical dark energy values is close to the scale of LIGO frequencies ($\sim 10^2$Hz), we have not imposed any observational constraint on $c_{\rm GW}$.
However, we argued in section~\ref{sec:cutoff} that the tunings $\beta_2 = \beta_3 = ... = \beta_N = 0$ (so that $G_4 \sim \beta_{N+1} X^{N+1}$) raises this cutoff---for instance, for $\Lambda^3 \sim M_P H_0^2$, the symmetry-breaking parameter $\delta = \Lambda/M_P \sim 10^{-40}$ and so the next-to-leading $\beta_3 X^3$ interaction becomes strongly coupled at $10^4 \Lambda \sim 10^6$ Hz. This is safely above LIGO frequencies, but let us stress that it is the strong coupling scale around a flat Minkowski background with $\phi = 0$. For a cosmological background in which $|X| \sim 1$, the strong coupling scale of \eqref{quarticA} remains close to $\Lambda$ for the $G_4(X)$ considered in \eqref{eqn:G4sq}.
One exception is the DBI Galileon tuning, $G_4 = \sqrt{1-\beta_1 X}$, which does not have any interactions at $\Lambda$ even on a background with $|X| \sim 1$, and in that case the multi-messenger detection of GW170817 can be reliably used to constrain the effective disformal coupling on this background. As commented in \cite{Sakstein:2017xjx}, this bound is not stronger than other constraints, such as from horizontal branch stars or from energy-loss by the Primakov process in the Sun. 
A key open question is whether the ladder screening mechanism near binary systems still takes place about cosmological backgrounds and could alleviate these constraints on the disformal coupling.

\paragraph{Additional EFT Interactions.}
We have considered scalar-tensor EFTs of the form \eqref{quarticA}, which is sufficiently general to capture a wide range of models, and yet also simple enough to allow straightforward analytical solutions to the equations of motion.  
But even within the context of weakly broken Galileon symmetry, there are additional interactions ($G_3$ and $G_5$) which could be included in this Lagrangian. 
Indeed, it was observed in \cite{deRham:2017imi} that while a quartic Galileon interaction alone cannot simultaneously provide Vainshtein screening and satisfy positivity bounds, a combination of cubic and quartic Galileon interactions may do so. In the context of Horndeski theories, adding an interaction $G_3  (X) \phi^\mu_\mu$ to \eqref{quarticA} leads to an additional contribution to the positivity bounds \cite{Melville:2019wyy},
\begin{align}
 \beta_2  >  - \bar{G}_{3,X}^2 \; . 
\end{align}
which now allows for a negative value of $\beta_2$. 
On the other hand, near a static, spherically symmetric point-like source, the scalar field profile is determined by the equation of motion, 
\begin{align}
 G_\phi - y G_\phi^2 - z G_{\phi}^{3} = 1 
\end{align}
where there are now two expansion parameters,
\begin{align}
 z = 8 \beta_2 \frac{R_V^6}{r^6} \;\;\;\; \text{and} \;\;\;\; y = 4 \bar{G}_{3,X} \frac{R_V^3}{r^3} \; , 
\end{align}
which can be varied independently. A smooth resummation of the perturbative series solution which can interpolate between the boundary condition at large $r$ and the screened profile at small $r$ can be found providing that $z < - y^2 /3$, or in terms of the Horndeski functions,
\begin{align}
\beta_2 \leq 0 \;\; \text{and} \;\;  \bar{G}_{3,X} <  \sqrt{ - \tfrac{3}{2} \beta_2} \; . 
\end{align}
which coincides with the analogous result of \cite{Nicolis:2008in} for the Galileon when the $\bar{G}_{4,X}$ mixing with gravity is turned off. 
Such a theory can therefore have Vainshtein screening from the $\beta_2 X^2$ interaction in $G_4$ providing that $\bar{G}_{3,X}$ is sufficiently large. 
This is just one example of a theory beyond the quartic $G_4$ action \eqref{quarticA} that we have focussed on, and it will be interesting in future to explore further the interplay between resummation and positivity bounds in different classes of modified gravity theories, containing both additional self-interactions like above and possibly also additional light degrees of freedom (vector-tensor theories, etc.), as well as the more general weakly broken Galileon theories of \cite{Santoni:2018rrx}.


\paragraph{Including Gravitational Effects.}
We have implicitly worked throughout in the decoupling limit (large $M_P$), so that gravitational effects can be neglected.
Although we have referred to diagrams in which a graviton is apparently exchanged, since these interactions can be shuffled into purely scalar vertices via an appropriate field redefinition, they are not ``gravitational'' in that sense (i.e. at the level of the scattering amplitude there is no pole associated with this exchange). 
Since gravitational contributions can affect the positivity bounds \cite{Alberte:2020jsk, Caron-Huot:2021rmr} (see also \cite{Tokuda:2020mlf, Herrero-Valea:2020wxz, Noumi:2021uuv}), it would be interesting in future to investigate whether this might open up new regions of parameter space in which screening can coexist with unitarity, causality and locality in a (gravitational) UV completion.

\paragraph{Other Dispersive Arguments.}
The positivity bounds that we have used follow from a dispersion relation for the on-shell $2\to2$ scattering amplitude, but analogous dispersion relations exist for the 2-body potential \cite{Feinberg:1988yw}. Developing positivity-type arguments directly for the potential would provide a very powerful, model-independent, way to analyse fifth forces and screening. Work in this direction was begun in \cite{Brax:2017xho, Banks:2020gpu}, in which the force mediated by a generic dark sector field is expressed in terms of a (positive) spectral density.

\subsubsection*{Acknowledgements}
\noindent We thank Claudia de Rham, Johannes Noller and Jeremy Sakstein for useful comments on a earlier version of this work. 
SM is supported by an UKRI Stephen Hawking Fellowship (EP/T017481/1) and partially by STFC consolidated grants ST/P000681/1 and ST/T000694/1.

\appendix
\section{Scattering Amplitudes}
\label{app:positivity}

In this Appendix, we describe the positivity constraints required of a low-energy EFT of $P(X)$ or weakly broken Galileon form if it is to admit a standard UV completion (i.e. one which is unitarity, causality and local). 
First we provide a succinct list of the UV properties that underpin these bounds (and which would have to be violated in order to access any region of EFT parameter space excluded by positivity), both in the traditional Lorentz-invariant case \cite{Adams:2006sv}, and in the case of spontaneously broken Lorentz invariance \cite{Grall:2021xxm}. 
Then in sections~\ref{app:pos_PX} and \ref{app:pos_WBG}, we collect the scattering amplitudes for $P(X)$ and weakly broken Galileons \eqref{quarticA} respectively. 
After briefly reviewing the known positivity bounds in each case from scattering on the Lorentz-invariant background $\phi = 0$, we derive new bounds from scattering fluctuations $\varphi$ about the boost-breaking background\footnote{
Due to the shift symmetry of the covariant $\phi$ theory, the effective interactions for $\varphi$ are time-translation invariant---only boosts are broken by this background, since $\partial_\mu \phi$ provides a preferred (time-like) direction. 
},
\begin{align}
 \phi = \alpha \LambdaP^2 t + \varphi \;\; , \quad g_{\mu\nu} = \eta_{\mu\nu} 
 \label{eqn:pos_background}
 \end{align}
where $\alpha$ is sufficiently small that we can neglect any change in the background Minkowski spacetime geometry.
These new bounds allow us to go beyond the quartic interactions and place constraints on the signs of higher order terms. In particular, for a $P(X)$ theory in which  $\lambda_N X^N$ is the dominant interaction, our bound \eqref{eqn:pos_PX} reduces to $(-1)^N \lambda_N > 0$, which reproduces the result of \cite{Chandrasekaran:2018qmx}. Analogously, for a Horndeski-type theory in which $\beta_N X^{N-1} \left( \phi^\mu_\mu \phi^\nu_\nu - \phi^{\mu}_\nu \phi^\nu_\mu \right)$ with $N > 1$ is the dominant interaction in the Lagrangian (e.g. the $G_4  (X)$ given in \eqref{eqn:G4sq}), we show that $(-1)^N \beta_N > 0$ is required for a unitary, causal, local UV completion.

\subsection{The UV Axioms that Lead to Positivity Bounds}
\label{app:pos_derivation}

Connecting the foundational properties of field theory---unitarity, causality and locality---to properties of scattering amplitudes lay at the heart of the $S$-matrix programme \cite{Chew, Eden}. 
Today, these ingredients are routinely combined with a modern Effective Field Theory perspective to place constraints on the EFT coefficients, following the seminal work of \cite{Adams:2006sv}. 
The following is not intended as a comprehensive review of the subject, but rather a short list of the properties which we are assuming in the main text when we impose positivity bounds. 

\paragraph{Lorentz-Invariant Positivity.}
For a Lorentz-invariant scattering process between two identical scalars, $\phi \phi \to \phi \phi$, the corresponding amplitude $\mathcal{A} (s,t)$ is a complex function of the two Mandelstam variables,
\begin{align}
 s = - \eta^{\mu\nu} ( p_1 + p_2 )_\mu (p_1 + p_2 )_\nu \;\; , \;\;\;\; t = - \eta^{\mu\nu} (p_1 + p_3 )_\mu (p_1 + p_3 )_\nu
 \label{eqn:Mand}
\end{align}
where $\eta_{\mu\nu} = \text{diag} ( -, + , + , + )$ is the flat Minkowski metric and the third variable $u = -\eta^{\mu\nu} ( p_1 + p_4 )_\mu (p_1 + p_4 )_\nu $ is given by $u = -s -t$ due to the on-shell relation for particle 4. For brevity we are going to neglect factors of the scalar field mass (which is very small if $\phi$ has an approximate shift symmetry).
We say that this amplitude corresponds to a ``standard'' UV completion if it obeys the following properties:
\begin{itemize}

\item \emph{Unitarity}. Conservation of probability (namely that time evolution is implemented by a unitary operator) leads to the optical theorem, which expresses $\text{Im} \, \mathcal{A}_{2 \to 2}$ as a sum over all other $\mathcal{A}_{2 \to n}$ amplitudes. This can be used (together with a partial wave expansion) to show that that any $t$ derivative of $\text{Im} \, \mathcal{A}$ is positive \cite{Nicolis:2009qm},
\begin{align}
\partial_t^j \,  \text{Im} \, \mathcal{A} (s,t) |_{t=0} > 0 
\label{eqn:unit_pos}
\end{align}
for any value of $s$ in the physical $s$-channel region (i.e. $s>0$ when $t=0$ and neglecting masses). This is the ``positive'' part of the positivity bounds.

\item \emph{Crossing symmetry}\footnote{
For local quantum theories with a mass gap, crossing has been rigorously proven from unitarity, causality and locality at the level of off-shell correlation functions, and see \cite{Mizera:2021ujs, Mizera:2021fap} for recent progress towards an entirely on-shell demonstration.
}. 
Crossing relates ingoing and outgoing states, and for identical scalar particles crossing particles 2 and 4 leads to the simple relation,
\begin{align}
\mathcal{A} (s,t) = \mathcal{A} (u , t)
\label{eqn:crossing}
\end{align}
between $s$- and $u$-channel amplitudes. This allows \eqref{eqn:unit_pos} to be applied also at negative values of $s$.  

\item \emph{Causality}.
A causal interaction produces an analytic response function, and for this 2-particle process this (together with crossing symmetry) amounts to $\mathcal{A} (s,t)$ being analytic in $s$ at fixed $t$ for any $\text{Im} \, s \neq 0$ on the physical sheet \cite{bogoliubov1959introduction, Hepp_1964, Bros:1964iho, Bremermann:1958zz, Martin:1965jj}. This allows the use of Cauchy's residue theorem,  
\begin{align}
\partial_s^j \mathcal{A} (s ,t) |_{s=s_0}  = \oint_C \frac{ds}{2 \pi i} \frac{ \mathcal{A} (s, t) }{ (s-s_0)^{j+1} }
\label{eqn:anal}
\end{align}
where the contour $C$ contains $s_0$, and modulo any poles and branch cuts on real axis (which are fixed by unitarity). The fact that the transition from the amplitude $\mathcal{A}_{\rm EFT}$ in the EFT to $\mathcal{A}_{\rm UV}$ in the UV completion must be smooth is the bridge that allows the positivity condition \eqref{eqn:unit_pos} to be applied in the EFT (which is not itself unitary at all scales). 

\item \emph{Locality}. 
Locality, or at least polynomial boundedness of the partial wave amplitudes, can be combined with unitarity and causality to give the Froissart bound \cite{PhysRev.123.1053, Martin:1962rt, Jin:1964zza}\footnote{
Note that the Froissart bound has not been proven with the same level of rigour in local quantum field theories without a mass gap---here, when we refer to ``locality'' of the UV, this corresponds to demanding that the high-energy growth amplitude do not exceed the Froissart bound of gapped theories, even when we neglect the mass of $\phi$. 
}, 
\begin{align}
\lim_{s\to \infty} | \mathcal{A} (s,t) | < s^2 \; . 
\label{eqn:Froissart}
\end{align} 
This allows us to discard any large $s$ contribution to \eqref{eqn:anal} providing $j \geq 2$.

\end{itemize}
Putting these ingredients together,
a standard (unitary, causal, local) UV completion at high energies therefore requires various bounds on $\mathcal{A}_{\rm EFT}$. For instance, the forward limit of $\partial_s^2 \mathcal{A}$ must be positive \cite{Adams:2006sv},
\begin{align}
 \partial_s^2 \mathcal{A}_{\rm EFT}  |_{\substack{s=0 \\ t=0 }} &=  \frac{2}{\pi} \int_{s_b}^{\infty} \frac{ds}{s^3}  \; \text{Im} \, \mathcal{A}_{\rm UV} |_{t=0}   > 0     \; , \label{eqn:posLO}   
\end{align}
and the $t$ derivatives must be bounded in terms of lower-order derivatives \cite{Vecchi:2007na, Nicolis:2009qm, deRham:2017avq}, 
\begin{align}
 \left( \partial_t + \tfrac{3}{2 s_b}   \right) \partial_s^2 \mathcal{A}_{\rm EFT}  |_{\substack{s=0 \\ t=0 }} &=  \frac{2}{\pi} \int_{s_b}^{\infty} \frac{ds}{s^3}  \left( \partial_t + \tfrac{3}{2} \tfrac{s - s_b}{s \, s_b}   \right) \text{Im} \, \mathcal{A}_{\rm UV} |_{t=0}   > 0  \; , \label{eqn:posNLO} 
\end{align}
where $s_b$ is the scale at which the branch cut on the positive real-axis begins, which generically is set by the mass gap, $4m^2$.
In weakly coupled theories, a portion of the branch cut between $4m^2$ and the EFT cutoff $\Lambda$ may be subtracted, allowing for a large $s_b$ and hence a stronger bound \eqref{eqn:posNLO}.

\paragraph{Positivity without Boosts.}
When considering fluctuations around a boost-breaking background, e.g. $\phi = \alpha t + \varphi$, the amplitude for scattering $\varphi$ fluctuations is constrained by three fewer symmetries and therefore can depend explicitly on three additional variables, which we take to be the energies $\omega_1, \omega_2 ,  \omega_3$ of the fluctuations (since time translations are unbroken $\omega_4 = - \omega_1 - \omega_2 - \omega_3$ is fixed by energy conservation). Furthermore, no symmetry connects the coefficients of $\dot \varphi^2$ and $(\partial_i \varphi)^2$ in the Lagrangian, and so $\varphi$ may have a non-trivial speed of sound. Focussing on theories in which the free propagation of the scalar field is determined by $\omega^2 = c_s^2 k^2$, where $k$ is the magnitude of the spatial momentum, we will abuse notation and define effective Mandelstam variables on this background using the effective metric of the free propagation,
\begin{align}
 s = (\omega_1 + \omega_2 )^2 - c_s^2 ( k_1 + k_2 )^2 \;\; , \;\;\;\; t = (\omega_1 + \omega_3 )^2 - c_s^2 ( k_1 + k_3 )^2 \; , 
 \label{stu}
\end{align}
which coincide with \eqref{eqn:Mand} when $c_s = 1$. This has the advantage that $u$ remains $-s-t$. 

The central distinction with Lorentz-invariant positivity bounds is that now some prescription must be provided for how to hold the three energy variables of $\mathcal{A} (s,t, \omega_1 ,\omega_2, \omega_3 )$ fixed when performing the partial derivatives and integration in any dispersion relation (including \eqref{eqn:posLO} and \eqref{eqn:posNLO}). 
In particular for $s$-channel scattering, since the spatial momenta $c_s | k_1 + k_2 | > \omega_1 - \omega_2$ on-shell, the Mandelstam $s$ defined in \eqref{stu} must obey,
\begin{align}
 s \leq ( \omega_1 + \omega_2 )^2 - (\omega_1 - \omega_2 )^2 \; .
 \label{sbound}
\end{align}
This means that holding $\omega_1$ and $\omega_2$ fixed is not an option, since \eqref{sbound} would always be violated at sufficiently large $s$, invalidating any unitarity bound on the UV amplitude (which only apply to physical on-shell momenta).

This problem was first considered in \cite{Baumann:2015nta}, where a convenient ``centre-of-mass-frame'' kinematics was used ($\omega_1 = \omega_2 = - \omega_3 = \sqrt{s}$), but this choice introduces unphysical branch cuts and spoils the crossing relation \eqref{eqn:crossing}. 
More recently, \cite{Grall:2021xxm} reconsidered the problem and showed that the analogous properties for the amplitude $\mathcal{A}_{\varphi \varphi \to \varphi \varphi}$ required for positivity bounds are\footnote{
Note that in the full UV theory, $\varphi$ will generally be replaced in the unbroken phase by some local operator $\mathcal{O}$, in which case it is the off-shell correlator of $\mathcal{O}$'s which must obey these properties. 
}, 
\begin{itemize}

\item \emph{Unitarity}. 
The optical theorem can again be used (via a suitable spherical wave expansion) to establish that the imaginary part is positive in the forward limit \cite{Grall:2020tqc}\footnote{
See also \cite{Baumann:2011su, Baumann:2014cja, Koehn:2015vvy, deRham:2017aoj} for earlier work applying unitarity to theories in which boosts are broken. 
},
\begin{align}
\partial_t^n \text{Im} \, \mathcal{A} (s, t, \omega_1 ,\omega_2, \omega_3 ) |_{ \substack{t=0 \\ \omega_1 = - \omega_3} } > 0 
\label{eqn:unitarity2}
\end{align}
for any physical value of $s$ and the energies. 

\item \emph{Analyticity}. 
Since traditional proofs of analyticity from causality for Lorentz-invariant amplitudes leverage the so-called Breit frame (in which the spatial part of $p_1 - p_3$ vanishes), it is convenient to change variables from $(\omega_1 ,\omega_2 ,\omega_3)$ to a new set of variables $(\gamma , M , \omega_t)$ which correspond to the three components of $(p_1 - p_3)_i$, 
\begin{align}
\omega_1 &= \gamma M + \frac{\omega_t}{2} \; , \;\;\;\; &\omega_2 &= \frac{s-u}{8M} - \frac{\omega_t}{2}  \nonumber \\
\omega_3 &= -\gamma M + \frac{\omega_t}{2} \; , \;\;\;\; &\omega_4 &= \frac{u-s}{8M} - \frac{\omega_t}{2}\; . 
\label{eqn:gMwt}
\end{align}
The choice $\gamma = 1$, $\omega_t = 0$ and $M = m$ corresponds to Breit-frame-kinematics, but in general any $\gamma > 1$ and $M > 0$ correspond to real physical momenta in the forward limit.   
The UV requirement that leads to positivity bounds in the IR is that $\mathcal{A}$ is an analytic function of $s$ at fixed $t, \gamma, M , \omega_t$, so that Cauchy's theorem may once again be applied, 
\begin{align}
\partial_s^j \mathcal{A} (s, t, \gamma , M , \omega_t ) |_{s=s_0} =  \oint_C \frac{ds}{2\pi i}  \frac{ \mathcal{A} (s, t, \gamma , M , \omega_t ) }{ (s - s_0 )^{j+1} } \; . 
\end{align}
The connection between analyticity and causality when boosts are spontaneously broken has yet to be put on the same footing as the Lorentz-invariant case, but at least heuristically (at large $s$) the same arguments seem to apply for a certain range of sound speeds \cite{Grall:2021xxm}, and the careful choice of variables \eqref{eqn:gMwt} guarantees analyticity at any order in perturbation theory.  

\item \emph{Crossing}. 
The analogue of the crossing relation \eqref{eqn:crossing}, 
\begin{align}
\mathcal{A} (s, t, \omega_1, \omega_2, \omega_3 ) = \mathcal{A} (u, t, \omega_1, \omega_4, \omega_3 ) \; , 
\end{align}
implies that $\mathcal{A} (s, t , \gamma, M , \omega_t ) = \mathcal{A} (u, t , \gamma, M , \omega_t )$. 
As with analyticity, this relation clearly holds at any order in perturbation theory (just by virtue of the fact that we symmetrise over the external kinematics of any Feynman diagram, so exchanging the labels of particles 2 and 4 leaves the amplitude unchanged), and also holds at high energies where Lorentz symmetry is restored.  

\item \emph{Boundedness}.
The final UV assumption that underpins boost-breaking positivity bounds is the analogue of the Froissart bound \eqref{eqn:Froissart}, 
\begin{align}
 \lim_{s\to \infty}  | \mathcal{A} (s,t, \gamma, M , \omega_t )   | < s^2  \; , 
\end{align}
which applies in the high-$s$ regime where Lorentz symmetry is restored and so follows from the usual arguments (one can also argue for this boundedness directly from the spherical wave expansion of the boost-breaking amplitude \cite{Grall:2021xxm}). 

\end{itemize}
Putting these UV properties together, the analogue of the forward limit positivity bound \eqref{eqn:posLO} is then,
\begin{align}
 \partial_s^2 A_{\rm EFT} ( s, t , \gamma,  M , \omega_t ) |_{\substack{ s = 0\\ t=0 \\ \omega_t = 0 } } = \frac{2}{\pi}  \int_{s_b}^{\infty} \frac{ds}{s^3}  \; \text{Im} \, \mathcal{A}_{\rm UV} (s, t, \gamma, M , \omega_t ) |_{ \substack{t=0 \\ \omega_t = 0} }  > 0  \; .
 \label{eqn:posLO2}
\end{align}

To go beyond the forward limit, one must correct for the fact that the Breit variables \eqref{eqn:gMwt} (which make analyticity and crossing manifest) depend explicitly on $t$, and therefore $t$ derivatives of $\mathcal{A} (s,t,\gamma, M ,\omega_t)$ are \emph{no longer} strictly positive, 
\begin{align}
 \partial_t \text{Im} \, \mathcal{A} (s, t, \gamma , M, \omega_t  ) = \left( \partial_t + \tfrac{1}{8M} \partial_{\omega_2} \right) \text{Im} \, \mathcal{A} \left( s, t, \omega_1, \omega_2, \omega_3 \right) \; . 
\label{eqn:dw2}
\end{align}
due to the $\partial_{\omega_2}$ term.
As described in the Appendix of \cite{Grall:2021xxm}, rather than fix $\omega_2$ using \eqref{eqn:gMwt} the key to going beyond the forward limit in a way which preserves good crossing behaviour is to consider an \emph{integral} of the amplitude over a small interval,
\begin{align}
\mathcal{I} ( s, t, \gamma , M ,  E_2 ) := \int_{ \frac{s-u}{8M} - E_2 }^{\frac{s-u}{8M} + E_2 } d \omega_2 \; \mathcal{A} ( s, t , \omega_1 , \omega_2 ,  \omega_3 ) |_{\omega_1 = -\omega_3 = \gamma M}  \; .
\end{align}
We assume that the constant $E_2$ can be chosen sufficiently small that this integral converges for any $s$ (which is certainly the case in perturbation theory, since  $\mathcal{A} (s, t, \omega_1 , \omega_2, \omega_3 )$ is analytic in $\omega_2$ at fixed $s$), and so provides a new complex function which shares the properties listed above---in particular, the integration limits have been chosen so that $\mathcal{I} ( u, t, \gamma, M , E_2) = \mathcal{I} ( s, t, \gamma, M , E_2)$ inherits the crossing relation of $\mathcal{A}$. 
The purpose of considering $\mathcal{I}$ is that now unitarity gives,
\begin{align}
\left( \partial_t + \tfrac{1}{8M} \partial_{E_2}   \right) \text{Im}\,  \mathcal{I} (s, t, \gamma , M , E_2 ) >  0  
\end{align}
for all $s > 0 , \;  M >0  $ and $\gamma \geq s / (s - 4 E_2 M) \geq 1$ (to be compatible with the condition \eqref{sbound} for real momenta in the $s$-channel), since the $\partial_{E_2}$ term compensates for the $\partial_{\omega_2}$ term in \eqref{eqn:dw2}.
This gives the analogue of the positivity bound \eqref{eqn:posNLO} for amplitudes with spontaneously broken Lorentz boosts,
\begin{align}
\left[ \partial_t + \tfrac{3}{2 s_b} + \tfrac{1}{8M} \partial_{E_2}   \right] \partial_s^2 \mathcal{I}_{\rm EFT} |_{\substack{s=0 \\ t=0 }}  
= \frac{2}{\pi} \int_{s_b}^{\infty} \frac{d s}{ s^3}  \;\left( \partial_t + \tfrac{1}{8M} \partial_{E_2}  + \tfrac{3}{2} \tfrac{ s - s_b}{ s \, s_b}  \right)  \text{Im} \, \mathcal{I}_{\rm UV}  |_{t=0}  
> 0 \;  .
\label{eqn:posNLO2}
\end{align}

~\\
The above lists of UV axioms are intended to make clear the physical meaning of the positivity bounds applied below and in the main text. 
If one were to construct a UV theory in which the $2 \to 2$ scattering amplitude violates one or more of the above assumptions, then this theory can give rise to a low-energy EFT which violates the positivity bounds. 
Such UV theories certainly exist (e.g. a simple two-scalar field model in which one scalar has the ``wrong sign'' kinetic term). 
The purpose of the positivity bounds is to provide a diagnostic of these UV features: we may of course continue to study interactions like $\lambda_2 X^2$ with $\lambda_2 < 0$, or anti-DBI, or any low-energy EFT which violates positivity, but we should be aware of the fact that they are implicitly committing us to a non-standard UV completion in which one of the above basic properties does not hold.

\subsection{$P(X)$ Amplitudes}
\label{app:pos_PX}

In the main text, our focus was on simple $P(X)$ theories \eqref{eqn:PN} in which a single $X^N$ interaction dominates. This makes for a cleaner comparison between the conditions for resummation (smooth screened solutions) and for positivity (standard UV completion). 
Here, we will consider the scattering amplitude and positivity bounds for a general $P(X)$ theory, and only specialise to \eqref{eqn:PN} at the end. 

\paragraph{Lorentz-Invariant Bounds.}
Expanding $\mathcal{L} = \Lambda^4 P(X)$ about the Lorentz-invariant vacuum $\phi = 0$ to quartic order,
\begin{align}
 \mathcal{L} = \bar{P}_{,X} ( \partial \phi )^2 + \frac{1}{2} \bar{P}_{,XX} \frac{ ( \partial \phi )^4}{\Lambda^4} \; .
\end{align}
where the overbar denotes that the function is evaluated at $X=0$. 
Classical stability requires that $\bar{P}_{,X} > 0$ (this can be viewed as positivity of the $1\to 1$ scattering process, i.e. fluctuations in the free theory have unitary propagation). 
A similar constraint can be placed on $P_{,XX}$ by considering the $2\to 2$ scattering amplitude,
\begin{align}
\mathcal{A} (s,t) = \frac{ \bar{P}_{,XX} }{ \bar{P}_{,X}^2} \;  \frac{s^2 + t^2 + u^2 }{ \Lambda^4 }
\end{align} 
where the factor of $\bar{P}_{,X}^2$ arises from the canonical normalisation of the field. 
For generic $P(X)$ theories, the Lorentz-invariant positivity bound \eqref{eqn:posLO} requires that $\bar{P}_{,XX} > 0$.

\paragraph{Positivity without Boosts.}
Expanding $\mathcal{L} = \LambdaP^4 P(X)$ about the boost-breaking vacuum $\phi = \alpha \LambdaP^2 t$ to quadratic order,
\begin{align}
\int d^4 x \,  \mathcal{L} = \int dt \, d^3 \mathbf{x} \; \frac{Z^2}{c_s^3} \left(   \dot \varphi^2 - c_s^2 ( \partial_i \varphi )^2     \right)
\end{align}
where $\dot \varphi = \partial_t \varphi$ and the wavefunction normalisation and sound speed are given by,
\begin{align}
 \frac{Z^2}{c_s^3} = -2 P_{,X} + 4 \alpha^2 P_{,XX} \;\; , \quad c_s^2  = \frac{ -2 P_{,X} }{ -2 P_{,X} + 4 \alpha^2 P_{,XX} } \; ,
\end{align} 
where each function is evaluated at $X=-\alpha^2$. 
Classical stability requires that these are both positive. 
Expanding to quartic order, the effective action for $\varphi$ is,
\begin{align} 
 S[ \varphi ]&= \int d^4 x \, c_\pi^{-3} \Big(
-\tfrac{1}{2} ( \hat{\partial} \varphi)^2
+ \frac{ \alpha_1 }{ \hat{\Lambda}^2 }   \;  \dot \varphi^3 
- \frac{ \alpha_2 }{ \hat{\Lambda}^2 }  \; \dot \varphi ( \hat{\partial} \varphi)^2   
+ \frac{ \beta_1 }{ \hat{\Lambda}^4 }  \; \dot \varphi^4
- \frac{ \beta_2 }{  \hat{\Lambda}^4 }   \; \dot \pi^2  ( \hat{\partial} \varphi)^2  
+  \frac{ \beta_3 }{ \hat{\Lambda}^4 }   \; ( \hat{\partial} \varphi)^4  
\Big)\,,
\label{eqn:EFT_action}
\end{align}
where $( \hat{\partial} \varphi )^2 = -\dot \varphi^2 + c_s^2 \delta^{ij} \partial_i \varphi \partial_j \varphi$ is contracted using the effective metric which determines the free propagation and the overall factor of $c_\pi^3$ ensures canonical normalisation \cite{deRham:2017aoj, Grall:2020tqc}.
$\{ \alpha_1, \alpha_2 ,\beta_1, \beta_2, \beta_3 \}$ are constant Wilson coefficients, which are partially fixed by the non-linearly realised boost symmetry,
\begin{align}
	\alpha_2= \frac{1-c_s^2}{2c_s^2}\,,\quad \beta_2=\frac{3}{2c_s^2}\alpha_1+\frac{(1-c_s^2)^2}{2c_s^4}\,,\quad \beta_3=\frac{1-c_s^2}{8c_s^4}\, .
	\label{eqn:superfluid}
\end{align} 
The remaining coefficients are given by,
\begin{align}
 \hat{\Lambda}^2 = \frac{\alpha}{Z} \Lambda^2 \;\; , \;\; \frac{\alpha_1}{c_s^3} =  \frac{4 \alpha^4}{3 Z^2} P_{,X}^3 \partial_X \left( \frac{ P_{,XX} }{P_{,X}^3 }  \right)  \;\; ,  \;\; \frac{\beta_1}{c_s^3} =  \frac{2 \alpha^6}{3 Z^2} \partial_X \left( P_{,X}^3  \partial_X \left( \frac{ P_{,XX} }{P_{,X}^3 }  \right) \right) \; ,
\end{align}
evaluated at $X=-\alpha^2$. 
The $2 \to 2$ scattering amplitude from the EFT \eqref{eqn:EFT_action} has been studied in \cite{Baumann:2015nta, Grall:2021xxm}, and is given explicitly by, 
\begin{align}
\hat{\Lambda}^4\mathcal{A}_s (s,t, \omega_s, \omega_t, \omega_u) &= 
2 \beta_3(s^2+t^2+u^2)  
+ (2 \beta_2-4 \alpha_2^2) (s\,\omega_s^2 + t\, \omega_t^2 +u\, \omega_u^2) \nonumber \\
&\quad+ 24 \omega_1 \omega_2 \omega_3 \omega_4  \left[   \left( \beta_1- 4 \alpha_1 \alpha_2  \right) 
-  \frac{3}{2} \alpha_1^2  \,   \left(\frac{\omega_s^2}{s}+\frac{\omega_t^2}{t}+\frac{\omega_u^2}{u}\right) \right] \; .   \nonumber 
\end{align}
Note that $\mathcal{A}$ now depends explicitly on the energies of the fluctuations, $\omega_s = \omega_1 + \omega_2$, $\omega_t = \omega_1 + \omega_3$ and $\omega_u = \omega_1 + \omega_4$, and we define $s$ and $t$ using the effective metric which governs the propagation of $\varphi$ \eqref{stu}.

The simplest positivity bound on this boost-breaking background \eqref{eqn:posLO2} gives the constraint\footnote{
The discrepancy with Appendix D of \cite{Ye:2019oxx} is due to the fact that we have evaluated the amplitude with the Breit kinematics \eqref{eqn:gMwt} to properly account for the breaking of boosts. 
},
\begin{align}
&\frac{P_{,XX} }{ P_{,X}^2 } - \alpha^2 \left( - \frac{ 2 P_{,XXX} }{ P_{,X}^2 } \right) + \alpha^4 
\left( 
 \frac{ 3 P_{,XX}^3 }{ 2 P_{,X}^4 } +  \frac{ P_{,XX} P_{,XXX} }{ 2 P_{,X}^3 } +  \frac{ P_{,XXXX} }{ 2 P_{,X}^2 }  \right)  \nonumber  \\
&\qquad\qquad\qquad\qquad\qquad\quad- \alpha^6 \left(
 \frac{2 P_{,XX}^4 }{ P_{,X}^5 } - \frac{ P_{,XXX}^2 }{ P_{,X}^3 } + \frac{  P_{,XX} P_{,XXXX} }{ P_{,X}^3 }  \right) > 0 \; , 
 \label{eqn:pos_PX}
\end{align}
where each function is evaluated at $X = - \alpha^2$. Although we have assumed that $\alpha$ is small enough to neglect any curvature sourced by $\phi$, at no point have we expanded in $\alpha \ll 1$. Also note that the tree-level Feynman diagrams encoded by each term can be inferred from the power of $P_{,X}$ in the numerator---a factor of $P_{,X}^{I+2}$ corresponds to $I$ internal lines.

For a $P(X)$ theory with power counting \eqref{eqn:Kexp}, tuned \eqref{eqn:KRaise} so that the dominant interaction is $\lambda_n X^n$ with $n > 1$, \eqref{eqn:pos_PX} becomes,
\begin{align}
(-\alpha^2)^{n} \frac{ \lambda_n }{ \alpha^2 } \left[ 
1 
+ 6n (- \alpha^2 )^{n} \frac{\lambda_n}{\alpha^2}
 + 12 n (2n -1 ) ( - \alpha^2 )^{2n }  \frac{ \lambda_n^2  }{\alpha^4}
 +  8 n^2  (1 - 2 n)^2 (-  \alpha^2 )^{3 n} \frac{ \lambda_n^3 }{\alpha^6}
 \right] > 0  \; .
 \label{eqn:pos_KX}
\end{align}
In particular, if we now consider small values of $\alpha \ll 1$, then this becomes $(-1)^n \lambda_n > 0$, as presented in the main text. In fact, from the full expression \eqref{eqn:pos_KX} we see that this positivity bound is satisfied for \emph{all} real values of $\alpha$ when $(-1)^n \lambda_n > 0$. For theories in which $(-1)^n \lambda_n < 0$, although the $\phi = 0$ background violates positivity, there is always an interval of non-zero values for $\alpha$ which satisfy \eqref{eqn:pos_KX} (determined by the real roots of this order $3n$ polynomial). 
In a theory in which $\lambda_n$ has the ``wrong'' sign according to Lorentz-invariant positivity arguments, it would be interesting to search for a UV completion for fluctuations about a non-trivial vacuum such as this $\phi = \alpha \Lambda t$, which can satisfy \eqref{eqn:pos_KX}. 
That positivity can be used to assess which vacua in the IR may have standard UV completions will be discussed in more detail elsewhere \cite{Noller:2021toappear}.

\subsection{Horndeski Amplitudes}
\label{app:pos_WBG}

\paragraph{Lorentz-Invariant Bounds.}
For the weakly broken Galileon~\eqref{quarticA}, the $\phi \phi \to \phi \phi$ scattering amplitude about the $\phi = 0$ background and its resulting positivity constraint on $G_4 (X)$ was computed in \cite{Melville:2019wyy}. 
Expanding the action to quartic order in the fields, the two dominant interactions are given in \eqref{eqn:Horndeski_vertex_LO}. 
The tree-level amplitude is then given by the two Feynman diagrams\footnote{
Note that this graviton exchange diagram does \emph{not} give rise to any $t$-channel pole, since the antisymmetric structure of the vertex exactly cancels the graviton propagator, which can be written symbolically as\begin{equation}
\tilde{h}_\mu^\nu
\; \feynmandiagram [horizontal=a to b] {
a -- [photon] b
}; \;
\tilde{h}_\rho^\sigma =  \left( \delta^{\mu \alpha \rho}_{\nu \beta \sigma} \partial_\alpha \partial^\beta \right)^{-1}
\end{equation}
} shown in Figure~\ref{fig:tree_level},
\begin{align}
 \mathcal{A}_{\phi \phi \to \phi \phi} (s,t) = 6 \left( \bar{G}_{4,XX} + \frac{\bar{G}_{4,X}^2}{\bar{G}_4}  \right) s t u \; . 
\end{align}
While the simplest bound \eqref{eqn:posLO} places a constraint on $G_2 (X)$, to bound $G_4$ one must go beyond forward limit scattering. The positivity bound \eqref{eqn:posNLO} leads to \cite{Melville:2019wyy},
\begin{equation}
 \G_{4,X}^2 +  \G_4 \G_{4,XX}  = -\beta_2 < 0 \, .
\end{equation}

\begin{figure}
\centering
$A_{4} $
		\qquad = \quad
		\begin{tikzpicture}[baseline=-0.6cm]
			\begin{feynman}
				\vertex (a1);
				\vertex [below=0.5cm of a1] (b1);
				\vertex [below=0.5cm of b1] (c1);
				\vertex [left=0.5cm of a1] (a2);
				\vertex [below=0.5cm of a2] (b2);
				\vertex [below=0.5cm of b2] (c2);
				\vertex [left=0.5cm of a2] (a3);
				\vertex [below=0.5cm of a3] (b3);
				\vertex [below=0.5cm of b3] (c3);
				\vertex [left=0.5cm of a3] (a4);
				\vertex [below=0.5cm of a4] (b4);
				\vertex [below=0.5cm of b4] (c4);
				\node at (-0.5,  -1.5) {$ \G_{4,XX} $};
				
				\diagram*{
				(a3) -- [scalar] (b2),
				(c3) -- [scalar] (b2),
				(a1) -- [scalar] (b2),
				(c1) -- [scalar] (b2),
				};				
			\end{feynman}		
		\end{tikzpicture}
		\qquad + \qquad
 		\begin{tikzpicture}[baseline=-0.6cm]
			\begin{feynman}
				\vertex (a1);
				\vertex [below=0.5cm of a1] (b1);
				\vertex [below=0.5cm of b1] (c1);
				\vertex [left=0.5cm of a1] (a2);
				\vertex [below=0.5cm of a2] (b2);
				\vertex [left=0.15cm of b2] (b2l);
				\vertex [above right=0.2cm of b2] (b2rt);
				\vertex [below right=0.2cm of b2] (b2rb);
				\vertex [below=0.5cm of b2] (c2);
				\vertex [left=0.8cm of a2] (a3);
				\vertex [below=0.5cm of a3] (b3);
				\vertex [right=0.15cm of b3] (b3r);
				\vertex [above left=0.2cm of b3] (b3lt);
				\vertex [below left=0.2cm of b3] (b3lb);
				\vertex [below=0.5cm of b3] (c3);
				\vertex [left=0.5cm of a3] (a4);
				\vertex [below=0.5cm of a4] (b4);
				\vertex [below=0.5cm of b4] (c4);
 				\node at (-0.8,  -1.5) {$ \G_{4,X}^2$};
				
				\diagram*{
				(a4) -- [scalar] (b3),
				(c4) -- [scalar] (b3),
				(b3) -- [photon] (b2),
				(b2) -- [scalar] (a1),
				(b2) -- [scalar] (c1),
				};				
			\end{feynman}		
		\end{tikzpicture}
 \caption{Diagrams contributing to the 4-point scattering amplitude for quartic Horndeski.
 \label{fig:tree_level}}
\end{figure}

\begin{figure}
 \centering
\qquad \qquad
     \begin{subfigure}[b]{0.3\textwidth}
         \centering
		\begin{tikzpicture}[baseline=-0.6cm]
			\begin{feynman}
				\vertex (a1);
				\vertex [below=0.5cm of a1] (b1);
				\vertex [below=0.5cm of b1] (c1);
				\vertex [left=0.5cm of a1] (a2);
				\vertex [below=0.5cm of a2] (b2);
				\vertex [left=0.15cm of b2] (b2l);
				\vertex [above right=0.2cm of b2] (b2rt);
				\vertex [below right=0.2cm of b2] (b2rb);
				\vertex [below=0.5cm of b2] (c2);
				\vertex [left=0.8cm of a2] (a3);
				\vertex [below=0.5cm of a3] (b3);
				\vertex [right=0.15cm of b3] (b3r);
				\vertex [above left=0.2cm of b3] (b3lt);
				\vertex [below left=0.2cm of b3] (b3lb);
				\vertex [below=0.5cm of b3] (c3);
				\vertex [left=0.5cm of a3] (a4);
				\vertex [below=0.5cm of a4] (b4);
				\vertex [below=0.5cm of b4] (c4);
 				\node at (0.6,  0.1) {$\phi (p_2)$};
  				\node at (0.6,  -1.1) {$\phi (p_4)$};
 				\node at (-2.4,  0.1) {$\psi (p_1)$};
 				\node at (-2.4,  -1.1) {$\psi (p_3)$};
    				\node at (-0.9,  -0.8) {\footnotesize $h_{\mu\nu}$};
				
				\diagram*{
				(a4) -- [] (b3),
				(c4) -- [] (b3),
				(b3) -- [photon] (b2),
				(b2) -- [scalar] (a1),
				(b2) -- [scalar] (c1),
				};				
			\end{feynman}		
		\end{tikzpicture}
         \caption{Horndeski frame.}
     \end{subfigure}
     \hfill
     \begin{subfigure}[b]{0.3\textwidth}
         \centering
			\begin{tikzpicture}[baseline=-0.6cm]
			\begin{feynman}
				\vertex (a1);
				\vertex [below=0.5cm of a1] (b1);
				\vertex [below=0.5cm of b1] (c1);
				\vertex [left=0.5cm of a1] (a2);
				\vertex [below=0.5cm of a2] (b2);
				\vertex [left=0.15cm of b2] (b2l);
				\vertex [above right=0.2cm of b2] (b2rt);
				\vertex [below right=0.2cm of b2] (b2rb);
				\vertex [below=0.5cm of b2] (c2);
				\vertex [left=0.5cm of a2] (a3);
				\vertex [below=0.5cm of a3] (b3);
				\vertex [right=0.15cm of b3] (b3r);
				\vertex [above left=0.2cm of b3] (b3lt);
				\vertex [below left=0.2cm of b3] (b3lb);
				\vertex [below=0.5cm of b3] (c3);
				\vertex [left=0.5cm of a3] (a4);
				\vertex [below=0.5cm of a4] (b4);
				\vertex [below=0.5cm of b4] (c4);
 				\node at (0.6,  0.1) {$\phi (p_2)$};
  				\node at (0.6,  -1.1) {$\phi (p_4)$};
 				\node at (-1.6,  0.1) {$\chi (p_1)$};
 				\node at (-1.6,  -1.1) {$\chi (p_3)$};
				
				\diagram*{
				(a3) -- [] (b2),
				(c3) -- [] (b2),
				(b2) -- [scalar] (a1),
				(b2) -- [scalar] (c1),
				};				
			\end{feynman}		
		\end{tikzpicture}
         \caption{Einstein frame.}
     \end{subfigure}
     \qquad\qquad
\caption{
Diagrams contributing to the $\phi \psi \to \phi \psi$ amplitude for quartic Horndeski, in both (a) the original frame \eqref{quarticA} and (b) following field redefinition \eqref{eqn:dis_redef_NL} (which replaces the scalar-tensor mixing with a direct disformal coupling between $\phi$ and matter).
}
\label{fig:phiMat}
\end{figure}
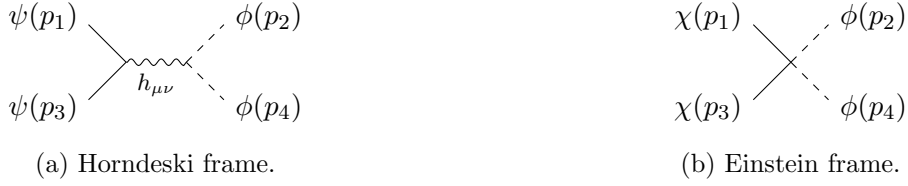

Furthermore, it was recently shown in \cite{deRham:2021fpu} that $\G_{4,X} = \beta_1/2$, the coefficient that controls the disformal coupling to matter, can be constrained by the positivity of scalar-matter scattering. 
For example for a scalar matter field $\psi$ (with canonical kinetic term and neglecting self-interactions),
\begin{align}
 T^{\mu\nu} &= \nabla^\mu \psi \nabla^\nu \psi - \frac{1}{2} g^{\mu\nu} \nabla^\alpha \psi \nabla_\alpha \psi \;\; ,
 \label{eqn:Tscalar}
\end{align} 
the $\phi \psi \to \phi \psi$ amplitude is given by the single Feynman diagram shown in Figure~\ref{fig:phiMat},
\begin{align}
\mathcal{A}_{\phi \psi \to \phi \psi} (s,t) =  \frac{ \beta_1  }{4 M_P \Lambda^3} ( s^2  + u^2 - t^2 ) \; , 
\end{align}
and so the forward-limit positivity bound \eqref{eqn:posLO} requires that $\beta_1 > 0$.

\paragraph{Positivity without Boosts.}
To constrain the higher order derivatives of $G_4$, one can consider scattering fluctuations about a non-trivial background, such as \eqref{eqn:pos_background}.
For a general $G_4 (X)$, this requires accounting for the gravitational mixing. 
For simplicity, we will first focus on the particular tuning \eqref{eqn:G4Raise}, for which the leading interaction is simply \eqref{eqn:SRaise}. 
Expanding around the background $\phi = \alpha \sqrt{M_P \Lambda^3}\,  t$ to quartic order in fluctuations, one finds the interactions,  
\begin{align}
\mathcal{L}_{N+1} \supset 
\left( 
c_4  ( \partial \varphi )^2   + d_4  \dot\varphi^2 
  \right) \frac{  \varphi^\mu_\mu \varphi^\nu_\nu - \varphi^\mu_\nu \varphi^\nu_\mu }{ \Lambda^6 }
\end{align}
where,
\begin{align}
c_4 = N (N+1)   \beta_{N+1} (- \alpha^2 )^{N-1}  \;\;\;\; \text{and}  \;\;\;\; d_4 =  - 2 N (N^2 - 1)  \beta_{N+1} (-  \alpha^2 )^{N-1} \;  , 
\label{eqn:c4d4}
\end{align}
for the leading interaction \eqref{eqn:SRaise}.
The corresponding $2\to2$ scattering amplitude is,
\begin{align}
\Lambda^6 \, \mathcal{A}_{\varphi \varphi \to \varphi \varphi} 
= 
- 3 c_4 s t u + 
 d_4 (  \omega_s^2 t u + s \omega_t^2 u + s t \omega_u^2) \; , 
\label{eqn:pos_AGal}
\end{align}
where we have used that\footnote{
In general, $Z$ and $c_s$ are set by the $G_2 (X)$ function, and are only sensitive to $G_4 (X)$ through the metric redefinition which removes the cubic $h_{\mu\nu} \varphi \varphi$ coupling since $\phi^\mu_\nu$ vanishes on this background. 
} $Z = 1 + \mathcal{O} (\alpha^2)$ and $c_s^2 = 1 + \mathcal{O} ( \alpha^2 )$ at small $\alpha$. 
Since $\partial_s^2 \mathcal{A} |_{\substack{t=0 \\ \omega_t =0 }} = 0$, the forward limit positivity bound \eqref{eqn:posLO2} constrains only subleading operators (as in the Lorentz-invariant case), and to place constraints on $\beta_{N+1}$ we much go beyond the forward limit and use positivity bound \eqref{eqn:posNLO2}. 
For the amplitude \eqref{eqn:pos_AGal}, the forward limits vanish, $\partial_s^2 \mathcal{I} |_{t=0} = \partial_\delta \partial_s^2 \mathcal{I} |_{t=0} = 0$, and so \eqref{eqn:posNLO2} gives,
\begin{align}
\partial_t \partial_s^2 \mathcal{I} |_{\substack{ t=0} } = 4 E_2  (3 c_4 - d_4 \gamma )  > 0 
\label{eqn:dt_pos}
\end{align}
which must be satisfied for all $\gamma \geq 1 + E_2/\omega_b \geq 1$, where $\omega_b$ is the energy scale up to which the branch cut can be subtracted within the EFT. 
Applied to \eqref{eqn:c4d4}, this bound becomes,
\begin{align}
 (-1)^{N+1} \beta_{N+1} > 0 \; , 
\end{align}
as reported in the main text.

\paragraph{General $G_4 (X)$ Bounds.}
Focussing on a Horndeski theory with the particularly simple $G_4 (X)$ given in \eqref{eqn:G4sq} and working to leading order in $\alpha^2$ has allowed us to completely remove the contributions from graviton exchange using a field redefinition.
For a general $G_4 (X)$ at finite $\alpha$, this is no longer possible: for instance, after the field redefinition,
\begin{align}
\tilde{g}_{\mu\nu} = g_{\mu\nu} + \frac{2 \hat{G}_{4,X} }{ \hat{G}_4 + 2 \alpha^2 \hat{G}_{4,X}  } \phi_\mu \phi_\nu
\end{align}
where the $\hat{G}_4$ functions are evaluated at $X = -\alpha^2$, the $h_{\mu\nu} \tilde{I}^{\mu\nu}$ interaction is removed but there remains a cubic $h \varphi \varphi$ vertex from the $h_{\mu\nu} \tilde{J}^{\mu\nu}$ interaction, which schematically looks like,
\begin{align}
 \mathcal{L} \supset \,   \alpha^2 \left( \frac{ \hat{G}_{4,X}^2 + \hat{G}_4 \hat{G}_{4,XX}  }{ \hat{G}_4 } \right) \,   \frac{ h \,  \ddot{\varphi} \, \partial^2 \varphi }{ \Lambda^3 }  \,  .
\end{align}
This will contribute to the $\varphi \varphi \to \varphi \varphi$ and $\varphi \psi \to \varphi \psi$ amplitudes via a graviton exchange diagram, which schematically will lead to corrections to the positivity bounds at finite $\alpha$,
\begin{align}
 \hat{G}_{4,XX}  +  \frac{ \hat{G}_{4,X}^2 }{ \hat{G}_4 }   <  \mathcal{O} \left(  \frac{ \alpha^4 }{ \hat{G}_4 } \left( \hat{G}_{4,XX}  +  \frac{ \hat{G}_{4,X}^2 }{ \hat{G}_4 }  \right)^2   \right) \; ,
 \label{eqn:pos_G4XX}  \\
 \frac{ \hat{G}_{4,X} }{ \hat{G}_4 }   < \mathcal{O} \left( \frac{\alpha^2}{\hat{G}_4} \left( \hat{G}_{4,XX}  +  \frac{ \hat{G}_{4,X}^2 }{ \hat{G}_4 }  \right) \right) \; . 
 \label{eqn:pos_G4X}
\end{align}
We do not compute these corrections explicitly because they are the same order as the corrections we have neglected from the spacetime curvature and from the $\varphi$ wavefunction normalisation $Z$ and sound speed $c_s$ on this background. 
However, crucially we see that in the $\varphi \psi \to \varphi \psi$ bound \eqref{eqn:pos_G4X} there are corrections which $\sim \alpha^2 \hat{G}_{4,XX}$ to the bound on $\hat{G}_{4,X}$. 
This resolves an otherwise puzzling discrepancy. Consider the theory \eqref{eqn:G4sq} with $\beta_1 = 0$, namely $G_4 = \sqrt{1- \beta_{N+1} X^{N+1}}$. 
Had we taken the flat space bounds $\bar{G}_{4,XX} + \bar{G}_{4,X}^2/\bar{G}_4 < 0$ and $\bar{G}_{4,X} < 0$ and naively applied them on the background $X= - \alpha^2$, we would have concluded that $(- 1)^{N+1} \beta_{N+1} > 0$ and $(- 1)^{N+1} \beta_{N+1} < 0$ respectively! We see from \eqref{eqn:pos_G4X} that in fact demanding $\hat{G}_{4,X} < 0$ is only justified when $\hat{G}_{4,X} \gg  \alpha^2 \hat{G}_{4,XX}$ at small $\alpha$, which is not true when $\beta_1 = 0$, and therefore actually the only positivity requirement in this case is $(- 1)^{N+1} \beta_{N+1} > 0$ from the bound \eqref{eqn:pos_G4XX}.

~\\
This concludes our discussion of the positivity bounds placed on $P(X)$ and quartic Horndeski theories from scattering fluctuations about the background $\phi \; \propto \; \alpha t$ for small $\alpha$. 
It would be particularly interesting to consider other backgrounds for $\phi$, such as the galileid \cite{Nicolis:2015sra} or a large value of $\alpha$ which drives an expanding cosmology, and use the positivity bounds \eqref{eqn:posLO2} and \eqref{eqn:posNLO2} to place further constraints on this scalar-tensor theory.

\section{Disformal Field Redefinition}
\label{app:disformal}

In the interest of a self-contained presentation, in this Appendix we describe the disformal field redefinition,
\begin{align}
 g_{\mu\nu} = \tilde{g}_{\mu\nu} - D \;  \frac{ \phi_\mu \phi_\nu}{ M_P \Lambda^3} \; ,  \label{eqn:dis_redef}
\end{align}
in more detail. We will only require results for constant $D$, and a more general ($X$ and $\phi$ dependent) transformation can be found in \cite{Zumalacarregui:2012us, Bettoni:2013diz, Zumalacarregui:2013pma, Achour:2016rkg}. 

The inverse metric is,
\begin{align}
g^{\mu\nu} =  \tilde{g}^{\mu\nu}  +   D \, ( 1 + D X )  \,  \frac{ \tilde{\phi}^\mu \tilde{\phi}^\nu}{ M_P \Lambda^3} \; ,
\end{align}
where the tilde on the $\phi$ emphasises that its index should be raised with the $\tilde{g}_{\mu\nu}$ metric, $\tilde{\phi}^\mu = \tilde{g}^{\mu\nu} \partial_\mu \phi = \phi^\mu / (1+ D X)$, and therefore\footnote{
Note that we will not introduce a $\tilde{X}$ variable, but rather express $\tilde{g}^{\mu\nu} \phi_\mu \phi_\nu$ where it appears in terms of the original $X = g^{\mu\nu} \phi_\mu \phi_\nu / M_P \Lambda^3$. 
} $\tilde{g}^{\mu\nu} \phi_\mu \phi_\nu = X / (1 + D X )$. 
The second derivatives are also simply rescaled, $\phi_{\mu\nu} = ( 1 + D X ) \tilde{\phi}_{\mu\nu}$, and the tensors appearing in the equations of motion can be written in terms of the $\tilde{g}_{\mu\nu}$ metric as,
\begin{align}
I^{\mu\nu} &=  ( 1 + D X )^2  \tilde{I}^{\mu\nu}  +   2 D ( 1 + D X )^3  \tilde{J}^{\mu\nu}  \;  ,  \nonumber \\
J^{\mu\nu} &=  ( 1 + D X )^4  \tilde{J}^{\mu\nu}  \; ,  \nonumber \\
G^{\mu\nu} &= \tilde{G}^{\mu\nu} -  \frac{D}{2} (1 + D X )  \tilde{I}^{\mu\nu} - \frac{D^2}{2}  (1  + D X )^2  \tilde{J}^{\mu\nu}  \; . 
\end{align}

\paragraph{Metric Equation of Motion.}
The metric equation of motion $\delta S / \delta g_{\mu\nu}$ can therefore be written as,
\begin{align}
\frac{1}{ M_P \Lambda^3} \frac{\delta S}{\delta g_{\mu \nu} } &=  - G_4  \frac{ M_P  \tilde{G}^{\mu\nu} }{\Lambda^3} 
+  \frac{G_2}{2} \tilde{g}^{\mu\nu} -  ( 1 + D X )^{5/2} \partial_X \left(  \frac{ G_2 }{ \sqrt{1 + D X} }   \right)  \frac{ \tilde{\phi}^\mu \tilde{\phi}^\nu}{ M_P \Lambda^3 }  \nonumber \\ 
& - \left(1 + D X \right)^{5/2}   \partial_{ X }  \sqrt{ \frac{G_4^2 }{ 1  + D  X } }   \frac{ \tilde{I}^{\mu\nu}}{\Lambda^6}  
  - 2 (1 + D X )^{7/2} \,  \partial_{ X }^2 \sqrt{  G_4^2  \left( 1 + D  X \right) }  \frac{ \tilde{J}^{\mu\nu}}{M_P \Lambda^{9}}  
\label{eqn:metric_eom_NL}
\end{align}
As claimed in the main text, the leading interactions at $\Lambda$ can be removed by fixing $D = 2 \bar{G}_{4,X} / \bar{G}_4 = - \beta_1$.

\paragraph{Scalar Equation of Motion.}
The scalar equation of motion $\delta S / \delta \phi = 0$ given in \eqref{eqn:phi_eom_NL} can be written as,
\begin{align}
\left[ Z \,  \tilde{g}^{\mu\nu} 
+  Z' \frac{ \tilde{\phi}^\mu \tilde{\phi}^\nu}{ M_P \Lambda^3}
 -   2  (1 + D X )^3 \partial_X^2 G_{4}^2  \frac{ \tilde{I}^{\mu\nu} }{\Lambda^6} 
 - \frac{4}{3}  (1 + D X)^4 \partial_X^3 \left[  G_4^2    (1 + D X ) \right]  \frac{ \tilde{J}^{\mu\nu}}{M_P \Lambda^{9}}   \right] \tilde{\phi}_{\mu\nu}  = 0 
 \label{eqn:scalar_eom_D}
\end{align}
where the coefficients of the quadratic terms are,
\begin{align}
Z &=  - 2  (1 + D X) G_4^2 \partial_X \left( \frac{G_2}{G_{4}}  \right)  \;\; , \\
Z' &=  - 4 (1 + D X)^3 \partial_X \left( G_2 G_{4,X} \right) + D (1 + D X ) Z  \; .
\end{align}
Using \eqref{eqn:scalar_eom_D}, it is straightforward to write down the next-to-leading order interactions at $\mathcal{O} \left( M_P \Lambda^9  \right)$, since when $D = -\beta_1$ the metric equation of motion sets $\tilde{G}^{\mu\nu} \sim \partial^6 \phi^4/ M_P^2 \Lambda^6$ and so we can simply replace $\nabla_\mu$ with $\partial_\mu$ at this order,
\begin{align}
\partial_\mu \left[ 
 \tilde{\phi}^\mu 
 - \frac{4}{\Lambda^6} \delta^{\mu \alpha  \rho}_{\nu \beta \sigma} \phi_\nu \phi_\alpha^\beta \phi_\rho^\sigma \left( 
  \beta_2  +  \left(  \beta_3  + 3 \beta_1 \beta_2  \right)  X  
 \right)
  \right] + \mathcal{O} \left( \frac{ \partial^{10} \phi^7 }{ M_P^2 \Lambda^{12} }  \right)  = 0  \; ,
  \label{eqn:phi_eom_NLO}
\end{align}
where we have canonically normalised $\bar{G}_{2,X} = - 1/2$.
Note that in terms of $G_4 (X)$, the coefficient of the next-to-leading interaction is,
\begin{align}
 \beta_3 + 3 \beta_1 \beta_2 = \frac{1}{3} \left(  \G_{4,XXX} + 9 \G_{4,X} \G_{4,XX} + 6 \G_{4,X}^3   \right) \; ,
\end{align} 
which corresponds to the three diagrams shown in Figure~\ref{fig:WBG_NLO}. 
We see that once $\beta_2 = 0$ is tuned to remove the leading interactions at $\Lambda^6$, then it is the $\beta_3$ coupling which determines the next-to-leading corrections at $M_P \Lambda^9$.

\begin{figure}
\centering
$A_6 $
		\qquad = \quad
		\begin{tikzpicture}[baseline=-0.6cm]
			\begin{feynman}
				\vertex (a1);
				\vertex [below=0.5cm of a1] (b1);
				\vertex [below=0.5cm of b1] (c1);
				\vertex [left=0.5cm of a1] (a2);
				\vertex [below=0.5cm of a2] (b2);
				\vertex [below=0.5cm of b2] (c2);
				\vertex [left=0.5cm of a2] (a3);
				\vertex [below=0.5cm of a3] (b3);
				\vertex [below=0.5cm of b3] (c3);
				\vertex [left=0.5cm of a3] (a4);
				\vertex [below=0.5cm of a4] (b4);
				\vertex [below=0.5cm of b4] (c4);
				\node at (-0.5,  -1.5) {$\G_{4,XXX}$};
				
				\diagram*{
				(a3) -- [scalar] (b2),
				(c3) -- [scalar] (b2),
				(a1) -- [scalar] (b2),
				(c1) -- [scalar] (b2),
				(b1) -- [scalar] (b2),
				(b3) -- [scalar] (b2)
				};				
			\end{feynman}		
		\end{tikzpicture}
		\qquad + \qquad
 		\begin{tikzpicture}[baseline=-0.6cm]
			\begin{feynman}
				\vertex (a1);
				\vertex [below=0.5cm of a1] (b1);
				\vertex [below=0.5cm of b1] (c1);
				\vertex [left=0.5cm of a1] (a2);
				\vertex [below=0.5cm of a2] (b2);
				\vertex [left=0.15cm of b2] (b2l);
				\vertex [above right=0.2cm of b2] (b2rt);
				\vertex [below right=0.2cm of b2] (b2rb);
				\vertex [below=0.5cm of b2] (c2);
				\vertex [left=0.8cm of a2] (a3);
				\vertex [below=0.5cm of a3] (b3);
				\vertex [right=0.15cm of b3] (b3r);
				\vertex [above left=0.2cm of b3] (b3lt);
				\vertex [below left=0.2cm of b3] (b3lb);
				\vertex [below=0.5cm of b3] (c3);
				\vertex [left=0.5cm of a3] (a4);
				\vertex [below=0.5cm of a4] (b4);
				\vertex [below=0.5cm of b4] (c4);
 				\node at (-0.8,  -1.5) {$  \G_{4,X} \G_{4,XX}$};
				
				\diagram*{
				(a4) -- [scalar] (b3),
				(c4) -- [scalar] (b3),
				(b3) -- [photon] (b2),
				(b2) -- [scalar] (a1),
				(b2) -- [scalar] (c1),
				(a3) -- [scalar] (b3),
				(c3) --[scalar] (b3)
				};				
			\end{feynman}		
		\end{tikzpicture}
		\qquad + \qquad 
		 		\begin{tikzpicture}[baseline=-0.6cm]
			\begin{feynman}
				\vertex (a1);
				\vertex [below=0.5cm of a1] (b1);
				\vertex [above =0.2cm of b1] (b1rt);
				\vertex [below =0.2cm of b1] (b1rb);
				\vertex [below=0.5cm of b1] (c1);
				\vertex [left=0.5cm of a1] (a2);
				\vertex [below=0.5cm of a2] (b2);
				\vertex [left=0.15cm of b2] (b2l);
				\vertex [above =0.2cm of b2] (b2rt);
				\vertex [below =0.2cm of b2] (b2rb);
				\vertex [below=0.5cm of b2] (c2);
				\vertex [left=0.8cm of a2] (a3);
				\vertex [below=0.5cm of a3] (b3);
				\vertex [right=0.15cm of b3] (b3r);
				\vertex [above left=0.2cm of b3] (b3lt);
				\vertex [below left=0.2cm of b3] (b3lb);
				\vertex [below=0.5cm of b3] (c3);
				\vertex [left=0.5cm of a3] (a4);
				\vertex [below=0.5cm of a4] (b4);
				\vertex [below=0.5cm of b4] (c4);
 				\node at (-0.8,  -1.5) {$  \G_{4,X}^3$};
				
				\diagram*{
				(a4) -- [scalar] (b3),
				(c4) -- [scalar] (b3),
				(b2rt) -- [scalar] (a1),
				(b2rt) -- [scalar] (b1rt),
				(b2rb) -- [scalar] (b1rb), 
				(b2rb) -- [scalar] (c1),
				(b3) -- [photon] (b2rt),
				(b3) -- [photon] (b2rb)
				};				
			\end{feynman}		
		\end{tikzpicture}
 \caption{Diagrams contributing to the 6-point scattering amplitude for quartic Horndeski.
 \label{fig:WBG_NLO}}
\end{figure}

\bibliographystyle{apsrev4-1}
\bibliography{DBI_Galileon_References}

\end{document}